\documentclass[aps,prd,twocolumn,showpacs,superscriptaddress,nofootinbib,floatfix,letterpaper]{revtex4-2}
\usepackage{multirow}
\usepackage{amsmath}
\usepackage{graphicx}
\usepackage{color}
\usepackage{hyperref}
\usepackage{tabularx}
\usepackage{lipsum}
\usepackage{float}
\usepackage{hhline}
\usepackage{caption}
\usepackage{mathrsfs}
\usepackage{pgffor} 
\usepackage{xparse} 
\usepackage{comment}
\usepackage{xcolor}
\usepackage{lineno}

\usepackage{hyperref}

\newcommand{\sizecheck}{0} 
\newcommand{\PRDsupp}{1}   
\ifnum\PRDsupp=0
  
\else
  
\fi

\newif\ifpdf
\ifx\pdfoutput\undefined
   \pdffalse
\else
   \pdfoutput=1
   \pdftrue
\fi
\ifpdf
   \usepackage{graphicx}
   \usepackage{epstopdf}
\else
   \usepackage{graphicx}
\fi

\begin{document}
\title{Sensitivity to Sterile Neutrinos with Elastic Neutrino-Electron and Highly Elastic Neutrino-Nucleus Scattering in the DUNE Near Detector}

%

\newcommand{\Albanysuny}{University of Albany, SUNY, Albany, NY 12222, USA}
\newcommand{\Aligarh}{Aligarh Muslim University, Aligarh-202002, India}
\newcommand{\Almaty}{Institute of Nuclear Physics at Almaty, Almaty 050032, Kazakhstan }
\newcommand{\Amsterdam}{University of Amsterdam, NL-1098 XG Amsterdam, The Netherlands}
\newcommand{\Antalya}{Antalya Bilim University, 07190 D{\"o}{\c{s}}emealt{\i}/Antalya, Turkey}
\newcommand{\Antananarivo}{University of Antananarivo, Antananarivo 101, Madagascar}
\newcommand{\Antioquia}{University of Antioquia, Medell{\'\i}n, Colombia}
\newcommand{\AntonioNarino}{Universidad Antonio Nari{\~n}o, Bogot{\'a}, Colombia}
\newcommand{\Argonne}{Argonne National Laboratory, Argonne, IL 60439, USA}
\newcommand{\Arizona}{University of Arizona, Tucson, AZ 85721, USA}
\newcommand{\Asuncion}{Universidad Nacional de Asunci{\'o}n, San Lorenzo, Paraguay}
\newcommand{\Athens}{University of Athens, Zografou GR 157 84, Greece}
\newcommand{\Atlantico}{Universidad del Atl{\'a}ntico, Barranquilla, Atl{\'a}ntico, Colombia}
\newcommand{\Augustana}{Augustana University, Sioux Falls, SD 57197, USA}
\newcommand{\Bern}{University of Bern, CH-3012 Bern, Switzerland}
\newcommand{\Beykent}{Beykent University, Istanbul, Turkey}
\newcommand{\BolognaUniversity}{Universit{\`a} di Bologna, 40127 Bologna, Italy}
\newcommand{\Boston}{Boston University, Boston, MA 02215, USA}
\newcommand{\Bristol}{University of Bristol, Bristol BS8 1TL, United Kingdom}
\newcommand{\Brookhaven}{Brookhaven National Laboratory, Upton, NY 11973, USA}
\newcommand{\Bucharest}{University of Bucharest, Bucharest, Romania}
\newcommand{\CalBerkeley}{University of California Berkeley, Berkeley, CA 94720, USA}
\newcommand{\CalDavis}{University of California Davis, Davis, CA 95616, USA}
\newcommand{\CalIrvine}{University of California Irvine, Irvine, CA 92697, USA}
\newcommand{\CalRiverside}{University of California Riverside, Riverside CA 92521, USA}
\newcommand{\CalSantabarbara}{University of California Santa Barbara, Santa Barbara, CA, 93106, USA}
\newcommand{\Caltech}{California Institute of Technology, Pasadena, CA 91125, USA}
\newcommand{\Cambridge}{University of Cambridge, Cambridge CB3 0HE, United Kingdom}
\newcommand{\Campinas}{Universidade Estadual de Campinas, Campinas - SP, 13083-970, Brazil}
\newcommand{\CataniaUniversitadi}{Universit{\`a} di Catania, 2 - 95131 Catania, Italy}
\newcommand{\Catolica}{Universidad Cat{\'o}lica del Norte, Antofagasta, Chile}
\newcommand{\CBPF}{Centro Brasileiro de Pesquisas F\'isicas, Rio de Janeiro, RJ 22290-180, Brazil}
\newcommand{\CEASaclay}{IRFU, CEA, Universit{\'e} Paris-Saclay, F-91191 Gif-sur-Yvette, France}
\newcommand{\CERN}{CERN, The European Organization for Nuclear Research, 1211 Meyrin, Switzerland}
\newcommand{\Charles}{Institute of Particle and Nuclear Physics of the Faculty of Mathematics and Physics of the Charles University, 180 00 Prague 8, Czech Republic }
\newcommand{\Chicago}{University of Chicago, Chicago, IL 60637, USA}
\newcommand{\ChungAng}{Chung-Ang University, Seoul 06974, South Korea}
\newcommand{\CIEMAT}{CIEMAT, Centro de Investigaciones Energ{\'e}ticas, Medioambientales y Tecnol{\'o}gicas, E-28040 Madrid, Spain}
\newcommand{\Cincinnati}{University of Cincinnati, Cincinnati, OH 45221, USA}
\newcommand{\Cinvestav}{Centro de Investigaci{\'o}n y de Estudios Avanzados del Instituto Polit{\'e}cnico Nacional (Cinvestav), Mexico City, Mexico}
\newcommand{\ColoradoBoulder}{University of Colorado Boulder, Boulder, CO 80309, USA}
\newcommand{\ColoradoState}{Colorado State University, Fort Collins, CO 80523, USA}
\newcommand{\Columbia}{Columbia University, New York, NY 10027, USA}
\newcommand{\conida}{Comisi{\'o}n Nacional de Investigaci{\'o}n y Desarrollo Aeroespacial, Lima, Peru}
\newcommand{\Cti}{Centro de Tecnologia da Informa{\c{c}}{\~a}o Renato Archer, Amarais - Campinas, SP - CEP 13069-901, Brazil}
\newcommand{\CUSB}{Central University of South Bihar, Gaya, 824236, India }
\newcommand{\CzechAcademyofSciences}{Institute of Physics, Czech Academy of Sciences, 182 00 Prague 8, Czech Republic}
\newcommand{\CzechTechnical}{Czech Technical University, 115 19 Prague 1, Czech Republic}
\newcommand{\DannecyleVieux}{Laboratoire d{\textquoteright}Annecy de Physique des Particules, Universit{\'e} Savoie Mont Blanc, CNRS, LAPP-IN2P3, 74000 Annecy, France}
\newcommand{\Daresbury}{Daresbury Laboratory, Cheshire WA4 4AD, United Kingdom}
\newcommand{\Dordt}{Dordt University, Sioux Center, IA 51250, USA}
\newcommand{\drew}{Drew University, Madison, NJ 07940, USA}
\newcommand{\Drexel}{Drexel University, Philadelphia, PA 19104, USA}
\newcommand{\Duke}{Duke University, Durham, NC 27708, USA}
\newcommand{\Durham}{Durham University, Durham DH1 3LE, United Kingdom}
\newcommand{\Edinburgh}{University of Edinburgh, Edinburgh EH8 9YL, United Kingdom}
\newcommand{\EIA}{Universidad EIA, Envigado, Antioquia, Colombia}
\newcommand{\Eotvos}{E{\"o}tv{\"o}s Lor{\'a}nd University, 1053 Budapest, Hungary}
\newcommand{\erciyes}{Erciyes University, Kayseri, Turkey}
\newcommand{\FCUL}{Faculdade de Ci{\^e}ncias da Universidade de Lisboa, 1749-016 Lisboa, Portugal}
\newcommand{\FederaldeAlfenas}{Universidade Federal de Alfenas, Po{\c{c}}os de Caldas - MG, 37715-400, Brazil}
\newcommand{\FederaldeGoias}{Universidade Federal de Goias, Goiania, GO 74690-900, Brazil}
\newcommand{\FederaldoABC}{Universidade Federal do ABC, Santo Andr{\'e} - SP, 09210-580, Brazil}
\newcommand{\FederaldoRio}{Universidade Federal do Rio de Janeiro,  Rio de Janeiro - RJ, 21941-901, Brazil}
\newcommand{\Fermi}{Fermi National Accelerator Laboratory, Batavia, IL 60510, USA}
\newcommand{\Ferrarauniv}{University of Ferrara, Ferrara, Italy}
\newcommand{\Florida}{University of Florida, Gainesville, FL 32611-8440, USA}
\newcommand{\Floridastate}{Florida State University, Tallahassee, FL, 32306, USA}
\newcommand{\Fluminense}{Fluminense Federal University, 9 Icara{\'\i} Niter{\'o}i - RJ, 24220-900, Brazil }
\newcommand{\GATech}{Georgia Institute of Technology, North Avenue Atlanta, GA 30332, USA}
\newcommand{\Genova}{Universit{\`a} degli Studi di Genova, Genova, Italy}
\newcommand{\Georgian}{Georgian Technical University, Tbilisi, Georgia}
\newcommand{\Granada}{University of Granada \& CAFPE, 18002 Granada, Spain}
\newcommand{\GranSasso}{Gran Sasso Science Institute, L'Aquila, Italy}
\newcommand{\GranSassoLab}{Laboratori Nazionali del Gran Sasso, L'Aquila AQ, Italy}
\newcommand{\Grenoble}{University Grenoble Alpes, CNRS, Grenoble INP, LPSC-IN2P3, 38000 Grenoble, France}
\newcommand{\Guanajuato}{Universidad de Guanajuato, Guanajuato, C.P. 37000, Mexico}
\newcommand{\Harish}{Harish-Chandra Research Institute, Jhunsi, Allahabad 211 019, India}
\newcommand{\Hawaii}{University of Hawaii, Honolulu, HI 96822, USA}
\newcommand{\hkust}{Hong Kong University of Science and Technology, Kowloon, Hong Kong, China}
\newcommand{\Houston}{University of Houston, Houston, TX 77204, USA}
\newcommand{\Hyderabad}{University of Hyderabad, Gachibowli, Hyderabad - 500 046, India}
\newcommand{\Idaho}{Idaho State University, Pocatello, ID 83209, USA}
\newcommand{\IFIC}{Instituto de F{\'\i}sica Corpuscular, CSIC and Universitat de Val{\`e}ncia, 46980 Paterna, Valencia, Spain}
\newcommand{\IGFAE}{Instituto Galego de F{\'\i}sica de Altas Enerx{\'\i}as, University of Santiago de Compostela, Santiago de Compostela, 15782, Spain}
\newcommand{\ihep}{Institute of High Energy Physics, Chinese Academy of Sciences, Beijing, China}
\newcommand{\IISc}{Indian Institute of Science, Bengaluru, CV Raman Road, Bengaluru, Karnataka, 560012, India}
\newcommand{\Iitk}{Indian Institute of Technology Kanpur, Uttar Pradesh 208016, India}
\newcommand{\Illinoisinstitute}{Illinois Institute of Technology, Chicago, IL 60616, USA}
\newcommand{\Imperial}{Imperial College of Science Technology and Medicine, London SW7 2BZ, United Kingdom}
\newcommand{\IndGuwahati}{Indian Institute of Technology Guwahati, Guwahati, 781 039, India}
\newcommand{\IndHyderabad}{Indian Institute of Technology Hyderabad, Hyderabad, 502285, India}
\newcommand{\Indiana}{Indiana University, Bloomington, IN 47405, USA}
\newcommand{\INFNBologna}{Istituto Nazionale di Fisica Nucleare Sezione di Bologna, 40127 Bologna BO, Italy}
\newcommand{\INFNCatania}{Istituto Nazionale di Fisica Nucleare Sezione di Catania, I-95123 Catania, Italy}
\newcommand{\INFNFerrara}{Istituto Nazionale di Fisica Nucleare Sezione di Ferrara, I-44122 Ferrara, Italy}
\newcommand{\INFNFrascati}{Istituto Nazionale di Fisica Nucleare Laboratori Nazionali di Frascati, Frascati, Roma, Italy}
\newcommand{\INFNGenova}{Istituto Nazionale di Fisica Nucleare Sezione di Genova, 16146 Genova GE, Italy}
\newcommand{\INFNLecce}{Istituto Nazionale di Fisica Nucleare Sezione di Lecce, 73100 - Lecce, Italy}
\newcommand{\INFNMilanBicocca}{Istituto Nazionale di Fisica Nucleare Sezione di Milano Bicocca, 3 - I-20126 Milano, Italy}
\newcommand{\INFNMilano}{Istituto Nazionale di Fisica Nucleare Sezione di Milano, 20133 Milano, Italy}
\newcommand{\INFNNapoli}{Istituto Nazionale di Fisica Nucleare Sezione di Napoli, I-80126 Napoli, Italy}
\newcommand{\INFNPadova}{Istituto Nazionale di Fisica Nucleare Sezione di Padova, 35131 Padova, Italy}
\newcommand{\INFNPavia}{Istituto Nazionale di Fisica Nucleare Sezione di Pavia,  I-27100 Pavia, Italy}
\newcommand{\INFNPisa}{Istituto Nazionale di Fisica Nucleare Laboratori Nazionali di Pisa, Pisa PI, Italy}
\newcommand{\INFNRoma}{Istituto Nazionale di Fisica Nucleare Sezione di Roma, 00185 Roma RM, Italy}
\newcommand{\INFNRomavergata}{Istituto Nazionale di Fisica Nucleare Roma Tor Vergata , 00133 Roma RM, Italy}
\newcommand{\INFNSud}{Istituto Nazionale di Fisica Nucleare Laboratori Nazionali del Sud, 95123 Catania, Italy}
\newcommand{\Infntorino}{Istituto Nazionale di Fisica Nucleare, Sezione di Torino, Turin, Italy}
\newcommand{\Ingenieria}{Universidad Nacional de Ingenier{\'\i}a, Lima 25, Per{\'u}}
\newcommand{\Insubria}{University of Insubria, Via Ravasi, 2, 21100 Varese VA, Italy}
\newcommand{\Iowa}{University of Iowa, Iowa City, IA 52242, USA}
\newcommand{\IowaState}{Iowa State University, Ames, Iowa 50011, USA}
\newcommand{\IPLyon}{Institut de Physique des 2 Infinis de Lyon, 69622 Villeurbanne, France}
\newcommand{\IPM}{Institute for Research in Fundamental Sciences, Tehran, Iran}
\newcommand{\IRLPPC}{Particle Physics and Cosmology International Research Laboratory	, CNRS - Universit{\'e} Paris Cit{\'e} - The University of Chicago, Chicago IL,  60637, USA}
\newcommand{\IST}{Instituto Superior T{\'e}cnico da Universidade de Lisboa, Universidade de Lisboa, 1049-001 Lisboa, Portugal}
\newcommand{\Ita}{Instituto Tecnol{\'o}gico de Aeron{\'a}utica, Sao Jose dos Campos, Brazil}
\newcommand{\ITPMYerevan}{Institute for Theoretical Physics and Modeling, Yerevan 0036, Armenia}
\newcommand{\Iwate}{Iwate University, Morioka, Iwate 020-8551, Japan}
\newcommand{\Jacksonstate}{Jackson State University, Jackson, MS 39217, USA}
\newcommand{\Jawaharlal}{Jawaharlal Nehru University, New Delhi 110067, India}
\newcommand{\Jeonbuk}{Jeonbuk National University, Jeonrabuk-do 54896, South Korea}
\newcommand{\Jyvaskyla}{Jyv{\"a}skyl{\"a} University, FI-40014 Jyv{\"a}skyl{\"a}, Finland}
\newcommand{\Kansasstate}{Kansas State University, Manhattan, KS 66506, USA}
\newcommand{\Kavli}{Kavli Institute for the Physics and Mathematics of the Universe, Kashiwa, Chiba 277-8583, Japan}
\newcommand{\KEK}{High Energy Accelerator Research Organization (KEK), Ibaraki, 305-0801, Japan}
\newcommand{\KISTI}{Korea Institute of Science and Technology Information, Daejeon, 34141, South Korea}
\newcommand{\Kyiv}{Taras Shevchenko National University of Kyiv, 01601 Kyiv, Ukraine}
\newcommand{\Lancaster}{Lancaster University, Lancaster LA1 4YB, United Kingdom}
\newcommand{\LawrenceBerkeley}{Lawrence Berkeley National Laboratory, Berkeley, CA 94720, USA}
\newcommand{\LIP}{Laborat{\'o}rio de Instrumenta{\c{c}}{\~a}o e F{\'\i}sica Experimental de Part{\'\i}culas, 1649-003 Lisboa, 3004-516 Coimbra, and 4710-057 Braga, Portugal}
\newcommand{\Liverpool}{University of Liverpool, L69 7ZE, Liverpool, United Kingdom}
\newcommand{\LNA}{National Laboratory for Astrophysics, Rua dos Estados Unidos, 154 Bairro das Na{\c{c}}{\~o}es Itajub{\'a} / MG - 37.504-364, Brasil}
\newcommand{\LosAlmos}{Los Alamos National Laboratory, Los Alamos, NM 87545, USA}
\newcommand{\Louisanastate}{Louisiana State University, Baton Rouge, LA 70803, USA}
\newcommand{\LpBordeaux}{Laboratoire de Physique des Deux Infinis Bordeaux - IN2P3, F-33175 Gradignan, Bordeaux, France, }
\newcommand{\Lucknow}{University of Lucknow, Uttar Pradesh 226007, India}
\newcommand{\Mainz}{Johannes Gutenberg-Universit{\"a}t Mainz, 55122 Mainz, Germany}
\newcommand{\Manchester}{University of Manchester, Manchester M13 9PL, United Kingdom}
\newcommand{\Marmara}{Marmara University, Marmara {\"U}niversitesi G{\"o}ztepe Yerle{\c{s}}kesi 34722 Kad{\i}k{\"o}y - {\.I}stanbul, Turkey}
\newcommand{\Massinsttech}{Massachusetts Institute of Technology, Cambridge, MA 02139, USA}
\newcommand{\Medellin}{University of Medell{\'\i}n, Medell{\'\i}n, 050026, Colombia }
\newcommand{\Michigan}{University of Michigan, Ann Arbor, MI 48109, USA}
\newcommand{\Michiganstate}{Michigan State University, East Lansing, MI 48824, USA}
\newcommand{\MilanoBicocca}{Universit{\`a} di Milano Bicocca , 20126 Milano, Italy}
\newcommand{\MilanoUniv}{Universit{\`a} degli Studi di Milano, I-20133 Milano, Italy}
\newcommand{\Minho}{Universidade do Minho, R. da Universidade, 4710-057 Braga, Portugal}
\newcommand{\Minnduluth}{University of Minnesota Duluth, Duluth, MN 55812, USA}
\newcommand{\Minntwin}{University of Minnesota Twin Cities, Minneapolis, MN 55455, USA}
\newcommand{\Mississippi}{University of Mississippi, University, MS 38677, USA}
\newcommand{\napoli}{Universit{\`a} degli Studi di Napoli Federico II , 80138 Napoli NA, Italy}
\newcommand{\Nikhef}{Nikhef National Institute of Subatomic Physics, 1098 XG Amsterdam, Netherlands}
\newcommand{\NISER}{National Institute of Science Education and Research, An OCC of Homi Bhabha National Institute, Bhubaneswar, Odisha 752050, India}
\newcommand{\Northdakota}{University of North Dakota, Grand Forks, ND 58202-8357, USA}
\newcommand{\Northernillinois}{Northern Illinois University, DeKalb, IL 60115, USA}
\newcommand{\Northwestern}{Northwestern University, Evanston, Il 60208, USA}
\newcommand{\NotreDame}{University of Notre Dame, Notre Dame, IN 46556, USA}
\newcommand{\NoviSad}{University of Novi Sad, 21102 Novi Sad, Serbia}
\newcommand{\Ohiostate}{Ohio State University, Columbus, OH 43210, USA}
\newcommand{\OregonState}{Oregon State University, Corvallis, OR 97331, USA}
\newcommand{\Oxford}{University of Oxford, Oxford, OX1 3RH, United Kingdom}
\newcommand{\PacificNorthwest}{Pacific Northwest National Laboratory, Richland, WA 99352, USA}
\newcommand{\Padova}{Universt{\`a} degli Studi di Padova, I-35131 Padova, Italy}
\newcommand{\Panjab}{Panjab University, Chandigarh, 160014, India}
\newcommand{\Parissaclay}{Universit{\'e} Paris-Saclay, CNRS/IN2P3, IJCLab, 91405 Orsay, France}
\newcommand{\Parisuniversite}{Universit{\'e} Paris Cit{\'e}, CNRS, Astroparticule et Cosmologie, Paris, France}
\newcommand{\Parma}{University of Parma,  43121 Parma PR, Italy}
\newcommand{\Pavia}{Universit{\`a} degli Studi di Pavia, 27100 Pavia PV, Italy}
\newcommand{\Penn}{University of Pennsylvania, Philadelphia, PA 19104, USA}
\newcommand{\PennState}{Pennsylvania State University, University Park, PA 16802, USA}
\newcommand{\PhysicalResearchLaboratory}{Physical Research Laboratory, Ahmedabad 380 009, India}
\newcommand{\Pisa}{Universit{\`a} di Pisa, I-56127 Pisa, Italy}
\newcommand{\Pitt}{University of Pittsburgh, Pittsburgh, PA 15260, USA}
\newcommand{\Pontificia}{Pontificia Universidad Cat{\'o}lica del Per{\'u}, Lima, Per{\'u}}
\newcommand{\PuertoRico}{University of Puerto Rico, Mayaguez 00681, Puerto Rico, USA}
\newcommand{\Punjab}{Punjab Agricultural University, Ludhiana 141004, India}
\newcommand{\QMUL}{Queen Mary University of London, London E1 4NS, United Kingdom }
\newcommand{\Radboud}{Radboud University, NL-6525 AJ Nijmegen, Netherlands}
\newcommand{\Rice}{Rice University, Houston, TX 77005, USA}
\newcommand{\Rochester}{University of Rochester, Rochester, NY 14627, USA}
\newcommand{\Royalholloway}{Royal Holloway College London, London, TW20 0EX, United Kingdom}
\newcommand{\Rutgers}{Rutgers University, Piscataway, NJ, 08854, USA}
\newcommand{\Rutherford}{STFC Rutherford Appleton Laboratory, Didcot OX11 0QX, United Kingdom}
\newcommand{\Salento}{Universit{\`a} del Salento, 73100 Lecce, Italy}
\newcommand{\Santacarina}{Universidade do Estado de Santa Catarina , Santa Catarina, 89219-710, Brazil}
\newcommand{\santamarta}{Universidad del Magdalena, Santa Marta, Colombia}
\newcommand{\Sapienza}{Sapienza University of Rome, 00185 Roma RM, Italy}
\newcommand{\SergioArboleda}{Universidad Sergio Arboleda, 11022 Bogot{\'a}, Colombia}
\newcommand{\Sheffield}{University of Sheffield, Sheffield S3 7RH, United Kingdom}
\newcommand{\SLAC}{SLAC National Accelerator Laboratory, Menlo Park, CA 94025, USA}
\newcommand{\Southcarolina}{University of South Carolina, Columbia, SC 29208, USA}
\newcommand{\SouthDakotaSchool}{South Dakota School of Mines and Technology, Rapid City, SD 57701, USA}
\newcommand{\SouthDakotaState}{South Dakota State University, Brookings, SD 57007, USA}
\newcommand{\StonyBrook}{Stony Brook University, SUNY, Stony Brook, NY 11794, USA}
\newcommand{\SURF}{Sanford Underground Research Facility, Lead, SD, 57754, USA}
\newcommand{\Sussex}{University of Sussex, Brighton, BN1 9RH, United Kingdom}
\newcommand{\Syracuse}{Syracuse University, Syracuse, NY 13244, USA}
\newcommand{\Tecnologica}{Universidade Tecnol{\'o}gica Federal do Paran{\'a}, Curitiba, Brazil}
\newcommand{\TelAviv}{Tel Aviv University, Tel Aviv-Yafo, Israel}
\newcommand{\TexasAMcollege}{Texas A\&M University, College Station, TX 77840, USA}
\newcommand{\TexasAMcorpuscristi}{Texas A\&M University - Corpus Christi, Corpus Christi, TX 78412, USA}
\newcommand{\TexasArlington}{University of Texas at Arlington, Arlington, TX 76019, USA}
\newcommand{\Texasaustin}{University of Texas at Austin, Austin, TX 78712, USA}
\newcommand{\Toronto}{University of Toronto, Toronto, Ontario M5S 1A1, Canada}
\newcommand{\Tufts}{Tufts University, Medford, MA 02155, USA}
\newcommand{\Unifesp}{Universidade Federal de S{\~a}o Paulo, 09913-030, S{\~a}o Paulo, Brazil}
\newcommand{\UniversityCollegeLondon}{University College London, London, WC1E 6BT, United Kingdom}
\newcommand{\univkansas}{University of Kansas, Lawrence, KS 66045, USA}
\newcommand{\UNMSM}{Universidad Nacional Mayor de San Marcos, Lima, Peru}
\newcommand{\ValleyCity}{Valley City State University, Valley City, ND 58072, USA}
\newcommand{\Vigo}{University of Vigo, E- 36310, Vigo, Spain}
\newcommand{\VirginiaTech}{Virginia Tech, Blacksburg, VA 24060, USA}
\newcommand{\Warsaw}{University of Warsaw, 02-093 Warsaw, Poland}
\newcommand{\Warwick}{University of Warwick, Coventry CV4 7AL, United Kingdom}
\newcommand{\Wellesley}{Wellesley College, Wellesley, MA 02481, USA}
\newcommand{\Wichita}{Wichita State University, Wichita, KS 67260, USA}
\newcommand{\WilliamMary}{William and Mary, Williamsburg, VA 23187, USA}
\newcommand{\Wisconsin}{University of Wisconsin Madison, Madison, WI 53706, USA}
\newcommand{\Yale}{Yale University, New Haven, CT 06520, USA}
\newcommand{\York}{York University, Toronto M3J 1P3, Canada}
\affiliation{\Albanysuny}
\affiliation{\Aligarh}
\affiliation{\Almaty}
\affiliation{\Amsterdam}
\affiliation{\Antalya}
\affiliation{\Antananarivo}
\affiliation{\Antioquia}
\affiliation{\AntonioNarino}
\affiliation{\Argonne}
\affiliation{\Arizona}
\affiliation{\Asuncion}
\affiliation{\Athens}
\affiliation{\Atlantico}
\affiliation{\Augustana}
\affiliation{\Bern}
\affiliation{\Beykent}
\affiliation{\BolognaUniversity}
\affiliation{\Boston}
\affiliation{\Bristol}
\affiliation{\Brookhaven}
\affiliation{\Bucharest}
\affiliation{\CalBerkeley}
\affiliation{\CalDavis}
\affiliation{\CalIrvine}
\affiliation{\CalRiverside}
\affiliation{\CalSantabarbara}
\affiliation{\Caltech}
\affiliation{\Cambridge}
\affiliation{\Campinas}
\affiliation{\CataniaUniversitadi}
\affiliation{\Catolica}
\affiliation{\CBPF}
\affiliation{\CEASaclay}
\affiliation{\CERN}
\affiliation{\Charles}
\affiliation{\Chicago}
\affiliation{\ChungAng}
\affiliation{\CIEMAT}
\affiliation{\Cincinnati}
\affiliation{\Cinvestav}
\affiliation{\ColoradoBoulder}
\affiliation{\ColoradoState}
\affiliation{\Columbia}
\affiliation{\conida}
\affiliation{\Cti}
\affiliation{\CUSB}
\affiliation{\CzechAcademyofSciences}
\affiliation{\CzechTechnical}
\affiliation{\DannecyleVieux}
\affiliation{\Daresbury}
\affiliation{\Dordt}
\affiliation{\drew}
\affiliation{\Drexel}
\affiliation{\Duke}
\affiliation{\Durham}
\affiliation{\Edinburgh}
\affiliation{\EIA}
\affiliation{\Eotvos}
\affiliation{\erciyes}
\affiliation{\FCUL}
\affiliation{\FederaldeAlfenas}
\affiliation{\FederaldeGoias}
\affiliation{\FederaldoABC}
\affiliation{\FederaldoRio}
\affiliation{\Fermi}
\affiliation{\Ferrarauniv}
\affiliation{\Florida}
\affiliation{\Floridastate}
\affiliation{\Fluminense}
\affiliation{\GATech}
\affiliation{\Genova}
\affiliation{\Georgian}
\affiliation{\Granada}
\affiliation{\GranSasso}
\affiliation{\GranSassoLab}
\affiliation{\Grenoble}
\affiliation{\Guanajuato}
\affiliation{\Harish}
\affiliation{\Hawaii}
\affiliation{\hkust}
\affiliation{\Houston}
\affiliation{\Hyderabad}
\affiliation{\Idaho}
\affiliation{\IFIC}
\affiliation{\IGFAE}
\affiliation{\ihep}
\affiliation{\IISc}
\affiliation{\Iitk}
\affiliation{\Illinoisinstitute}
\affiliation{\Imperial}
\affiliation{\IndGuwahati}
\affiliation{\IndHyderabad}
\affiliation{\Indiana}
\affiliation{\INFNBologna}
\affiliation{\INFNCatania}
\affiliation{\INFNFerrara}
\affiliation{\INFNFrascati}
\affiliation{\INFNGenova}
\affiliation{\INFNLecce}
\affiliation{\INFNMilanBicocca}
\affiliation{\INFNMilano}
\affiliation{\INFNNapoli}
\affiliation{\INFNPadova}
\affiliation{\INFNPavia}
\affiliation{\INFNPisa}
\affiliation{\INFNRoma}
\affiliation{\INFNRomavergata}
\affiliation{\INFNSud}
\affiliation{\Infntorino}
\affiliation{\Ingenieria}
\affiliation{\Insubria}
\affiliation{\Iowa}
\affiliation{\IowaState}
\affiliation{\IPLyon}
\affiliation{\IPM}
\affiliation{\IRLPPC}
\affiliation{\IST}
\affiliation{\Ita}
\affiliation{\ITPMYerevan}
\affiliation{\Iwate}
\affiliation{\Jacksonstate}
\affiliation{\Jawaharlal}
\affiliation{\Jeonbuk}
\affiliation{\Jyvaskyla}
\affiliation{\Kansasstate}
\affiliation{\Kavli}
\affiliation{\KEK}
\affiliation{\KISTI}
\affiliation{\Kyiv}
\affiliation{\Lancaster}
\affiliation{\LawrenceBerkeley}
\affiliation{\LIP}
\affiliation{\Liverpool}
\affiliation{\LNA}
\affiliation{\LosAlmos}
\affiliation{\Louisanastate}
\affiliation{\LpBordeaux}
\affiliation{\Lucknow}
\affiliation{\Mainz}
\affiliation{\Manchester}
\affiliation{\Marmara}
\affiliation{\Massinsttech}
\affiliation{\Medellin}
\affiliation{\Michigan}
\affiliation{\Michiganstate}
\affiliation{\MilanoBicocca}
\affiliation{\MilanoUniv}
\affiliation{\Minho}
\affiliation{\Minnduluth}
\affiliation{\Minntwin}
\affiliation{\Mississippi}
\affiliation{\napoli}
\affiliation{\Nikhef}
\affiliation{\NISER}
\affiliation{\Northdakota}
\affiliation{\Northernillinois}
\affiliation{\Northwestern}
\affiliation{\NotreDame}
\affiliation{\NoviSad}
\affiliation{\Ohiostate}
\affiliation{\OregonState}
\affiliation{\Oxford}
\affiliation{\PacificNorthwest}
\affiliation{\Padova}
\affiliation{\Panjab}
\affiliation{\Parissaclay}
\affiliation{\Parisuniversite}
\affiliation{\Parma}
\affiliation{\Pavia}
\affiliation{\Penn}
\affiliation{\PennState}
\affiliation{\PhysicalResearchLaboratory}
\affiliation{\Pisa}
\affiliation{\Pitt}
\affiliation{\Pontificia}
\affiliation{\PuertoRico}
\affiliation{\Punjab}
\affiliation{\QMUL}
\affiliation{\Radboud}
\affiliation{\Rice}
\affiliation{\Rochester}
\affiliation{\Royalholloway}
\affiliation{\Rutgers}
\affiliation{\Rutherford}
\affiliation{\Salento}
\affiliation{\Santacarina}
\affiliation{\santamarta}
\affiliation{\Sapienza}
\affiliation{\SergioArboleda}
\affiliation{\Sheffield}
\affiliation{\SLAC}
\affiliation{\Southcarolina}
\affiliation{\SouthDakotaSchool}
\affiliation{\SouthDakotaState}
\affiliation{\StonyBrook}
\affiliation{\SURF}
\affiliation{\Sussex}
\affiliation{\Syracuse}
\affiliation{\Tecnologica}
\affiliation{\TelAviv}
\affiliation{\TexasAMcollege}
\affiliation{\TexasAMcorpuscristi}
\affiliation{\TexasArlington}
\affiliation{\Texasaustin}
\affiliation{\Toronto}
\affiliation{\Tufts}
\affiliation{\Unifesp}
\affiliation{\UniversityCollegeLondon}
\affiliation{\univkansas}
\affiliation{\UNMSM}
\affiliation{\ValleyCity}
\affiliation{\Vigo}
\affiliation{\VirginiaTech}
\affiliation{\Warsaw}
\affiliation{\Warwick}
\affiliation{\Wellesley}
\affiliation{\Wichita}
\affiliation{\WilliamMary}
\affiliation{\Wisconsin}
\affiliation{\Yale}
\affiliation{\York}
\author{S.~Abbaslu} \affiliation{\IPM}
\author{F.~Abd Alrahman} \affiliation{\Houston}
\author{A.~Abed Abud} \affiliation{\CERN}
\author{R.~Acciarri} \affiliation{\CERN}
\author{M.~A.~Acero} \affiliation{\Atlantico}
\author{M.~R.~Adames} \affiliation{\Tecnologica}
\author{G.~Adamov} \affiliation{\Georgian}
\author{M.~Adamowski} \affiliation{\Fermi}
\author{K.~Adhikari} \affiliation{\Lancaster}
\author{C.~Adriano} \affiliation{\Campinas}
% ...
\collaboration{The DUNE Collaboration}
\noaffiliation

\date{\today}

\begin{abstract}
The Deep Underground Neutrino Experiment (DUNE) is a next-generation neutrino experiment designed for precision measurements of long-baseline three-flavor neutrino oscillations. The DUNE Near Detector (ND) is also sensitive to short-baseline oscillations arising from mixing with an eV-scale sterile neutrino. In this paper, we present a sensitivity study for sterile-neutrino oscillations in the $3+1$ framework using three complementary event samples in the simulated DUNE ND: highly elastic $\nu_\mu$ charged-current (CC), highly elastic $\nu_e$ CC, and neutrino-electron elastic-scattering ($\nu+e$) events. These highly exclusive samples are chosen to reduce neutrino-interaction and flux uncertainties: the low-hadronic-energy CC selections enhance the correlation between reconstructed lepton energy and neutrino energy, while the precisely calculable $\nu+e$ cross section provides an independent constraint on the neutrino flux. Residual interaction-model dependence remains in the CC samples. The statistical analysis considers $|U_{e4}|^2$, $|U_{\mu4}|^2$, $|U_{\tau4}|^2$, and $\Delta m^2$, with sensitivity results presented as fixed-$\Delta m^2$ slices for $|U_{\tau4}|^2=0$. The strongest sensitivity occurs for $4 \lesssim \Delta m^2 \lesssim 10~\mathrm{eV}^2$, where oscillations produce resolvable distortions of the reconstructed-energy spectra. For one year of on-axis forward-horn-current running, the 90\% confidence-level sensitivity reaches approximately $|U_{\mu4}|^2 \simeq 0.01$ and $|U_{e4}|^2 \simeq 0.015$, depending on the value of the other mixing parameter.
\end{abstract}

\ifnum\sizecheck=0  
\maketitle
\fi

\section{Introduction}
\label{sec:intro}

Experimental evidence has firmly established that neutrinos undergo oscillations, implying that they have nonzero mass and that there is nontrivial mixing between mass and flavor eigenstates~\cite{PDG}. This mixing is described by the $3\times3$ Pontecorvo-Maki-Nakagawa-Sakata (PMNS) matrix. The three-flavor framework provides a successful description of the large body of existing neutrino oscillation data. Measurements of the invisible decay width of the \(Z\) boson~\cite{Zres} further constrain the number of light active neutrinos participating in Standard Model weak interactions to three. Any additional light neutrino state must therefore be predominantly sterile with respect to the Standard Model weak interaction, but could still mix with the active neutrino states. Three anomalous experimental results have motivated searches for sterile-neutrino mixing:
\begin{itemize}
    \item \textbf{Gallium anomaly}: Radiochemical gallium experiments using intense radioactive calibration sources observed a deficit in the measured electron-neutrino rate relative to expectation~\cite{GalliumAnomaly, GalliumAnomalyStats}. This tension was later reinforced by the BEST experiment, which reported results consistent with the earlier gallium deficit~\cite{BEST}. More recently, a revised theoretical calculation of the neutrino-capture cross section on $^{71}$Ga finds a significantly reduced predicted rate, potentially accounting for the anomaly without invoking new physics~\cite{Cadeddu:2026Gallium}.

    \item \textbf{LSND and MiniBooNE anomalies}: The LSND and MiniBooNE experiments reported excesses of electron-neutrino-like events in predominantly muon-(anti)neutrino beams~\cite{LSND, MiniBooNE:excess}. MicroBooNE, located at a similar baseline to MiniBooNE and exposed to the same Booster Neutrino Beam, provides a particularly direct test of the MiniBooNE anomaly. Subsequent results using MicroBooNE data found no evidence for either $\nu_\mu \to \nu_e$ appearance or $\nu_e$ disappearance consistent with a single light sterile-neutrino interpretation, excluding the parameter region favored by the LSND/MiniBooNE anomalies within the $3+1$ framework at the 95\% confidence level~\cite{Rodrigues:2025,MicroBooNE_nature}.

    \item \textbf{Reactor short-baseline results}: A deficit in the detected $\bar{\nu}_e$ rate relative to flux predictions was observed by reactor antineutrino experiments~\cite{reactorAnomaly1, reactorAnomaly2, reactorAnomaly3}. In addition, some short-baseline measurements have reported indications of baseline-dependent spectral distortions~\cite{Neutrino4, DANSS:2018fnn, NEOS:2016wee}. At the same time, several other short-baseline reactor experiments have observed no comparable distortion, placing strong constraints on the relevant parameter space~\cite{PROSPECT, STEREO}. More recent isotope-dependent measurements and reactor-flux calculations have shown that deficiencies in the predicted antineutrino yields, particularly that of $^{235}$U, may account for at least part of the original rate anomaly rather than requiring an oscillation-only explanation~\cite{DayaBay, RENO:2018pwo, Giunti:2017yid, Estienne:2019ujo, Kopeikin:2021ugh}.
\end{itemize}

Despite these anomalies, global fits to short-baseline oscillation data have long exhibited significant tension within the $3+1$ sterile-neutrino framework, particularly between appearance and disappearance measurements~\cite{Kopp:2013vaa, Dentler:2018sju, Acero:2022wqg}. This tension, already evident in global analyses incorporating early MiniBooNE data, disfavors a simultaneous interpretation of all anomalous observations in terms of oscillations involving a single additional sterile-neutrino state. More recent appearance and disappearance constraints, including those from MicroBooNE~\cite{MicroBooNE}, further restrict the viable $3+1$ parameter space. The $3+1$ framework nevertheless remains a well-defined and widely used benchmark for short-baseline searches, providing the simplest extension of the three-neutrino model that introduces oscillations at an additional mass-squared splitting and allowing experimental sensitivities to be compared directly across different oscillation channels.

Accelerator-based neutrino experiments can search for sterile neutrinos through oscillation signatures inconsistent with the three-flavor framework. The primary charged-current channels include $\nu_\mu$ disappearance, $\nu_e$ disappearance, and $\nu_e$ appearance, for which CC events are measured as a function of reconstructed neutrino energy. Neutral-current (NC) event rates provide an additional disappearance channel: oscillations from an active neutrino flavor into a sterile state reduce the total NC interaction rate because sterile neutrinos do not participate in weak NC interactions. This channel is, in principle, sensitive to sterile mixing involving all three active flavors. In practice, its sensitivity is limited by the poor reconstruction of the incident neutrino energy because the final-state neutrino is undetected.

Long-baseline neutrino experiments employ near detectors (NDs) to measure neutrino interactions and constrain uncertainties in flux predictions and neutrino interaction cross sections. For eV$^2$-scale mass-squared splittings and GeV-scale neutrino energies, sterile-neutrino oscillations can become appreciable over baselines ranging from several hundred meters to kilometers, depending on $\Delta m^2$ and neutrino energy. Near detectors can therefore provide direct sensitivity to short-baseline oscillations rather than serving only as unoscillated reference measurements. MINOS/MINOS+~\cite{MINOS:sterile}, NO$\nu$A~\cite{NOvA:2024imi}, and T2K~\cite{T2K:2019efw} have used near- and far-detector data to search for $\nu_\mu$ disappearance and NC disappearance. The Fermilab Short-Baseline Neutrino (SBN) program employs multiple detectors and has strong sensitivity in the region $\Delta m^2 \sim 1~\mathrm{eV}^2$, where common flux and cross-section uncertainties are reduced through detector comparisons~\cite{SBNProgram}.

In the absence of a far-detector (FD) comparison, an ND-only sterile-neutrino search requires precise control of flux and cross-section systematic uncertainties while exploiting the large event statistics available at the ND. In the $3+1$ framework, the leading short-baseline oscillation probabilities are
\begin{equation}
\begin{split}
P_{\alpha\alpha} &= 1 - \sin^2 2\theta_{\alpha\alpha}\sin^2\Delta, \\
P_{\alpha\beta} &= \sin^2 2\theta_{\alpha\beta}\sin^2\Delta,
\qquad \alpha\neq\beta,
\end{split}
\label{eqn:OscProb}
\end{equation}
with $\Delta \equiv \Delta m^2_{41}L/(4E_\nu)$ and
\begin{equation}
\begin{split}
\sin^2 2\theta_{\alpha\alpha} &\equiv 4|U_{\alpha4}|^2\left(1-|U_{\alpha4}|^2\right), \\
\sin^2 2\theta_{\alpha\beta} &\equiv 4|U_{\alpha4}|^2|U_{\beta4}|^2,
\qquad \alpha\neq\beta,
\end{split}
\label{eqn:MixingAngles}
\end{equation}
where $\alpha,\beta=e,\mu,\tau$.

The observable event rate for a selected final-state flavor $\beta$ can be written schematically as
\begin{equation}
    R_\beta \sim \sum_{\alpha=e,\mu,\tau}
    \Phi_\alpha(E_\nu)
    \otimes P_{\alpha\beta}(E_\nu)
    \otimes \sigma_\beta(k,k')
    \otimes \varepsilon_\beta(E_\nu),
    \label{eqn:EvtRate}
\end{equation}
where $\Phi_\alpha$ is the incident neutrino flux of flavor $\alpha$, $P_{\alpha\beta}$ is the oscillation probability from flavor $\alpha$ to $\beta$, $\sigma_\beta$ is the interaction cross section for the selected final state, and $\varepsilon_\beta$ denotes the energy-dependent reconstruction and selection efficiency. The oscillation sensitivity therefore depends on uncertainties in the flux, interaction model, and detector response. Accelerator neutrino fluxes are typically known at the $\mathcal{O}(10\%)$ level, primarily because of uncertainties in meson production in the proton target~\cite{nueScattering}, while neutrino-nucleus cross sections carry uncertainties of comparable scale~\cite{nuInt,RPA}.

While neutrino-nucleus interactions are affected by complex nuclear effects, elastic neutrino-electron scattering ($\nu e^- \to \nu e^-$) is a purely electroweak process that can be calculated with high precision. At DUNE neutrino energies, the cross section can be written at tree level as
\begin{equation}
    \frac{d\sigma(\nu e^-\to\nu e^-)}{dy}
    =
    \frac{G_F^2 s}{\pi}
    \left[
    C_{LL}^2 + C_{LR}^2(1-y)^2
    \right],
    \label{eqn:nuexSec}
\end{equation}
where $y=T_e/E_\nu$ is the ratio of the electron kinetic energy to the incident neutrino energy, $G_F$ is the Fermi constant, and $s$ is the Mandelstam invariant. The coefficients $C_{LL}$ and $C_{LR}$ describe the left- and right-handed electroweak couplings of the electron. This process has a flavor dependence because $\nu_e e^-$ scattering receives contributions from both charged-current $W$-boson exchange and neutral-current $Z$-boson exchange, whereas $\nu_\mu e^-$ and $\nu_\tau e^-$ scattering proceed only through neutral-current exchange. Consequently, the $\nu_e e^-$ cross section is larger than those for the other neutrino flavors.

The use of the $\nu e^- \to \nu e^-$ sample to constrain the neutrino flux has been studied in Ref.~\cite{nueScattering}. At neutrino energies near $1~\mathrm{GeV}$, however, the total cross section for this process, $\sim10^{-42}~\mathrm{cm}^2$, is approximately 2000 times smaller than the neutrino-nucleon cross section at the same energy. The resulting low event rate has limited the statistical precision of this technique in previous experiments, preventing percent-level flux constraints~\cite{MINERvA:2015nqi, MINERvA:2022vmb}.

Unlike neutrino-electron scattering, CC neutrino-nucleus scattering involves interactions with bound nucleons and is subject to both initial- and final-state nuclear effects. As the incident neutrino energy increases, a broader range of energy transfers and hadronic final states becomes kinematically accessible, including resonance production and multiparticle final states whose modeling introduces additional uncertainties. Restricting the energy transfer to the hadronic system suppresses contributions from these more complex final states and reduces the energy dependence of the inclusive CC cross section, thereby providing sensitivity to the incident neutrino flux. This is the basis of the low-$\nu$ method, originally developed for high-energy neutrino scattering~\cite{Mishra:1990ax,Auchincloss:1990tu,Seligman:1997fe} and closely related to the low-$y$ approach~\cite{Belusevic:1987rn,Belusevic:1988ab}. The method was subsequently extended to the few-GeV regime, where quasielastic and resonance processes become increasingly important, by using substantially lower values of the maximum allowed energy transfer~\cite{Bodek:2012uu}. It has also been applied experimentally, including by MINERvA, which used a maximum energy transfer of $0.3~\mathrm{GeV}$ for its lowest-energy low-$\nu$ sample~\cite{MINERvA:lownu}. In this context, $\nu$ denotes the energy transferred from the neutrino to the nucleus and is defined for inclusive CC interactions as $\nu \equiv E_\nu-E_\ell$, where $E_\nu$ is the incident neutrino energy and $E_\ell$ is the energy of the outgoing charged lepton. The inclusive neutrino-nucleus cross section can then be expressed as
\begin{equation}
\begin{aligned}
\frac{d\sigma}{d\nu}
=
\frac{G_F^2 M}{\pi}
\int_0^1
\Bigg(
F_2
& - \frac{\nu}{E_\nu}[F_2 \mp xF_3]
\\
& + \frac{\nu}{2E_\nu^2}
\Bigg[
\frac{Mx(1-R_L)}{1+R_L}
F_2
\Bigg]
\\
& + \frac{\nu^2}{2E_\nu^2}
\Bigg[
\frac{F_2}{1+R_L} \mp xF_3
\Bigg]
\Bigg)\,dx .
\end{aligned}
\label{eqn:CCxS}
\end{equation}
where $G_F$ is the Fermi constant, $M$ is the mass of the struck nucleon, $F_2$ and $xF_3$ are structure functions describing the nucleon response to neutrino interactions, $R_L$ denotes the ratio of the longitudinal to transverse response, $x$ is the Bjorken scaling variable, and the upper and lower signs correspond to neutrinos and antineutrinos, respectively.

For events with $\nu$ below a cutoff $\nu_0$, the energy-dependent correction terms are suppressed, reducing the dependence of the cross section on the incident neutrino energy~\cite{Belusevic:1987rn,Belusevic:1988ab,Bodek:2012uu}. In the ideal low-$\nu$ limit, the cross section approaches an energy-independent form. At finite $\nu_0$ and few-GeV neutrino energies, however, residual energy dependence and interaction-model dependence remain~\cite{SubstandardCandle}. Because only a small amount of energy is transferred to the nucleus, $E_\nu \approx E_\ell$, and the charged-lepton energy therefore provides a useful estimator of the neutrino energy. Oscillations can consequently produce distortions directly in the reconstructed lepton-energy spectrum. The charged leptons in the selected low-$\nu$ sample are also predominantly forward, resulting in high reconstruction efficiency.

DUNE is well suited to search for short-baseline oscillations with its ND. Previous studies have investigated the sensitivity of DUNE to sterile-neutrino oscillations using the near and far detectors~\cite{DUNE:2020ypp}. The high intensity of the Long-Baseline Neutrino Facility (LBNF) neutrino beam~\cite{LBNF_Beam} is essential for obtaining the statistics required for percent-level flux constraints with $\nu e^- \rightarrow \nu e^-$. The highly granular ND enables selections of charged-current interactions with low reconstructed hadronic energy in both the $\nu_\mu$ and $\nu_e$ samples. In this paper, we study the sensitivity of the DUNE ND to sterile-neutrino oscillations using a multi-channel strategy that combines highly elastic $\nu_\mu$-CC and $\nu_e$-CC samples with neutrino-electron elastic scattering.

Although the results presented here are interpreted within a $3+1$ sterile-neutrino framework, the analysis strategy is more general. The combination of high-statistics charged-current samples with neutrino-electron scattering provides sensitivity to flavor- and energy-dependent distortions in the near-detector data while reducing reliance on hadronic-energy reconstruction. The $3+1$ model therefore serves as a concrete benchmark for illustrating the broader potential of the DUNE ND as a precision probe of unexpected short-baseline phenomena. Additional information from DUNE-PRISM could further strengthen such searches by providing complementary measurements across different neutrino-energy spectra~\cite{Hernandez-Garcia:2026asl}.

This paper is organized as follows: Section~\ref{sec:data} describes the neutrino samples used in this study and the corresponding reconstruction assumptions. Section~\ref{sec:uncertainties} summarizes the treatment of systematic uncertainties, and Section~\ref{sec:technique} describes the oscillation analysis methodology. The sensitivity of the DUNE ND to sterile-neutrino oscillations is presented in Section~\ref{sec:results}.

\section{DUNE ND Detector and Simulations}
\label{sec:data}

\subsection{The DUNE ND}

DUNE is a long-baseline neutrino experiment with an ND located approximately 574~m downstream of the proton target and an FD located 1285~km downstream~\cite{DUNE:2016rla}. The high interaction rate at the ND permits the selection of relatively rare interaction channels, including neutrino-electron elastic scattering.

The DUNE ND includes a liquid-argon time projection chamber (LArTPC, called ND-LAr) functionally coupled to the Temporary Muon Spectrometer (TMS), as well as a second measurement system not used in this analysis. Unlike conventional wire-readout LArTPCs, ND-LAr uses a pixelated anode that provides native three-dimensional imaging of neutrino interactions. The pixel location determines two spatial coordinates of each charge deposit, while the third is obtained from the drift time. This enables high-resolution reconstruction of particle tracks, energy depositions, and interaction topologies without combining multiple two-dimensional projections.

ND-LAr has an active argon volume measuring $7 \times 3 \times 5~\mathrm{m}^3$ in the horizontal, vertical, and beam directions, respectively, which is large enough to contain hadronic activity as well as wide-angle muons. Owing to its proximity to the beam source, the DUNE ND experiences an interaction rate of approximately one neutrino interaction per beam spill per 10 tons of detector mass, corresponding to approximately 15 interactions per spill in the active LAr volume. Beam spills occur approximately once every 1.2~s, with each spill lasting about 10~$\mu$s. To accommodate this rate, ND-LAr employs a modular design~\cite{DUNE:Mx2}. The detector is segmented into 35 $1 \times 3 \times 1~\mathrm{m}^3$ modules, with dimensions quoted in the horizontal, vertical, and beam directions, respectively, each consisting of a central cathode and two pixelated anode planes.

TMS is a magnetized steel range stack instrumented with scintillator planes~\cite{DUNE:OfflineCDR}. The instrumented region is $7 \times 3~\mathrm{m}^2$ in the horizontal and vertical directions, matching the transverse cross section of ND-LAr, and the total volume is $7.4 \times 5 \times 7~\mathrm{m}^3$ in the horizontal, vertical, and beam directions, respectively. The 1~cm-thick scintillator planes consist of strips in two orientations, either vertical or horizontal, which measure the location of energy deposits in two dimensions. Combining the horizontal and vertical planes gives 3D images of through-going particles. The steel is 1.5~cm thick in the upstream region, with 4~cm and 8~cm thick plates in the downstream part of the detector. TMS is able to range out muons up to 4.5~GeV/c momentum; including energy loss in ND-LAr, this places an upper limit on the energy of muons that can be reconstructed at about 5~GeV. The 1~T magnetic field in TMS is sufficient to measure the charge by curvature but does not provide a precise momentum measurement.

\begin{figure}[t]
\centering
\includegraphics[width=0.48\textwidth]{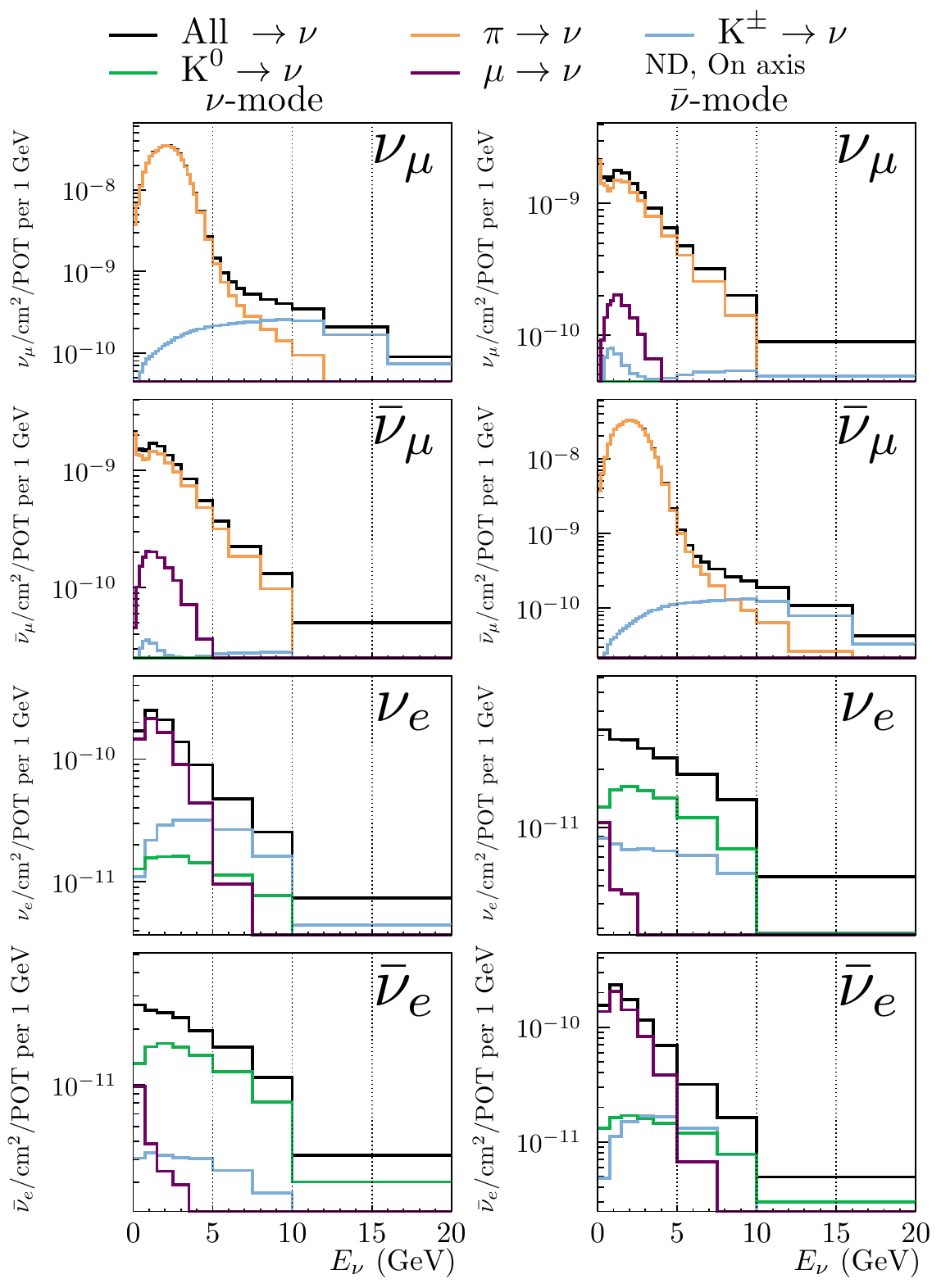}
\caption{The neutrino flux at the DUNE ND, taken from the DUNE Technical Design Report, Volume~II~\cite{DUNE:TDR}.}
\label{fig:NDflux}
\end{figure}

The neutrino flux in forward horn current (FHC) mode (“neutrino mode”) is shown in the left column of Fig.~\ref{fig:NDflux}. The $\nu_\mu$ flux has a broad peak between 1--3 GeV, with the maximum at 2.5 GeV corresponding to the oscillation maximum at the FD. The $\nu_e$ flux arises primarily from the $\pi \rightarrow \mu \rightarrow e$ decay chain at low energies and is dominated by kaon decays at higher energies.

The analysis uses the standard ND-LAr fiducial volume defined in the DUNE Technical Design Report (TDR)~\cite{DUNE:TDR}, corresponding to a $6\times2\times3~\mathrm{m}^3$ region in the horizontal, vertical, and beam directions, respectively, or approximately 50~t of liquid argon. The fiducial requirement removes regions near the detector boundaries to reduce edge effects and improve final-state containment and reconstruction. For nominal FHC beam conditions corresponding to $1.1\times10^{21}$ protons on target (POT) per year, the 50~t ND-LAr fiducial volume contains approximately $5.9\times10^7$ $\nu_\mu$-CC interactions and $4.5\times10^5$ electron-flavor CC interactions in the oscillation region $0.5<E_\nu<4~\mathrm{GeV}$~\cite{DUNE:CDR}. The latter rate corresponds to the combined $\nu_e+\bar{\nu}_e$ sample quoted in the DUNE CDR and is included here only to indicate the overall interaction-rate scale; the analysis itself uses only CC $\nu_e$ events. For the same nominal exposure, approximately 120 neutrino-electron elastic-scattering interactions per ton of argon are expected~\cite{nueScattering}.

\subsection{Simulation Samples}

\begin{figure*}[t]
\centering 
    \includegraphics[width=0.48\textwidth]{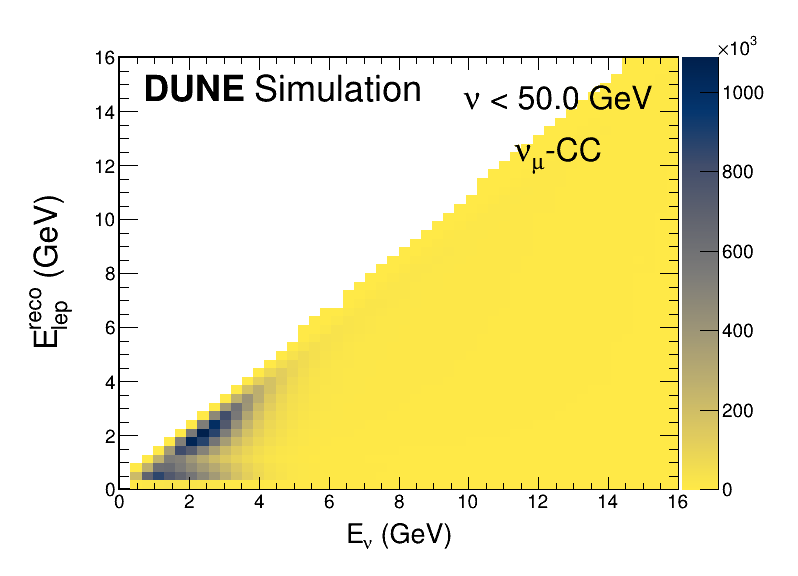} \includegraphics[width=0.48\textwidth]{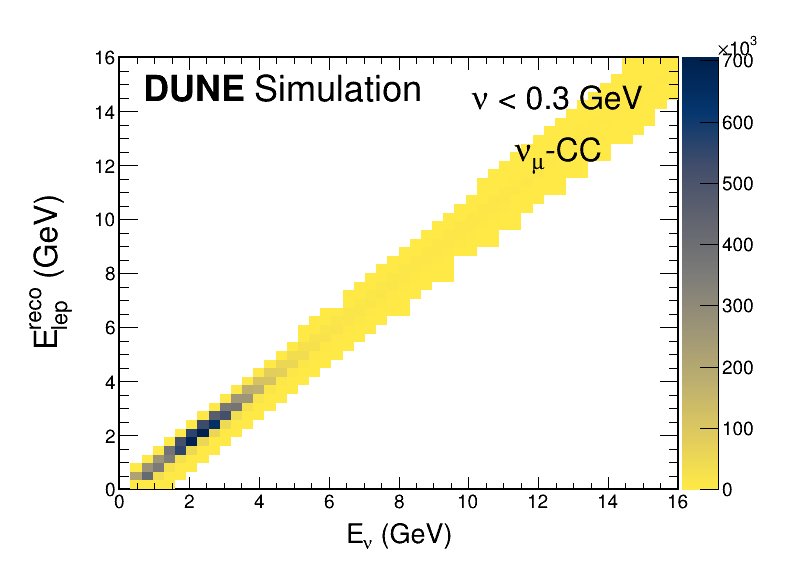} 
\caption{Correlation between true neutrino energy and reconstructed muon energy for the $\nu_\mu$-CC sample without a requirement on reconstructed energy transfer (left) and with the requirement $\nu^{\mathrm{reco}}<0.3~\mathrm{GeV}$ (right), for the nominal interaction and detector model used in this analysis.}
\label{fig:mECorrel} 
\end{figure*}

The analysis uses two sets of simulated samples. The $\nu_\mu$- and $\nu_e$-CC samples are taken from the official DUNE simulation used for the TDR long-baseline analysis~\cite{DUNE:TDR}. Because of the small $\nu e^-\to\nu e^-$ scattering cross section, a separate signal-only sample is used for this channel. This is the same sample used in the published DUNE $\nu e^-\to\nu e^-$ analysis~\cite{nueScattering}.

The neutrino flux is predicted using a Geant4-based simulation of proton interactions in the target and the subsequent propagation of secondary particles through the magnetic focusing system. Neutrinos are produced primarily through pion and kaon decays in a cylindrical decay pipe 200~m long and 4~m in diameter. Because neutrino production occurs over an extended region of approximately 220~m from the target through the decay volume, the neutrino flight distance to the detector spans approximately 350--570~m. This extended-source geometry smears the $L/E_\nu$ distribution and reduces sensitivity to narrow oscillation features relative to the point-source approximation. The impact of the extended neutrino source on the $L/E_\nu$ dependence is discussed further in Appendix~\ref{apdx:LEdep}.

Neutrino interactions in the LAr active volume are simulated using GENIE~2.12.10, with details provided in Ref.~\cite{GENIE,DUNE:TDR}. Final-state particles are propagated using Geant4~\cite{GEANT4:2002zbu} through an ND-LAr detector geometry based on the DUNE Near Detector Conceptual Design Report reference design~\cite{DUNE:CDR}. Events are classified as \(\nu_\mu\)-CC candidates if they contain a minimum-ionizing track, consistent with a muon or pion, with a length of at least 1~m and an average energy deposition below 3~MeV/cm. Events are classified as \(\nu_e\)-CC candidates using a parameterized electron-reconstruction model, with an energy-dependent reconstruction efficiency applied to the outgoing electron. The dominant background arises from NC \(\pi^0\) production, particularly when the leading photon converts within 2~cm of the interaction vertex and the second photon has an energy below 50~MeV, or when the two photons are nearly collinear, with an opening angle below 10~mrad.

This analysis uses the dedicated high-statistics $\nu e^-\to\nu e^-$ sample generated with GENIE on an argon target for Ref.~\cite{nueScattering}. Rather than propagating each event through the full detector simulation, the detector response is modeled using the parameterized electron-energy and angular resolutions developed in Ref.~\cite{nueScattering}, which were derived from dedicated Geant4 studies of forward-going electrons in liquid argon. The angular response is energy dependent and includes non-Gaussian tails associated with hard scattering.

\begin{figure*}[t]
\centering
  \includegraphics[width=0.48\textwidth]{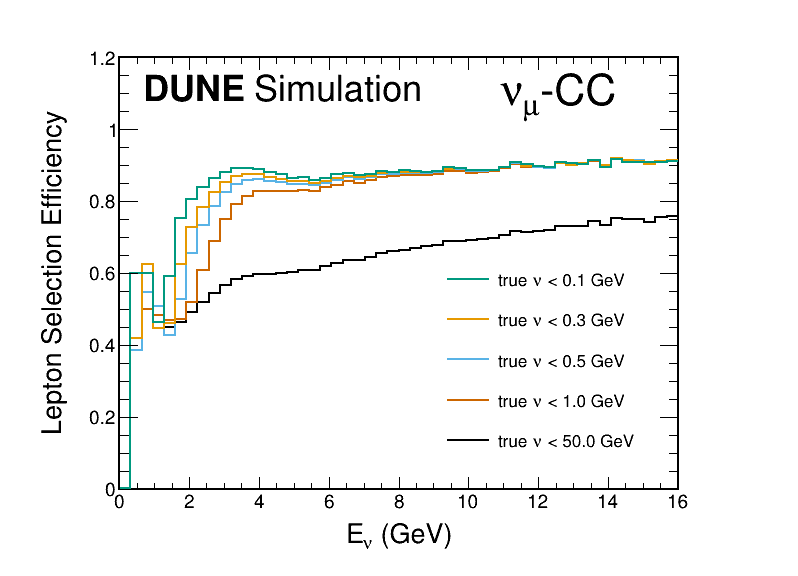}
  \includegraphics[width=0.48\textwidth]{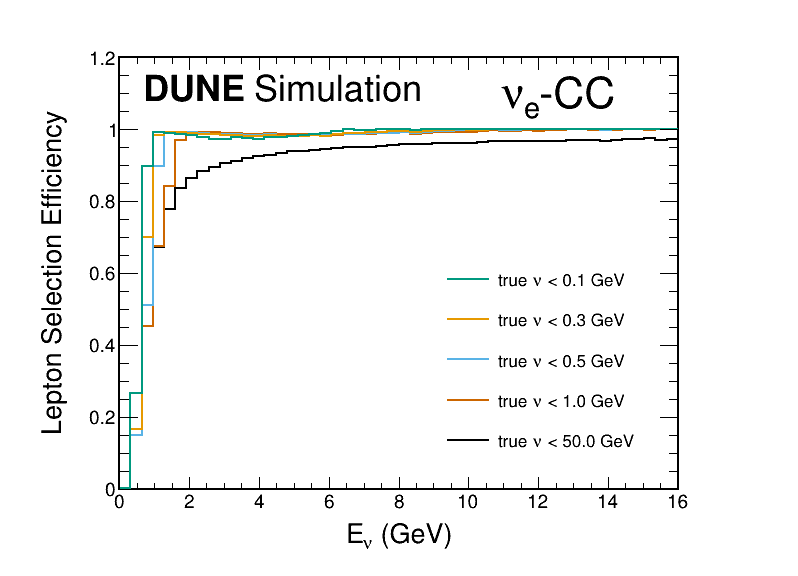}
\caption{Muon (left) and electron (right) CC selection efficiencies as functions of true neutrino energy for different requirements on the true energy transfer $\nu^{\mathrm{true}}$. For each threshold $\nu_0$, the efficiency is defined relative to CC events that satisfy the corresponding requirement $\nu^{\mathrm{true}}<\nu_0$.}
\label{fig:CCEff}
\end{figure*}

\subsection{Selection Criteria}

In addition to the fiducial-volume requirement described above, further selection criteria are applied based on event topology and energy deposition. A muon candidate is required to stop within the active volume of either ND-LAr or TMS, restricting the reconstructed muon energy to approximately 5~GeV. For samples containing a final-state electron, the energy-dependent reconstruction efficiency described above is applied.

The reconstructed energy transfer, $\nu^{\mathrm{reco}}$, is defined in this analysis as the total visible hadronic energy deposited in ND-LAr and therefore serves as a proxy for the true energy transfer $\nu$:
\begin{equation}
    \nu^{\mathrm{reco}}
    =
    E_P^{\mathrm{reco}}
    + E_N^{\mathrm{reco}}
    + E_{\pi^\pm}^{\mathrm{reco}}
    + E_{\pi^0}^{\mathrm{reco}}
    + E_{\mathrm{others}}^{\mathrm{reco}}.
    \label{eqn:nuReco}
\end{equation}
Here, $E_P^{\mathrm{reco}}$, $E_N^{\mathrm{reco}}$, $E_{\pi^\pm}^{\mathrm{reco}}$, $E_{\pi^0}^{\mathrm{reco}}$, and $E_{\mathrm{others}}^{\mathrm{reco}}$ denote the visible deposited energies assigned to the corresponding particle categories in the simulation. These categories are used only to construct the total visible hadronic energy; no reconstructed particle-identification requirement is applied. No corrections are applied for pion masses.

Figure~\ref{fig:mECorrel} shows, for the $\nu_\mu$-CC sample, the correlation between true neutrino energy and reconstructed muon energy with and without the nominal requirement $\nu^{\mathrm{reco}}<0.3~\mathrm{GeV}$. Within the nominal interaction and detector model used in this analysis, requiring small $\nu^{\mathrm{reco}}$ preferentially selects events in which only a small fraction of the neutrino energy is transferred to the hadronic system. The improved correlation motivates the use of reconstructed lepton energy as the oscillation-sensitive observable for the highly elastic CC samples. The strength of this correlation and the mapping between true and reconstructed quantities are interaction-model dependent, and alternative models can predict different behavior~\cite{SubstandardCandle}.

Figure~\ref{fig:CCEff} shows the muon (left) and electron (right) selection efficiencies of the DUNE ND. For each true-$\nu$ threshold $\nu_0$, the efficiency is defined as the fraction of events satisfying $\nu<\nu_0$ that also pass the corresponding lepton reconstruction and selection requirements. Less restrictive $\nu$ selections include events with lower-energy and less forward-going charged leptons, reducing the reconstruction and acceptance efficiency. The dip in the muon efficiency near $E_\nu\sim1.5~\mathrm{GeV}$ arises from the combined ND-LAr and downstream-spectrometer acceptance. In addition to satisfying the minimum-ionizing-track requirements, the muon must either stop within ND-LAr or be reconstructed in the downstream spectrometer so that its momentum can be measured. Muons that exit ND-LAr at large angles or stop before reaching the spectrometer therefore reduce the selection efficiency. The electron efficiency is negligible below 300~MeV, as in the electron-reconstruction efficiency model used in the DUNE long-baseline analysis~\cite{DUNE:TDR}.

\begin{figure*}[t]
\centering
  \includegraphics[width=0.48\textwidth]{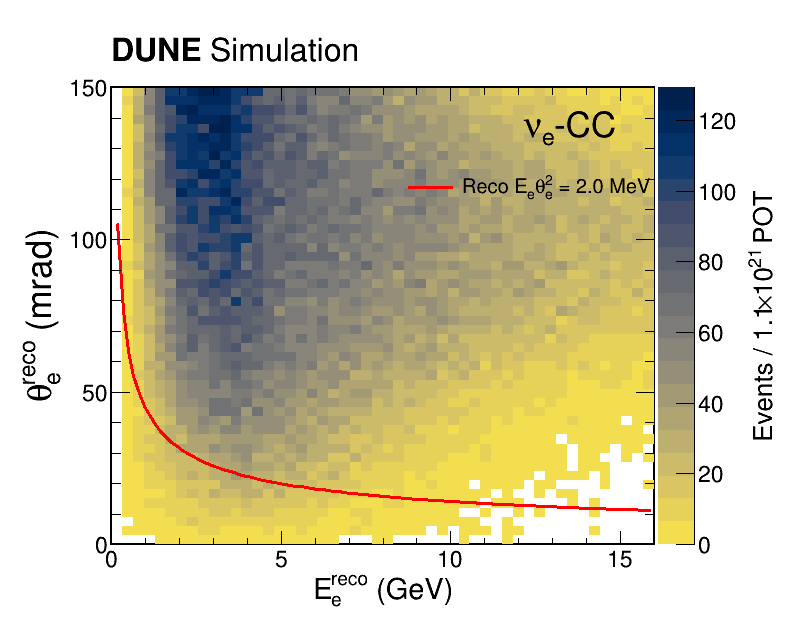}
  \includegraphics[width=0.48\textwidth]{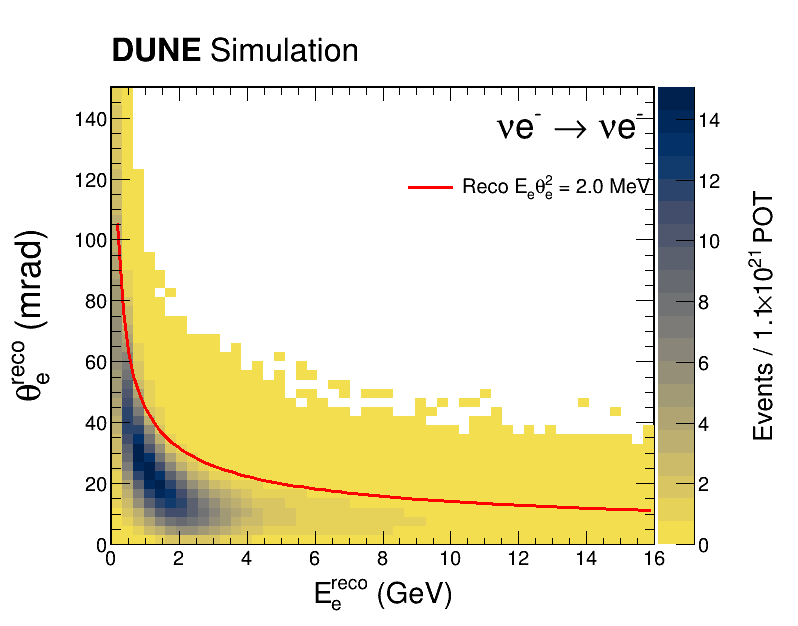}
\caption{Reconstructed electron scattering angle as a function of reconstructed electron energy for the $\nu_e$-CC (left) and $\nu e^-\to\nu e^-$ (right) samples, normalized to $1.1\times10^{21}$ POT. The color scales are independently autoscaled in the two panels. The red line indicates the selection requirement $E_e^{\mathrm{reco}}(\theta_e^{\mathrm{reco}})^2=2.0~\mathrm{MeV}$.}
\label{fig:EeTheta2}
\end{figure*}

The final selection criterion distinguishes the two samples containing an outgoing electron. For $\nu e^-\to\nu e^-$ scattering, the underlying kinematics satisfy
\begin{equation}
    1 - \cos\theta_e = \frac{m_e(1-y)}{E_e},
    \label{eqn:Etheta}
\end{equation}
where $y$ is the fractional kinetic-energy transfer defined in Eq.~\ref{eqn:nuexSec}. For the characteristic DUNE neutrino energies of $\mathcal{O}(1~\mathrm{GeV})$, the scattered electron is highly forward-going. Using the small-angle approximation $\cos\theta_e \approx 1-\theta_e^2/2$, Eq.~\ref{eqn:Etheta} becomes
\begin{equation}
    E_e\theta_e^2 \approx 2m_e(1-y).
\end{equation}
Since $0 \leq y \leq 1$, $\nu e^-\to\nu e^-$ events satisfy the approximate kinematic bound
\begin{equation}
    E_e\theta_e^2 \lesssim 2m_e \approx 1.02~\mathrm{MeV}.
\end{equation}
This strong forward-scattering constraint motivates the use of the reconstructed quantity $E_e^{\mathrm{reco}}(\theta_e^{\mathrm{reco}})^2$ to discriminate between the $\nu e^-\to\nu e^-$ and $\nu_e$ CC samples.

Figure~\ref{fig:EeTheta2} shows the distribution of reconstructed electron energy as a function of reconstructed scattering angle for the $\nu e^-\to\nu e^-$ and $\nu_e$ CC samples. The red line corresponds to the selection requirement $E_e^{\mathrm{reco}}(\theta_e^{\mathrm{reco}})^2 = 2.0~\mathrm{MeV}$. Events below this threshold are selected as $\nu e^-\to\nu e^-$ candidates, while events above the threshold are assigned to the $\nu_e$ CC sample.

CC~$\nu_e$ events with $E_e^{\mathrm{reco}}(\theta_e^{\mathrm{reco}})^2 < 2.0~\mathrm{MeV}$ are treated as background to the $\nu e^-\to\nu e^-$ sample. The 2.0~MeV threshold, which lies above the approximate signal kinematic limit of 1~MeV, is primarily determined by the finite angular resolution. For these background events, the reconstructed-energy-transfer requirement is tightened to $\nu^{\mathrm{reco}}<80~\mathrm{MeV}$. An additional requirement is imposed on $\nu e^-\to\nu e^-$ candidates to reject events with visible final-state hadronic activity. Events containing a $\pi^0$ in addition to an electron are rejected, as the $\pi^0$ cannot be collinear with the electron in the absence of other final-state particles while conserving four-momentum. Events containing a charged hadron with kinetic energy greater than 20~MeV are assumed to be identifiable, as a 20~MeV proton typically traverses more than one pixel in ND-LAr. Because the signal process contains no hadronic activity at the interaction vertex, the expected energy deposition is that of a single minimum-ionizing particle, approximately 1~MeV per 5~mm pixel. Although a 20~MeV proton is strongly quenched by electron-ion recombination, it typically deposits several MeV in one pixel or across two adjacent pixels, allowing such events to be rejected. Further details are given in Ref.~\cite{nueScattering}. No explicit reconstructed track- or shower-multiplicity requirement is imposed; instead, the selection relies on the electron candidate together with the kinematic requirement and hadronic-activity vetoes.

The LBNF beam contains an approximately 1\% intrinsic electron-neutrino component, while the $\nu e^-\to\nu e^-$ elastic-scattering cross section is significantly smaller than the CC cross section~\cite{nueScattering,nuInt}. Consequently, the $\nu_\mu$-CC sample has the largest event yield, followed by the $\nu_e$-CC sample, with the $\nu e^-\to\nu e^-$ sample having the lowest statistics. In principle, the $\nu e^-\to\nu e^-$ sample also provides sensitivity to $\nu_\mu\to\nu_\tau$ oscillations through the NC contribution, which is sensitive to all three active neutrino flavors. However, the individual flavor contributions to this sample cannot be distinguished, and the resulting sensitivity to the $\nu_\tau$ channel is negligible. Including this oscillation channel in the simulation therefore slightly weakens the constraint from $\nu_\mu$ disappearance.

\begin{figure*}
\centering
  \includegraphics[width=0.48\textwidth]{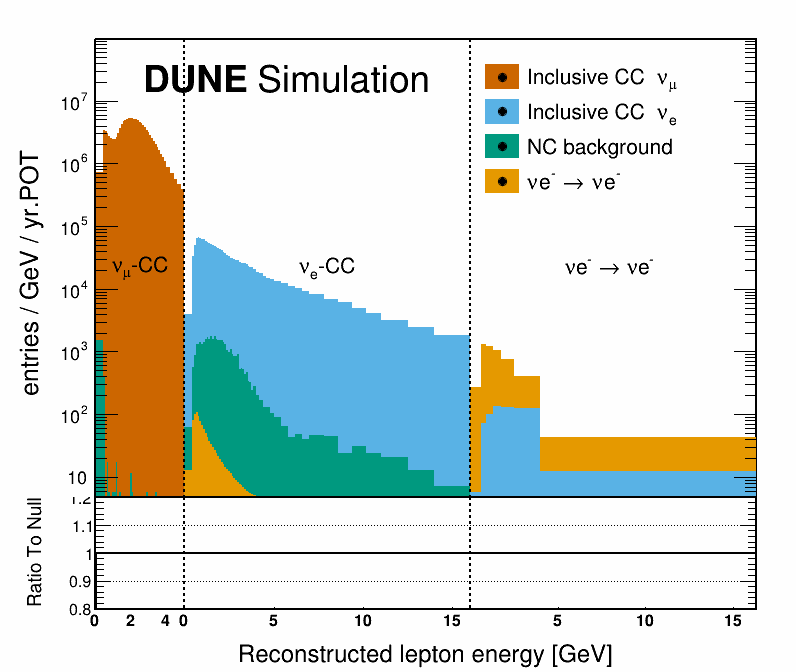}
  \includegraphics[width=0.48\textwidth]{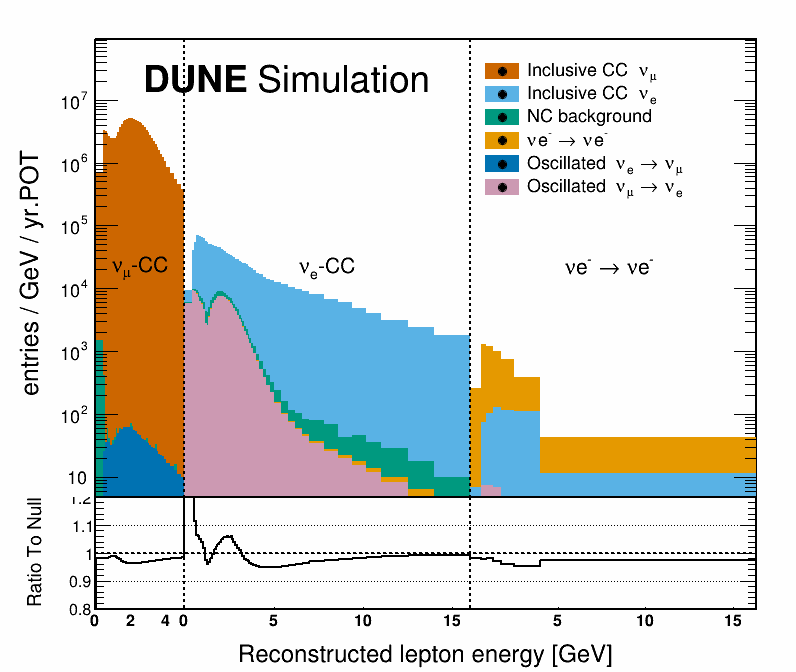}
\caption{Expected event spectra for one year-equivalent of on-axis exposure, corresponding to $1.1\times10^{21}$ POT, under the null hypothesis (left) and an example sterile-neutrino oscillation hypothesis with $(|U_{e4}|^2, |U_{\mu4}|^2, |U_{\tau4}|^2, \Delta m^2)=(0.04,\,0.01,\,0.2,\,6.0~\mathrm{eV}^2)$ (right), using the nominal selection requirements described in the text.}
\label{fig:tgt}
\end{figure*}

The final selected event spectra for one year-equivalent of on-axis DUNE ND exposure in FHC mode at an average beam power of 1.2~MW, corresponding to $1.1\times10^{21}$ POT, are shown in Fig.~\ref{fig:tgt} (left). The nominal running plan assumes that approximately 50\% of the exposure is taken on axis, although this fraction may differ during the initial running period. The $\nu_\mu$-CC and $\nu_e$-CC samples are dominated by their respective CC signals, with NC interactions providing the leading background in both samples. The $\nu e^-\to\nu e^-$ sample is dominated by elastic-scattering signal, with $\nu_e$ CC interactions providing the largest background contribution. The selected samples contain approximately $1.4\times10^{7}$ $\nu_\mu$-CC events, $2.1\times10^{5}$ $\nu_e$-CC events, and $2.6\times10^{3}$ $\nu e^-\to\nu e^-$ elastic-scattering events. Details of the sample statistics and analysis binning are given in Appendix~\ref{apdx:binning}.

\section{Uncertainties}
\label{sec:uncertainties}

In this work, systematic uncertainties from the neutrino flux, neutrino-argon interaction cross sections, and detector response are incorporated into a covariance matrix spanning the three analysis samples. The systematic covariances are constructed from parameter variations in fractional form and converted to absolute covariances using the predicted event rates under the null oscillation hypothesis. Statistical uncertainties are included as diagonal contributions based on the corresponding event counts. The resulting total covariance matrix is held fixed during the oscillation fit rather than rescaled at each trial parameter point. This approximation neglects the dependence of the covariance on the oscillation parameters and is expected to be adequate in the small-mixing region relevant to the sensitivity limits.

\begin{figure}[t]
\centering
  \includegraphics[width=0.48\textwidth]{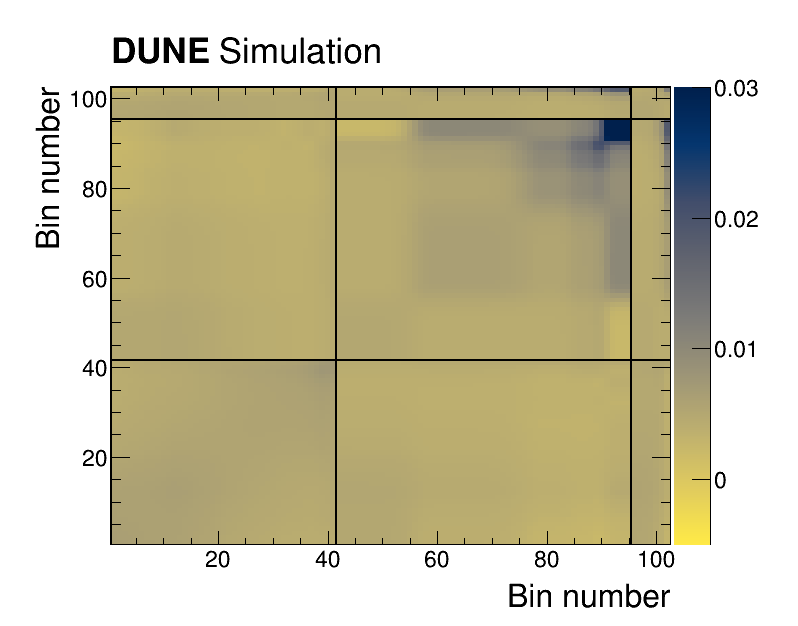}
\caption{Fractional covariance matrix associated with neutrino flux uncertainties for the nominal selection requirements described in Sec.~\ref{sec:data}. Analysis bins 1--41 correspond to the CC$\nu_\mu$ sample, bins 42--95 to the CC$\nu_e$ sample, and bins 96--102 to the $\nu e^-\to\nu e^-$ sample.}
\label{fig:fluxCov}
\end{figure}

Flux uncertainties are taken from the DUNE long-baseline oscillation sensitivity studies and are described in detail in Ref.~\cite{LBLpotential}. The dominant contribution arises from uncertainties in hadron production in the target, with additional contributions from the magnetic focusing system. The corresponding flux covariance matrix is defined in true neutrino-energy space. For this analysis, correlated flux variations derived from this covariance are propagated through the simulated true-neutrino-energy versus reconstructed-lepton-energy distributions to construct the covariance in the reconstructed-lepton-energy space of the three analysis samples. Figure~\ref{fig:fluxCov} shows the resulting fractional covariance matrix for the nominal selection requirements described in Sec.~\ref{sec:data}. The dependence of the flux uncertainties on the $\nu^{\mathrm{reco}}$ requirement is discussed further in Appendix~\ref{apdx:fluxCov}.

Another important source of systematic uncertainty arises from the modeling of neutrino-nucleus interactions. The selected CC samples receive substantial contributions from quasi-elastic (QE) scattering, with smaller contributions from single-pion production and shallow-inelastic scattering. Interaction-model uncertainties are evaluated using 68 GENIE systematic parameter variations adopted from the DUNE long-baseline sensitivity studies~\cite{LBLpotential}.

The low-$\nu$ selection reduces reliance on hadronic-energy reconstruction in the oscillation observable and can reduce sensitivity to some interaction-model effects, but it does not eliminate model dependence. The normalization and shape components of the GENIE uncertainty for the selected low-$\nu$ CC samples are discussed further in Appendix~\ref{apdx:xSModel}. In particular, uncertainties in the low-energy-transfer cross section and in the mapping between true and reconstructed energy transfer may not be fully spanned by the present nuisance-parameter model. Comparisons among alternative interaction models have shown potentially larger variations in this kinematic region~\cite{SubstandardCandle}.

\begin{figure}[t]
\centering
  \includegraphics[width=0.48\textwidth]{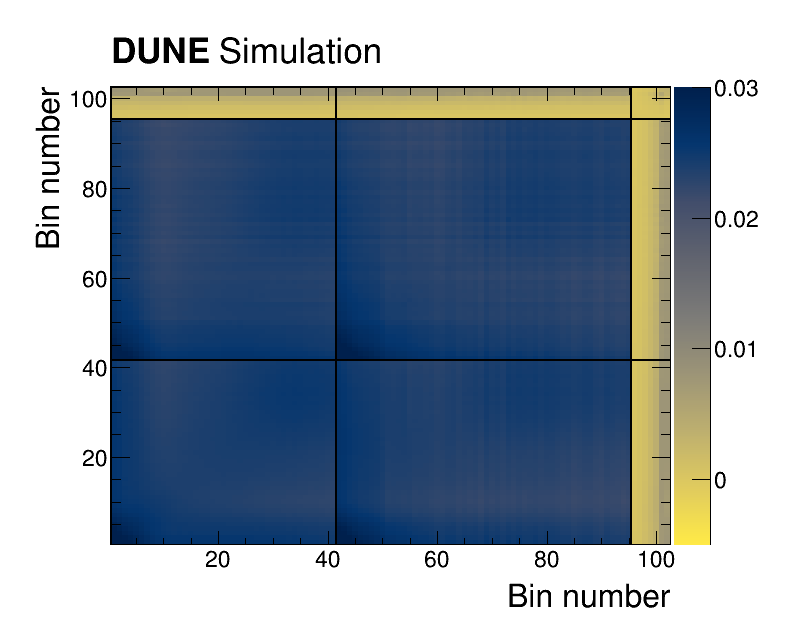}
\caption{Fractional covariance matrix associated with neutrino interaction cross-section uncertainties for the nominal selection requirements described in Sec.~\ref{sec:data}. Analysis bins 1--41 correspond to the CC$\nu_\mu$ sample, bins 42--95 to the CC$\nu_e$ sample, and bins 96--102 to the $\nu e^-\to\nu e^-$ sample.}
\label{fig:xScov_nue}
\end{figure}

A separate 0.5\% normalization uncertainty, taken to be fully correlated among reconstructed-energy bins, is assigned to neutrino-electron elastic scattering. This accounts for the residual theoretical uncertainty in the precisely calculable electroweak cross section, including radiative corrections~\cite{nueScat}, and contributes directly to the uncertainty on the absolute flux normalization inferred from the $\nu e^-\to\nu e^-$ sample. The total interaction cross-section covariance matrix, including both neutrino-nucleus and neutrino-electron scattering uncertainties, is shown in Fig.~\ref{fig:xScov_nue}. A view of the covariance matrix restricted to the two CC samples, with an independently scaled color axis to highlight their covariance structure, is provided in Appendix~\ref{apdx:xSModel} (Fig.~\ref{fig:xScov_CC}).

Detector-response uncertainties affect both lepton-energy reconstruction and the hadronic-energy response entering the $\nu^{\mathrm{reco}}$ selection. Following Ref.~\cite{LBLpotential}, systematic variations are applied to the nominal detector response described in Sec.~\ref{sec:data}. Uncertainties of 2\% and 2.5\% are assigned to the reconstructed muon and electron energy scales, respectively, and are treated as uncorrelated between the two lepton flavors because they rely on distinct reconstruction methods. A 10\% uncertainty is assigned to the reconstructed energy resolution for each lepton flavor and is also treated as uncorrelated between muons and electrons. The scale uncertainties produce coherent shifts in the reconstructed lepton energies, whereas the resolution uncertainties modify the widths of the reconstructed energy response distributions.

For the hadronic response, a 2\% uncertainty is assigned to the overall energy scale. An additional 5\% systematic variation is assigned to the reconstructed-energy response of protons and charged pions, following Ref.~\cite{LBLpotential} and motivated in part by the proton energy-scale uncertainty achieved by NOvA~\cite{NOvA:2018gge}. This response variation is distinct from the substantially larger event-by-event hadronic energy resolution. These uncertainties affect the selected event spectra through migration across the $\nu^{\mathrm{reco}}$ requirement. The hadronic-response variations are treated independently of the electron energy-scale uncertainty; the former modify the visible hadronic energy entering $\nu^{\mathrm{reco}}$, while the latter modifies the reconstructed electromagnetic-shower energy.

\begin{figure}[t]
\centering
  \includegraphics[width=0.48\textwidth]{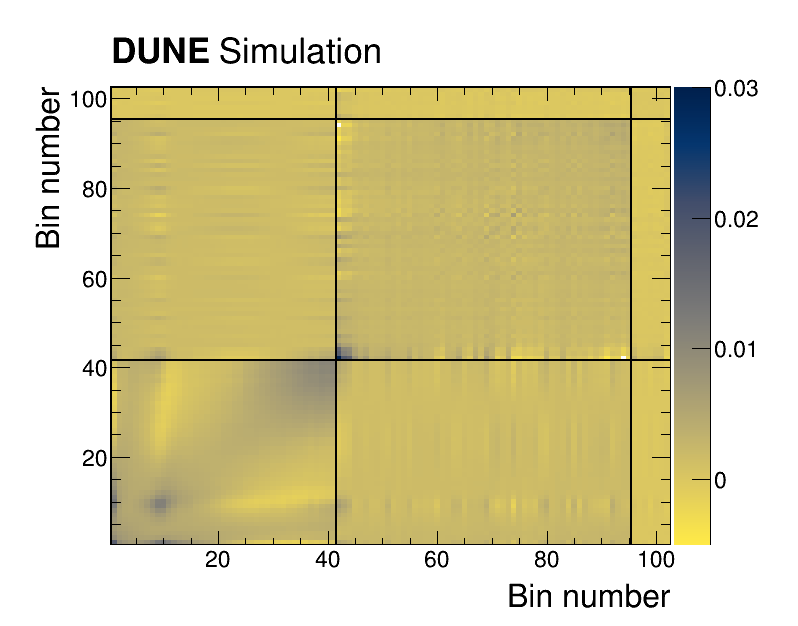}
\caption{Fractional covariance matrix associated with detector-response uncertainties for the nominal selection requirements described in Sec.~\ref{sec:data}. Analysis bins 1--41 correspond to the CC$\nu_\mu$ sample, bins 42--95 to the CC$\nu_e$ sample, and bins 96--102 to the $\nu e^-\to\nu e^-$ sample.}
\label{fig:detCov}
\end{figure}

Neutrons are challenging to reconstruct because only a fraction of their energy is typically visible, and the detector response depends sensitively on neutron propagation and secondary interactions in liquid argon. In this analysis, visible energy from neutron-induced charged particles and photons, including de-excitation and capture $\gamma$ rays that produce MeV-scale energy deposits, is included in the reconstructed $\nu$~\cite{Friedland:2018vry,Castiglioni:2020tsu,HernandezMorquecho:2026hdk}. A 30\% systematic uncertainty is applied to the nominal reconstructed neutron-energy response, motivated by the neutron-response uncertainties considered in Ref.~\cite{LBLpotential}. The uncertainty is implemented as two discrete shifts,
\[
E^{N\,\prime}_{\mathrm{reco}}
=
(1\pm0.3)E^{N}_{\mathrm{reco}}.
\]
For neutrons with true energy below 200~MeV, an additional stochastic response variation is included by replacing the nominal reconstructed neutron energy with a value sampled uniformly between zero and the corresponding true neutron energy,
\[
E^{N\,\prime}_{\mathrm{reco}}
\sim
\mathcal{U}\!\left(0,E^{N}_{\mathrm{true}}\right).
\]
Figure~\ref{fig:detCov} shows the total detector-response covariance matrix. More details about the detector response uncertainties are given in Appendix~\ref{apdx:detCov}.

\section{Analysis Technique}
\label{sec:technique}

Fits are performed as a function of reconstructed lepton energy in the three analysis samples described in Sec.~\ref{sec:data}. The sterile-neutrino model is parameterized by $|U_{e4}|^2$, $|U_{\mu4}|^2$, $|U_{\tau4}|^2$, and $\Delta m^2$. These parameters govern the $\nu_{\mu} \rightarrow \nu_{\mu}$, $\nu_{\mu} \rightarrow \nu_e$, $\nu_{\mu} \rightarrow \nu_{\tau}$, $\nu_e \rightarrow \nu_e$, $\nu_e \rightarrow \nu_{\mu}$, and $\nu_e \rightarrow \nu_{\tau}$ oscillation channels.

\begin{figure*}[t]
\centering
\includegraphics[width=0.48\textwidth]{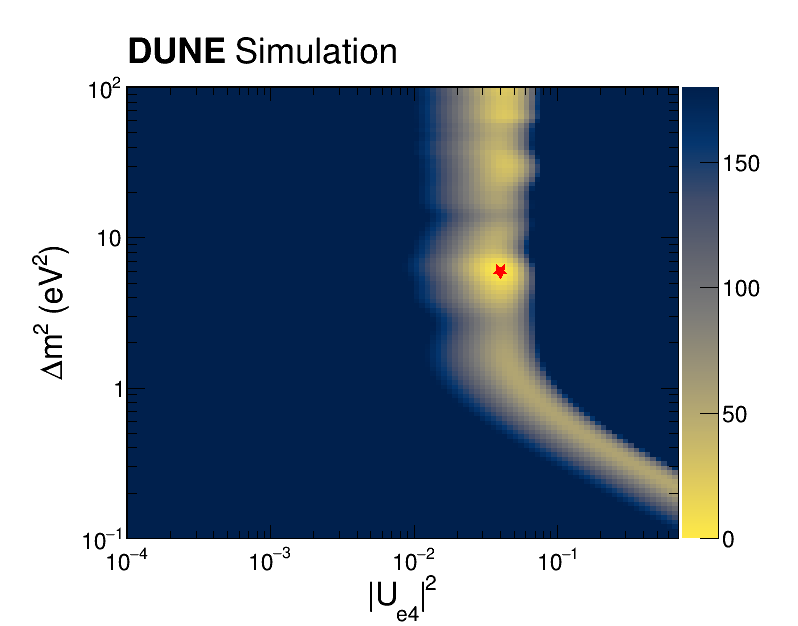}\hfill
\includegraphics[width=0.48\textwidth]{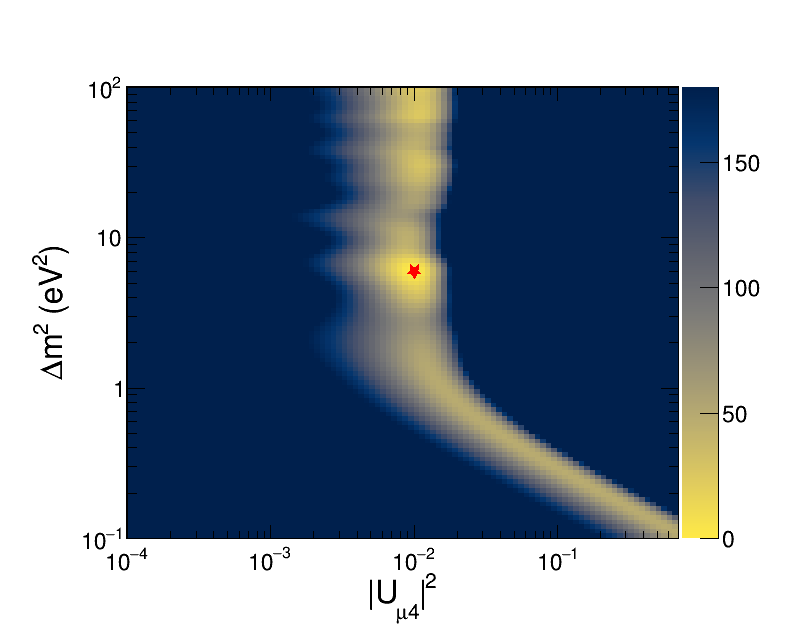}\\
\includegraphics[width=0.48\textwidth]{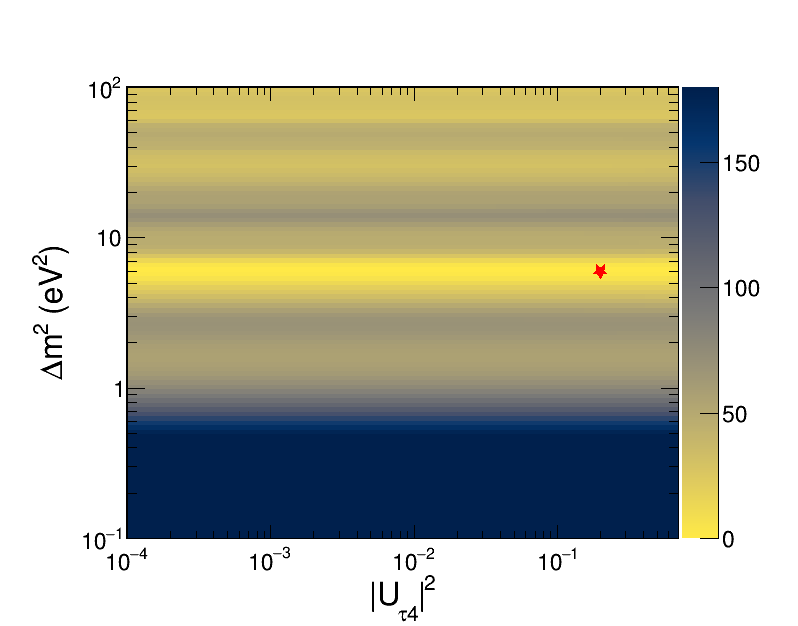}\hfill
\includegraphics[width=0.48\textwidth]{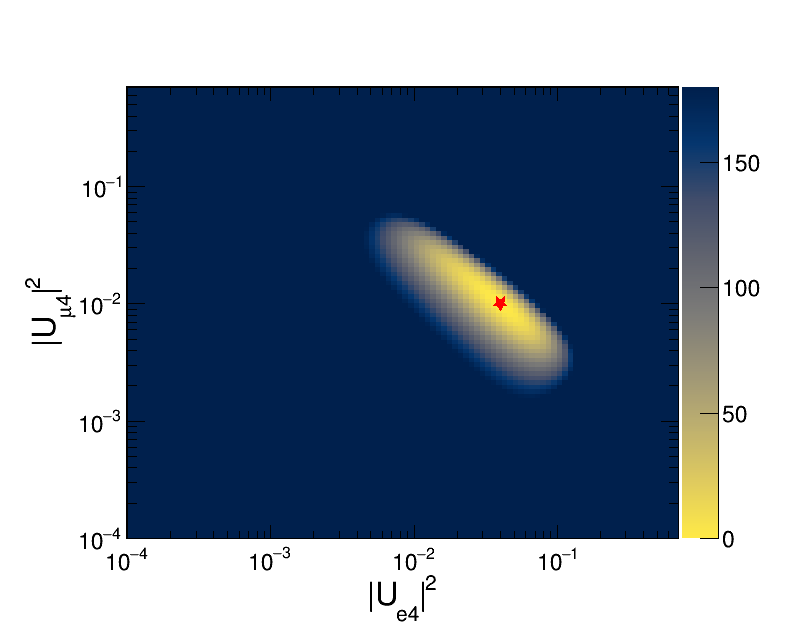}
\caption{Two-dimensional slices of the four-dimensional $\Delta\chi^2$ surface in the $(\Delta m^2,|U_{e4}|^2)$ (top left), $(\Delta m^2,|U_{\mu4}|^2)$ (top right), $(\Delta m^2,|U_{\tau4}|^2)$ (bottom left), and $(|U_{\mu4}|^2,|U_{e4}|^2)$ (bottom right) planes. In each slice, the two parameters not shown are fixed to their injected values. The injected oscillation hypothesis is $\vec{x}_0=(|U_{e4}|^2,|U_{\mu4}|^2,|U_{\tau4}|^2,\Delta m^2)=(0.04,\,0.01,\,0.2,\,6.0~\mathrm{eV}^2)$, indicated by the red star.}
\label{fig:dchi2Surfaces}
\end{figure*}

The three samples provide complementary sensitivity to sterile-neutrino oscillations. The $\nu_\mu$-CC sample primarily probes $\nu_\mu$ disappearance, while the $\nu_e$-CC sample is sensitive to both $\nu_e$ disappearance and $\nu_\mu\rightarrow\nu_e$ appearance. The $\nu e^-\to\nu e^-$ sample primarily constrains the neutrino flux, while retaining direct oscillation sensitivity through the flavor dependence of the elastic-scattering cross section. Oscillation channels involving $\nu_\tau$ are also included in the fit.

Figure~\ref{fig:tgt} (right) shows the reconstructed lepton-energy distributions for the three analysis samples under the illustrative oscillation hypothesis
\[
(|U_{e4}|^2, |U_{\mu4}|^2, |U_{\tau4}|^2, \Delta m^2)
=
(0.04,\,0.01,\,0.2,\,6.0~\mathrm{eV}^2),
\]
for an exposure of $1.1\times10^{21}$ POT. This benchmark is chosen to illustrate a clearly visible oscillation signal in the DUNE ND rather than a currently allowed parameter point, as recent experimental results have excluded this particular hypothesis~\cite{KATRIN,PROSPECT}. The $\nu_\mu$-CC spectrum is dominated by $\nu_\mu$ disappearance, while the $\nu_e$-CC spectrum reflects the interplay of $\nu_e$ disappearance and $\nu_\mu\rightarrow\nu_e$ appearance. The $\nu e^-\to\nu e^-$ sample also retains sensitivity to the appearance contribution through the flavor dependence of the elastic-scattering cross section.

The minimization process is performed using the \textsc{ROOT Minuit2} package~\cite{Minuit2}, which determines the best-fit oscillation parameters by minimizing
\begin{equation}
    \chi^2(\vec{x})
    =
    \sum_{i=1}^{N_{\mathrm{bins}}}
    \sum_{j=1}^{N_{\mathrm{bins}}}
    \left[M_i(\vec{x}) - D_i\right]
    V^{-1}_{ij}
    \left[M_j(\vec{x}) - D_j\right],
    \label{eqn:chi2}
\end{equation}
where $\vec{x} = (|U_{e4}|^2, |U_{\mu4}|^2, |U_{\tau4}|^2, \Delta m^2)$ denotes the oscillation parameters varied in the fit, $M_i(\vec{x})$ is the simulated prediction in bin $i$ evaluated at the trial parameter point $\vec{x}$, and $D_i$ is the corresponding observed or pseudo-data count. The covariance matrix $V_{ij}$ encodes the total uncertainty and correlations among bins, as described in Sec.~\ref{sec:uncertainties}. The combined fit contains 102 bins across the three analysis samples.

The degree to which a particular oscillation hypothesis $\vec{x}_0$ is disfavored is quantified by comparing the $\chi^2$ evaluated at the fixed test point with the minimum $\chi^2$ obtained by varying the oscillation parameters:
\begin{equation}
    \Delta\chi^2
    =
    \chi^2(\vec{x}_0)
    -
    \chi^2(\vec{x}_{\mathrm{BF}}),
    \label{eq:deltachi2}
\end{equation}
where $\vec{x}_{\mathrm{BF}}$ denotes the best-fit oscillation parameters obtained from the global minimization. By definition, $\chi^2(\vec{x}_{\mathrm{BF}})\leq\chi^2(\vec{x}_0)$, therefore $\Delta\chi^2\geq0$.

To illustrate the structure of the multidimensional $\Delta\chi^2$ surface, Fig.~\ref{fig:dchi2Surfaces} shows two-dimensional slices constructed using the illustrative injected oscillation hypothesis introduced above. These slices are used only to illustrate the parameter dependence and are not used to determine the expected sensitivity. Four slices through the four-dimensional parameter space are shown: $\Delta m^2$ versus each of $|U_{e4}|^2$, $|U_{\mu4}|^2$, and $|U_{\tau4}|^2$, and the $|U_{\mu4}|^2$--$|U_{e4}|^2$ plane. In each slice, the two parameters not shown are fixed to their injected values. The injected point corresponds to $\Delta\chi^2=0$ in each slice.

Localized low-$\Delta\chi^2$ regions are observed in both $|U_{e4}|^2$ and $|U_{\mu4}|^2$. The two parameters exhibit the expected anti-correlation, driven in part by the $\nu_\mu\rightarrow\nu_e$ appearance probability, whose amplitude depends on the product $|U_{e4}|^2|U_{\mu4}|^2$ as shown in Eq.~\ref{eqn:MixingAngles}. In $\Delta m^2$, the surface exhibits the expected periodic structure arising from the sinusoidal dependence of the oscillation probability on $\Delta m^2$, with the lowest $\Delta\chi^2$ region centered near the injected value. This structure produces multiple regions of reduced $\Delta\chi^2$ whose precise locations depend on correlations among the oscillation parameters. By contrast, the $\Delta\chi^2$ surface is largely flat in $|U_{\tau4}|^2$, indicating little sensitivity to this parameter over the physically allowed range. The limited sensitivity to $|U_{\tau4}|^2$ arises primarily through the flavor-dependent NC contribution to the $\nu e^-\to\nu e^-$ sample. The parameter is nevertheless retained in the fit because allowing $|U_{\tau4}|^2$ to vary slightly weakens the constraint on $|U_{\mu4}|^2$.

To determine DUNE's expected sensitivity to sterile-neutrino oscillations for one year-equivalent of on-axis ND exposure, corresponding to $1.1\times10^{21}$ POT, a separate $\Delta\chi^2$ surface is constructed using an Asimov data set generated under the null hypothesis,
\[
|U_{e4}|^2 = |U_{\mu4}|^2 = |U_{\tau4}|^2 = 0.
\]
For this data set, the global best fit corresponds to the no-oscillation hypothesis and has $\chi^2(\vec{x}_{\mathrm{BF}})=0$. The test statistic at each oscillation hypothesis $\vec{x}_0$ therefore reduces to
\[
\Delta\chi^2(\vec{x}_0)=\chi^2(\vec{x}_0).
\]
The resulting $\Delta\chi^2$ surface describes the expected incompatibility of each oscillation hypothesis with data generated in the absence of sterile-neutrino oscillations.

The confidence-level thresholds used to interpret this $\Delta\chi^2$ surface are determined separately using a Feldman--Cousins (FC) construction~\cite{Feldman:1997qc}. An ensemble of pseudo-experiments is generated under the no-oscillation hypothesis. For each pseudo-experiment, a correlated Gaussian fluctuation is drawn from the total covariance matrix, including statistical, flux, interaction-model, and detector-response uncertainties. The statistical contribution is included as the diagonal counting covariance in the Gaussian approximation, allowing statistical and systematic fluctuations to be generated simultaneously. The test statistic in Eq.~\ref{eq:deltachi2} is then evaluated with $\vec{x}_0$ fixed to the no-oscillation hypothesis, while the oscillation parameters are allowed to vary in determining the global best fit. Statistical and systematic fluctuations can therefore shift the best-fit point away from the no-oscillation hypothesis. The resulting ensemble distribution of $\Delta\chi^2$ is used to determine the empirical critical values for the desired confidence levels.

As a computational approximation, the FC construction is performed only for the no-oscillation hypothesis rather than repeated at every point in the four-dimensional oscillation parameter space. The resulting null-hypothesis critical values are used throughout the sensitivity calculation. This approximation is expected to be most appropriate near the sensitivity boundary, where the oscillation amplitudes are small, but does not guarantee exact point-by-point coverage and may become less accurate for large oscillation amplitudes.

Bounds are imposed on the mixing parameters following Ref.~\cite{nonUnitary}:
\begin{equation}
\begin{split}
    |U_{e4}|^2 &\in [0,\,0.394), \\
    |U_{\mu4}|^2 &\in [0,\,0.489), \\
    |U_{\tau4}|^2 &\in [0,\,0.718), \\
    |U_{e4}|^2 + |U_{\mu4}|^2 + |U_{\tau4}|^2 &\leq 1, \\
    \Delta m^2 &\in [0,\,\infty).
\end{split}
\label{eq:OscPara_Ranges_dchi2Fit}
\end{equation}
The final condition on the mixing parameters enforces the requirement implied by normalization of the fourth mass-eigenstate column and excludes unphysical regions in which the total active component exceeds unity. The mass-squared splitting is restricted to nonnegative values. These limits do not affect the sensitivity results but improve computational efficiency by preventing the minimization from exploring unphysical or weakly constrained regions where the $\chi^2$ surface becomes nearly flat in one or more parameters.

\section{Results}
\label{sec:results}

\begin{table*}[t]
\centering
    \caption{Feldman--Cousins critical $\Delta\chi^2$ values obtained under the no-oscillation hypothesis, together with Wilks expectations for comparison. The four-degree-of-freedom expectation corresponds to the four oscillation parameters varied in the global fit. The three-degree-of-freedom expectation is also shown as a reference because the fit has weak sensitivity to $|U_{\tau4}|^2$.}
    {\large
    \begin{tabular}{|c|c|c|c|}
        \hline
        \hspace{15pt} Confidence Level (C.L.) \hspace{15pt} & \hspace{15pt} FC $\Delta\chi^2$ \hspace{15pt} & \hspace{15pt} Wilks $\Delta\chi^2$ 3dof \hspace{15pt} & \hspace{15pt} Wilks $\Delta\chi^2$ 4dof \hspace{15pt} \\
        \hhline{|=|=|=|=|}
        $1\sigma$ ($68.2\%$) & 6.71  & 3.53  & 4.72  \\ \hline
        $90\%$               & 10.06 & 6.25  & 7.78  \\ \hline
        $2\sigma$ ($95.4\%$) & 12.25 & 8.02  & 9.72  \\ \hline
        $3\sigma$ ($99.7\%$) & 21.05 & 14.16 & 16.25 \\ \hline
    \end{tabular}
    }
    \label{tab:FCdchi2}
\end{table*}

\begin{figure}[t]
\centering
  \includegraphics[width=0.48\textwidth]{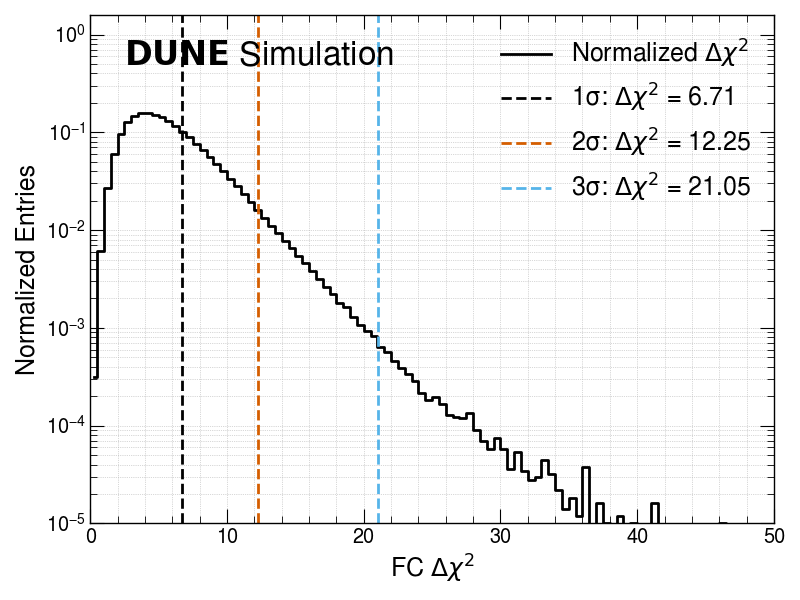}
\caption{Feldman--Cousins $\Delta\chi^2$ distribution obtained under the no-oscillation hypothesis, with the corresponding empirical $1\sigma$, $2\sigma$, and $3\sigma$ critical values.}
\label{fig:FCdchi2}
\end{figure}

The FC construction described in Sec.~\ref{sec:technique} is performed using the three-stage fitting procedure detailed in Appendix~\ref{apdx:FCfit}, with all four oscillation parameters allowed to vary in determining the global best fit. Figure~\ref{fig:FCdchi2} shows the $\Delta\chi^2$ distribution obtained from pseudo-experiments generated under the no-oscillation hypothesis. At 90\% C.L., the FC construction gives $\Delta\chi^2_{\mathrm{FC}}=10.06$, compared with the four-degree-of-freedom Wilks expectation of $\Delta\chi^2_{\mathrm{Wilks}}=7.78$. The larger FC critical values may reflect departures from the regularity conditions underlying the asymptotic approximation, including physical parameter boundaries, degeneracies in the oscillation parameter space, and weak sensitivity to $|U_{\tau4}|^2$. The critical values at additional confidence levels are summarized in Table~\ref{tab:FCdchi2}.

\begin{figure}[t]
\centering
  \includegraphics[width=0.48\textwidth]{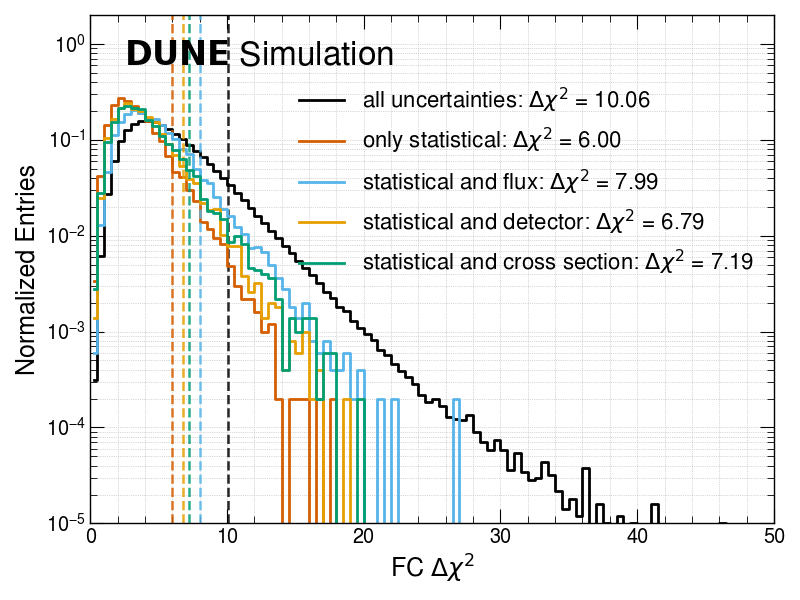}
\caption{Feldman--Cousins $\Delta\chi^2$ distributions obtained under the no-oscillation hypothesis for different combinations of systematic uncertainties, with the corresponding 90\% C.L. critical values.}
\label{fig:FCdchi2_uncertainties}
\end{figure}

Figure~\ref{fig:FCdchi2_uncertainties} shows the effect of the different systematic-uncertainty categories on the FC $\Delta\chi^2$ distribution. Among the uncertainties considered, the neutrino-flux uncertainty produces the largest broadening of the distribution and the largest increase in the corresponding 90\% C.L. critical value. The interaction-model and detector-response uncertainties have smaller effects individually, even though their bin-by-bin fractional uncertainties can be larger in parts of the analysis phase space.

\begin{figure*}[t]
\centering
  \includegraphics[width=0.33\textwidth]{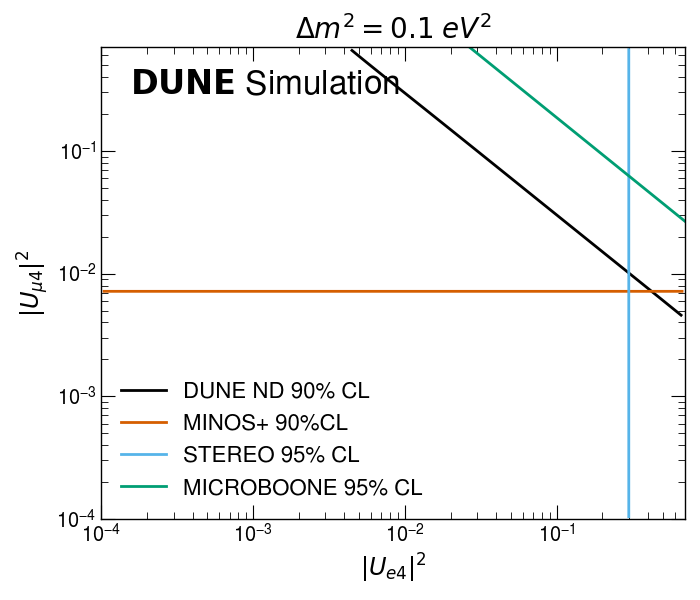}%
  \includegraphics[width=0.33\textwidth]{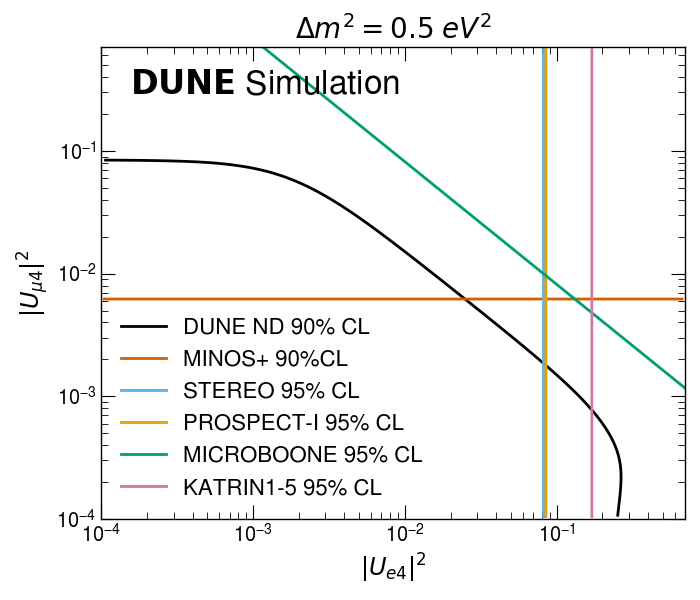}%
  \includegraphics[width=0.33\textwidth]{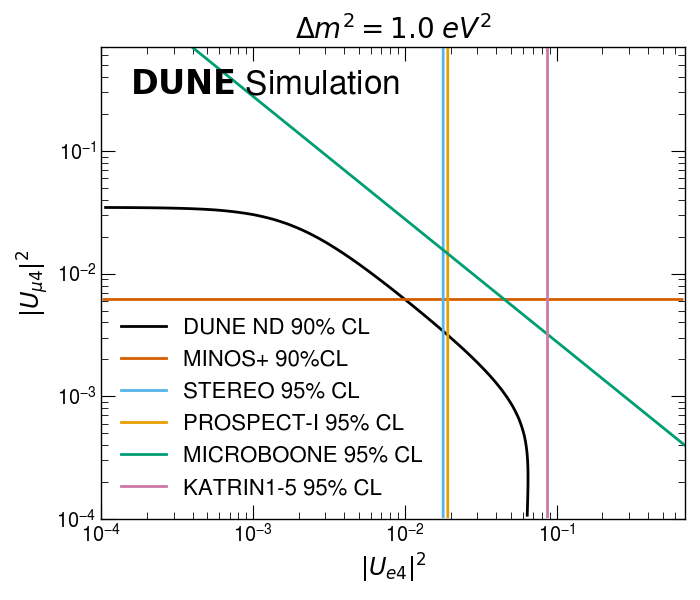}\\
  \includegraphics[width=0.33\textwidth]{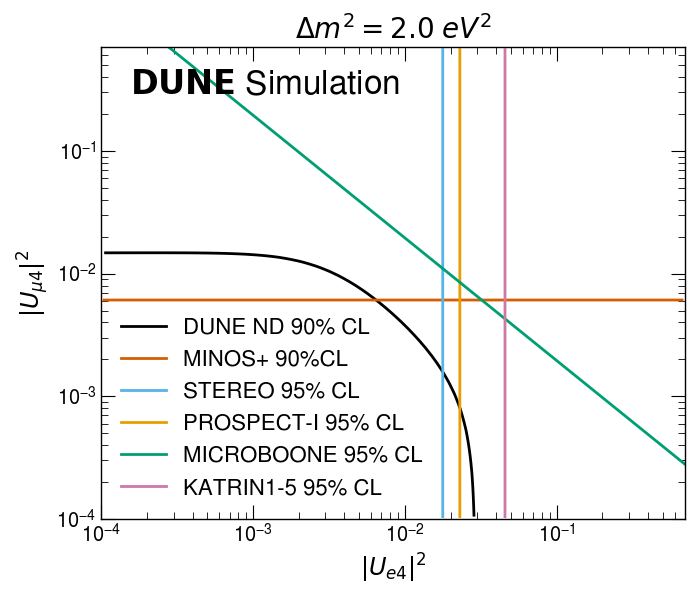}%
  \includegraphics[width=0.33\textwidth]{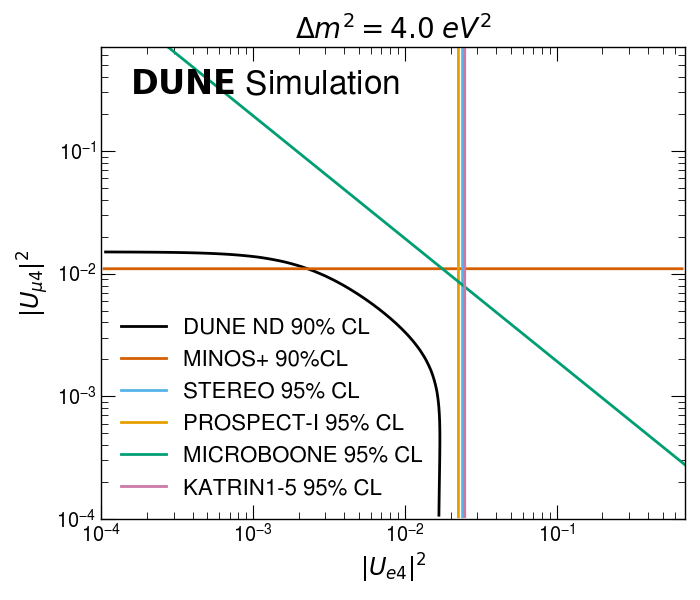}%
  \includegraphics[width=0.33\textwidth]{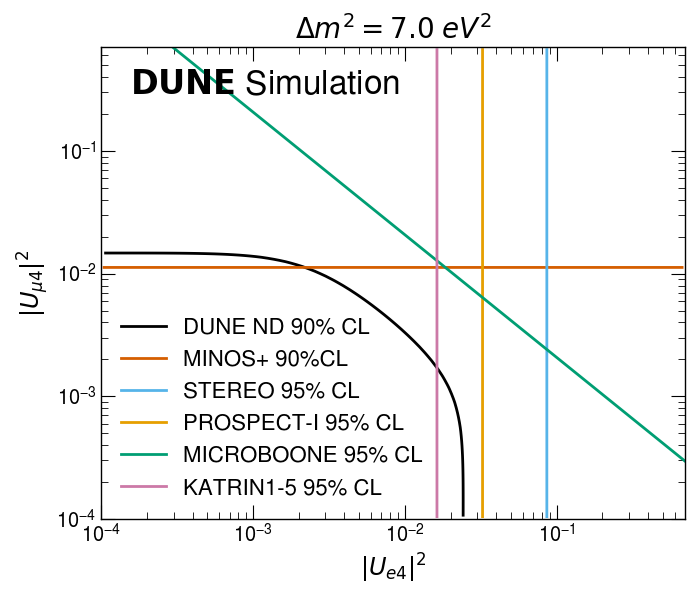}\\
  \includegraphics[width=0.33\textwidth]{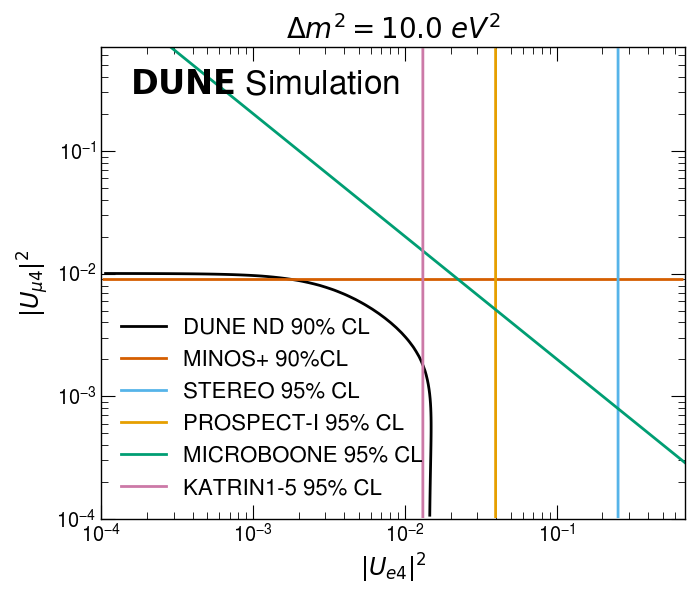}%
  \includegraphics[width=0.33\textwidth]{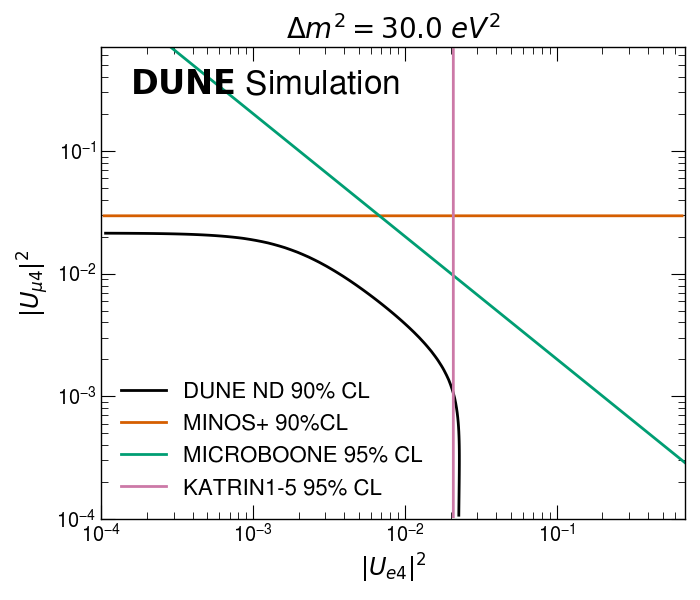}%
  \includegraphics[width=0.33\textwidth]{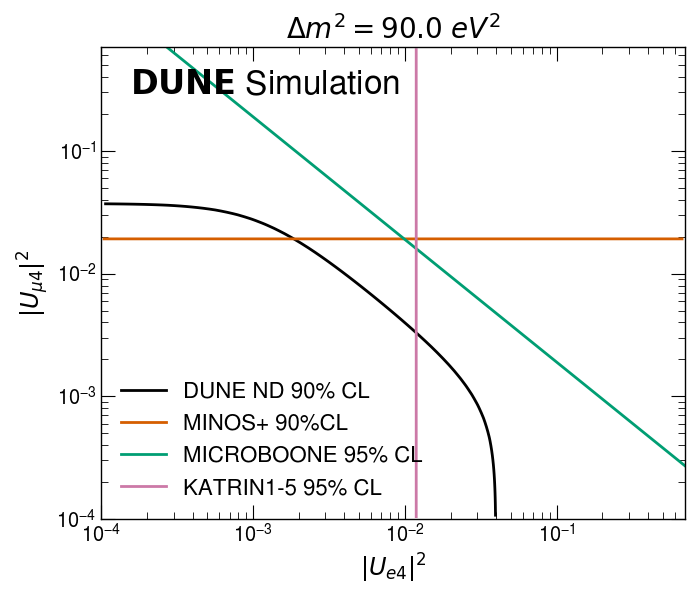}\\
\caption{DUNE 90\% C.L. sensitivity contours in fixed $\Delta m^2$ slices with $|U_{\tau4}|^2=0$, shown together with existing constraints from other experiments~\cite{KATRIN,MicroBooNE_nature,MINOS:sterile,PROSPECT,STEREO}. The confidence levels of the external constraints are those reported by the respective experiments.}
\label{fig:90CL_mue}
\end{figure*}

Figure~\ref{fig:90CL_mue} shows the DUNE 90\% C.L. sensitivity in fixed-$\Delta m^2$ slices of the multidimensional Asimov sensitivity surface, with $|U_{\tau4}|^2=0$, together with existing experimental constraints. These slices illustrate the expected sensitivity across different values of $\Delta m^2$ and should not be interpreted as independent two-parameter confidence regions. A common 90\% C.L. critical value, $\Delta\chi^2_{\mathrm{FC}}=10.06$, obtained from the null-hypothesis FC ensemble, is used for all slices as an approximation to a full point-by-point FC construction. Additional two-dimensional sensitivity contours in complementary parameter planes are shown in Appendix~\ref{apdx:contours}.

The sensitivity varies substantially with $\Delta m^2$. For $\Delta m^2\lesssim1~\mathrm{eV}^2$, only a small oscillation phase develops over the ND baseline for most of the relevant neutrino-energy range, resulting in limited sensitivity. The strongest sensitivity occurs for $\Delta m^2\sim4$--$10~\mathrm{eV}^2$, where both absolute event-rate changes and resolvable energy-dependent distortions contribute. In this region, the sensitivity reaches approximately $|U_{\mu4}|^2\sim0.01$ and $|U_{e4}|^2\sim0.015$, depending on the value of the other mixing parameter. At larger $\Delta m^2$, the oscillatory spectral structure becomes increasingly averaged over the neutrino production region, detector response, and analysis binning, and the sensitivity becomes dominated primarily by absolute event-rate changes relative to the null prediction. Representative spectra illustrating these different $\Delta m^2$ regimes are shown in Appendix~\ref{apdx:binning}.

Removing the $\nu e^-\to\nu e^-$ sample from the combined analysis results in a modest degradation in the expected sensitivity. Its contribution is particularly relevant at large $\Delta m^2$, where oscillatory spectral features become averaged and the sensitivity relies more strongly on absolute-rate information. In addition to constraining the incident neutrino flux, this sample retains direct sensitivity to changes in the beam flavor composition through the flavor dependence of the elastic-scattering cross section.

The sensitivity presented here is based on the nominal on-axis ND configuration. The DUNE-PRISM program will provide measurements at different off-axis positions, sampling neutrino spectra with different energy and flavor compositions. These measurements can provide additional data-driven constraints on flux and neutrino-interaction uncertainties and may improve sterile-neutrino sensitivity, particularly in regions where spectral systematics limit the measurement~\cite{Hernandez-Garcia:2026asl,SBND:2025pjc}.

\section{Conclusion}
\label{chap:conclusion}

We have presented a sterile-neutrino sensitivity study at the DUNE Near Detector in the $3+1$ framework using highly elastic $\nu_\mu$-CC and $\nu_e$-CC interactions together with neutrino-electron elastic scattering. The highly elastic CC selections exploit the small reconstructed hadronic energy to enhance the correlation between the incident neutrino energy and the reconstructed charged-lepton energy, reducing reliance on hadronic-energy reconstruction in the oscillation observable. The $\nu e^-\to\nu e^-$ sample provides a complementary constraint on the incident neutrino flux while retaining sensitivity to changes in its flavor composition.

The combined analysis uses disappearance, appearance, and flux-constraining information to probe a broad region of sterile-neutrino parameter space. The strongest sensitivity occurs for $\Delta m^2\sim4$--$10~\mathrm{eV}^2$, where both spectral distortions and absolute-rate changes contribute. At lower $\Delta m^2$, only a small oscillation phase develops over the accessible $L/E_\nu$ range, reducing the sensitivity. At larger $\Delta m^2$, the oscillations become increasingly averaged, washing out the spectral structure and leaving the sensitivity dominated primarily by absolute-rate information. In this regime, the flux constraint provided by neutrino-electron scattering helps preserve sensitivity to oscillation-induced rate changes.

In the parameter slices considered, the sensitivity reaches approximately $|U_{\mu4}|^2\sim0.01$ and $|U_{e4}|^2\sim0.015$ at 90\% C.L. for one year-equivalent of on-axis DUNE ND exposure in FHC mode, depending on the value of the other mixing parameter. The reach reflects the complementary sensitivity of the electron- and muon-flavor samples to disappearance and appearance channels. These results demonstrate the potential of the DUNE ND to perform an independent sterile-neutrino search using near-detector data alone, with sensitivity arising from both spectral and absolute-rate information.

The sensitivity reported here includes the neutrino-interaction uncertainties represented by the systematic model used in the analysis. Although the low-$\nu$ selection can reduce sensitivity to some interaction-model effects, additional model dependence may arise from the low-energy-transfer cross section and from the mapping between true and reconstructed energy transfer, which can alter migration into and out of the selected sample. Comparisons with alternative interaction models will be important for assessing the robustness of these sensitivity estimates. Targeted theoretical work in this restricted low-energy-transfer regime could help reduce these uncertainties and better constrain the residual model dependence of low-$\nu$-based measurements.

Near-detector sterile-neutrino measurements also provide an important cross-check for the standard DUNE oscillation program, since short-baseline oscillations that affect the interpretation of far-detector data could produce observable distortions in the near-detector samples.

\vspace{30pt}

\ifnum\sizecheck=0
  \begin{acknowledgments}

This document was prepared by the DUNE collaboration using the resources of the Fermi National Accelerator Laboratory (Fermilab), a U.S. Department of Energy, Office of Science, Office of High Energy Physics HEP User Facility. Fermilab is managed by Fermi Forward Discovery Group, LLC, acting under Contract No.~89243024CSC000002. This work was supported by CNPq, FAPERJ, FAPEG, FAPESP, and Funda{\c{c}}{\~a}o Arauc{\'a}ria, Brazil; CFI, IPP, and NSERC, Canada; CERN; ANID-FONDECYT, Chile; MSMT, Czech Republic; ERDF, FSE+, Horizon Europe, MSCA, and NextGenerationEU, European Union; CNRS/IN2P3 and CEA, France; PRISMA+, Germany; INFN, Italy; FCT, Portugal; CERN-RO/CDI, Romania; NRF, South Korea; Generalitat Valenciana, Junta de Andaluc{\'\i}a, MICINN, and Xunta de Galicia, Spain; SERI and SNSF, Switzerland; T{\"U}B{\.I}TAK, Turkey; The Royal Society and UKRI/STFC, United Kingdom; DOE and NSF, United States of America.

\end{acknowledgments}
  \bibliography{biblio}

@article{PDG,
    author = "Takahashi, F. and others",
    collaboration = "Particle Data Group",
    title = "{Review of Particle Physics}",
    doi = "10.1142/S0217751X26300115",
    journal = "Int. J. Mod. Phys. A",
    volume = "41",
    pages = "2630011",
    year = "2026"
}

@article{MINERvA:lownu,
    author = "Devan, J. and others",
    collaboration = "MINERvA",
    title = "{Measurements of the Inclusive Neutrino and Antineutrino Charged Current Cross Sections in MINERvA Using the Low-$\nu$ Flux Method}",
    eprint = "1610.04746",
    archivePrefix = "arXiv",
    primaryClass = "hep-ex",
    reportNumber = "FERMILAB-PUB-16-368-ND",
    doi = "10.1103/PhysRevD.94.112007",
    journal = "Phys. Rev. D",
    volume = "94",
    number = "11",
    pages = "112007",
    year = "2016"
}

@article{SubstandardCandle,
    author = "Wilkinson, Callum and others",
    title = "{A substandard candle: the low-$\nu $ method at few-GeV neutrino energies}",
    eprint = "2203.11821",
    archivePrefix = "arXiv",
    primaryClass = "hep-ph",
    doi = "10.1140/epjc/s10052-022-10754-7",
    journal = "Eur. Phys. J. C",
    volume = "82",
    number = "9",
    pages = "808",
    year = "2022"
}

@article{nueScattering,
    author = "Marshall, Chris M. and others",
    title = "{Neutrino-electron elastic scattering for flux determination at the DUNE oscillation experiment}",
    eprint = "1910.10996",
    archivePrefix = "arXiv",
    primaryClass = "hep-ex",
    doi = "10.1103/PhysRevD.101.032002",
    journal = "Phys. Rev. D",
    volume = "101",
    number = "3",
    pages = "032002",
    year = "2020"
}

@article{LBLpotential,
    author = "Abi, B. and others",
    collaboration = "DUNE",
    title = "{Long-baseline neutrino oscillation physics potential of the DUNE experiment}",
    eprint = "2006.16043",
    archivePrefix = "arXiv",
    primaryClass = "hep-ex",
    reportNumber = "FERMILAB-PUB-20-251-E-LBNF-ND-PIP2-SCD, PUB-20-251-E-LBNF-ND-PIP2-SCD",
    doi = "10.1140/epjc/s10052-020-08456-z",
    journal = "Eur. Phys. J. C",
    volume = "80",
    number = "10",
    pages = "978",
    year = "2020"
}

@article{Zres,
    author = "Schael, S. and others",
    collaboration = "ALEPH, DELPHI, L3, OPAL, SLD, LEP Electroweak Working Group, SLD Electroweak Group, SLD Heavy Flavour Group",
    title = "{Precision electroweak measurements on the $Z$ resonance}",
    eprint = "hep-ex/0509008",
    archivePrefix = "arXiv",
    reportNumber = "SLAC-R-774",
    doi = "10.1016/j.physrep.2005.12.006",
    journal = "Phys. Rept.",
    volume = "427",
    pages = "257--454",
    year = "2006"
}

@misc{DUNE:TDR,
    author = "Abi, Babak and others",
    collaboration = "DUNE",
    title = "{Deep Underground Neutrino Experiment (DUNE), Far Detector Technical Design Report, Volume II: DUNE Physics}",
    eprint = "2002.03005",
    archivePrefix = "arXiv",
    primaryClass = "hep-ex",
    reportNumber = "FERMILAB-PUB-20-025-ND, FERMILAB-DESIGN-2020-02",
    month = "2",
    year = "2020"
}

@article{DUNE:CDR,
    author = "Abed Abud, A. and others",
    collaboration = "DUNE",
    title = "{Deep Underground Neutrino Experiment (DUNE) Near Detector Conceptual Design Report}",
    eprint = "2103.13910",
    archivePrefix = "arXiv",
    primaryClass = "physics.ins-det",
    reportNumber = "FERMILAB-PUB-21-067-E-LBNF-PPD-SCD-T",
    doi = "10.3390/instruments5040031",
    journal = "Instruments",
    volume = "5",
    number = "4",
    pages = "31",
    year = "2021"
}

@misc{DUNE:OfflineCDR,
    author = "Abed Abud, A. and others",
    collaboration = "DUNE",
    title = "{DUNE Offline Computing Conceptual Design Report}",
    eprint = "2210.15665",
    archivePrefix = "arXiv",
    primaryClass = "physics.data-an",
    reportNumber = "FERMILAB-DESIGN-2022-01",
    doi = "10.2172/1895403",
    year = "2022"
}

@article{RPA,
    author = "Morfin, Jorge G. and others",
    title = "{Recent Developments in Neutrino/Antineutrino - Nucleus Interactions}",
    eprint = "1209.6586",
    archivePrefix = "arXiv",
    primaryClass = "hep-ex",
    reportNumber = "FERMILAB-PUB-12-529-PPD",
    doi = "10.1155/2012/934597",
    journal = "Adv. High Energy Phys.",
    volume = "2012",
    pages = "934597",
    year = "2012"
}

@misc{GENIE,
    author = "Andreopoulos, Costas and others",
    title = "{The GENIE Neutrino Monte Carlo Generator: Physics and User Manual}",
    eprint = "1510.05494",
    archivePrefix = "arXiv",
    primaryClass = "hep-ph",
    reportNumber = "FERMILAB-FN-1004-CD",
    month = "10",
    year = "2015"
}

@inproceedings{nuInt,
    author = "McFarland, Kevin Scott",
    title = "{Neutrino Interactions}",
    eprint = "0804.3899",
    archivePrefix = "arXiv",
    primaryClass = "hep-ex",
    pages = "65--90",
    month = "4",
    year = "2008"
}

@article{MINOS:sterile,
    author = "Adamson, P. and others",
    collaboration = "MINOS+",
    title = "{Search for sterile neutrinos in MINOS and MINOS+ using a two-detector fit}",
    eprint = "1710.06488",
    archivePrefix = "arXiv",
    primaryClass = "hep-ex",
    reportNumber = "FERMILAB-PUB-17-430-ND",
    doi = "10.1103/PhysRevLett.122.091803",
    journal = "Phys. Rev. Lett.",
    volume = "122",
    number = "9",
    pages = "091803",
    year = "2019"
}

@article{LSND,
    author = "Aguilar, A. and others",
    collaboration = "LSND",
    title = "{Evidence for neutrino oscillations from the observation of $\bar{\nu}_e$ appearance in a $\bar{\nu}_\mu$
 beam}",
    eprint = "hep-ex/0104049",
    archivePrefix = "arXiv",
    doi = "10.1103/PhysRevD.64.112007",
    journal = "Phys. Rev. D",
    volume = "64",
    pages = "112007",
    year = "2001"
}

@inproceedings{MiniBooNE:excess,
    author = "Aguilar-Arevalo, A. A. and others",
    collaboration = "MiniBooNE",
    title = "{A Combined $\nu_\mu \rightarrow \nu_e$ and $\bar \nu_\mu \rightarrow \bar \nu_e$ Oscillation Analysis of the MiniBooNE Excesses}",
    eprint = "1207.4809",
    archivePrefix = "arXiv",
    primaryClass = "hep-ex",
    reportNumber = "LA-UR-12-23041; Fermilab-PUB-12-394-AD-PPD, FERMILAB-PUB-12-394-AD-PPD, LA-UR-12-23041",
    month = "7",
    year = "2012"
}

@article{GalliumAnomaly,
    author = "Acero, Mario A. and others",
    title = "{Limits on nu(e) and anti-nu(e) disappearance from Gallium and reactor experiments}",
    eprint = "0711.4222",
    archivePrefix = "arXiv",
    primaryClass = "hep-ph",
    doi = "10.1103/PhysRevD.78.073009",
    journal = "Phys. Rev. D",
    volume = "78",
    pages = "073009",
    year = "2008"
}

@article{GalliumAnomalyStats,
    author = "Giunti, Carlo and others",
    title = "{Statistical Significance of the Gallium Anomaly}",
    eprint = "1006.3244",
    archivePrefix = "arXiv",
    primaryClass = "hep-ph",
    doi = "10.1103/PhysRevC.83.065504",
    journal = "Phys. Rev. C",
    volume = "83",
    pages = "065504",
    year = "2011"
}

@article{reactorAnomaly1,
    author = "Mueller, Th. A. and others",
    title = "{Improved Predictions of Reactor Antineutrino Spectra}",
    eprint = "1101.2663",
    archivePrefix = "arXiv",
    primaryClass = "hep-ex",
    reportNumber = "IRFU-10-280",
    doi = "10.1103/PhysRevC.83.054615",
    journal = "Phys. Rev. C",
    volume = "83",
    pages = "054615",
    year = "2011"
}

@article{reactorAnomaly2,
    author = "Huber, Patrick",
    title = "{On the determination of anti-neutrino spectra from nuclear reactors}",
    eprint = "1106.0687",
    archivePrefix = "arXiv",
    primaryClass = "hep-ph",
    doi = "10.1103/PhysRevC.85.029901",
    journal = "Phys. Rev. C",
    volume = "84",
    pages = "024617",
    year = "2011",
    note = "[Erratum: Phys.Rev.C 85, 029901 (2012)]"
}

@article{reactorAnomaly3,
    author = "Mention, G. and others",
    title = "{The Reactor Antineutrino Anomaly}",
    eprint = "1101.2755",
    archivePrefix = "arXiv",
    primaryClass = "hep-ex",
    doi = "10.1103/PhysRevD.83.073006",
    journal = "Phys. Rev. D",
    volume = "83",
    pages = "073006",
    year = "2011"
}

@article{nonUnitary,
    author = "Parke, Stephen and Ross-Lonergan, Mark",
    title = "{Unitarity and the three flavor neutrino mixing matrix}",
    eprint = "1508.05095",
    archivePrefix = "arXiv",
    primaryClass = "hep-ph",
    reportNumber = "IPPP-15-53, DCPT-15-106, FERMILAB-PUB-15-363-T",
    doi = "10.1103/PhysRevD.93.113009",
    journal = "Phys. Rev. D",
    volume = "93",
    number = "11",
    pages = "113009",
    year = "2016"
}

@article{nueScat,
    author = "Tomalak, Oleksandr and others",
    title = "{Theory of elastic neutrino-electron scattering}",
    eprint = "1907.03379",
    archivePrefix = "arXiv",
    primaryClass = "hep-ph",
    reportNumber = "FERMILAB-PUB-19-076-T",
    doi = "10.1103/PhysRevD.101.033006",
    journal = "Phys. Rev. D",
    volume = "101",
    number = "3",
    pages = "033006",
    year = "2020"
}

@article{STEREO,
    author = "Almaz{\'a}n, H. and others",
    collaboration = "STEREO",
    title = "{STEREO neutrino spectrum of $^{235}$U fission rejects sterile neutrino hypothesis}",
    eprint = "2210.07664",
    archivePrefix = "arXiv",
    primaryClass = "hep-ex",
    doi = "10.1038/s41586-022-05568-2",
    journal = "Nature",
    volume = "613",
    number = "7943",
    pages = "257--261",
    year = "2023"
}

@article{MicroBooNE,
    author = "Abratenko, P. and others",
    collaboration = "MicroBooNE",
    title = "{First Constraints on Light Sterile Neutrino Oscillations from Combined Appearance and Disappearance Searches with the MicroBooNE Detector}",
    eprint = "2210.10216",
    archivePrefix = "arXiv",
    primaryClass = "hep-ex",
    reportNumber = "FERMILAB-PUB-22-750-ND",
    doi = "10.1103/PhysRevLett.130.011801",
    journal = "Phys. Rev. Lett.",
    volume = "130",
    number = "1",
    pages = "011801",
    year = "2023"
}

@article{MicroBooNE_nature,
    author = "Abratenko, P. and others",
    collaboration = "MicroBooNE",
    title = "{Search for light sterile neutrinos with two neutrino beams at MicroBooNE}",
    eprint = "2512.07159",
    archivePrefix = "arXiv",
    primaryClass = "hep-ex",
    reportNumber = "FERMILAB-PUB-24-0865-PPD",
    doi = "10.1038/s41586-025-09757-7",
    journal = "Nature",
    volume = "648",
    number = "8092",
    pages = "64--69",
    year = "2025"
}

@article{PROSPECT,
    author = "Andriamirado, M. and others",
    collaboration = "PROSPECT, (PROSPECT Collaboration)*",
    title = "{Final Search for Short-Baseline Neutrino Oscillations with the PROSPECT-I Detector at HFIR}",
    eprint = "2406.10408",
    archivePrefix = "arXiv",
    primaryClass = "hep-ex",
    doi = "10.1103/PhysRevLett.134.151802",
    journal = "Phys. Rev. Lett.",
    volume = "134",
    number = "15",
    pages = "151802",
    year = "2025"
}

@misc{KATRIN,
    author = "Acharya, Himal and others",
    collaboration = "KATRIN",
    title = "{Sterile-neutrino search based on 259 days of KATRIN data}",
    eprint = "2503.18667",
    archivePrefix = "arXiv",
    primaryClass = "hep-ex",
    month = "3",
    year = "2025"
}

@inproceedings{LBNF_Beam,
    author = "Papadimitriou, Vaia and others",
    title = "{Design of the LBNF Beamline}",
    booktitle = "{7th International Particle Accelerator Conference}",
    eprint = "1704.04471",
    archivePrefix = "arXiv",
    primaryClass = "physics.acc-ph",
    reportNumber = "FERMILAB-CONF-16-163-AD",
    doi = "10.18429/JACoW-IPAC2016-TUPMR025",
    pages = "TUPMR025",
    year = "2016"
}

@article{BEST,
    author = "Barinov, V. V. and others",
    title = "{Results from the Baksan Experiment on Sterile Transitions (BEST)}",
    eprint = "2109.11482",
    archivePrefix = "arXiv",
    primaryClass = "nucl-ex",
    doi = "10.1103/PhysRevLett.128.232501",
    journal = "Phys. Rev. Lett.",
    volume = "128",
    number = "23",
    pages = "232501",
    year = "2022"
}

@article{DayaBay,
    author = "An, F. P. and others",
    collaboration = "Daya Bay",
    title = "{Comprehensive Measurement of the Reactor Antineutrino Spectrum and Flux at Daya Bay}",
    eprint = "2501.00746",
    archivePrefix = "arXiv",
    primaryClass = "nucl-ex",
    doi = "10.1103/PhysRevLett.134.201802",
    journal = "Phys. Rev. Lett.",
    volume = "134",
    number = "20",
    pages = "201802",
    year = "2025"
}

@article{Neutrino4,
    author = "Serebrov, A. P. and others",
    title = "{Search for sterile neutrinos with the Neutrino-4 experiment and measurement results}",
    eprint = "2005.05301",
    archivePrefix = "arXiv",
    primaryClass = "hep-ex",
    doi = "10.1103/PhysRevD.104.032003",
    journal = "Phys. Rev. D",
    volume = "104",
    number = "3",
    pages = "032003",
    year = "2021"
}

@article{DUNE:Mx2,
    author = "Abbaslu, Saeed and others",
    collaboration = "DUNE,",
    title = "{Operation of a Modular 3D-Pixelated Liquid Argon Time-Projection Chamber in a Neutrino Beam}",
    eprint = "2509.07012",
    archivePrefix = "arXiv",
    primaryClass = "physics.ins-det",
    reportNumber = "FERMILAB-PUB-25-0537-LBNF",
    doi = "10.3390/instruments10010018",
    journal = "Instruments",
    volume = "10",
    number = "1",
    pages = "18",
    year = "2026"
}

@manual{Minuit2,
  author       = {James, Fred and others},
  title        = {{MINUIT User's Guide}},
  organization = {CERN},
  note         = {MINUIT2 documentation},
  url          = {https://root.cern.ch/root/htmldoc/guides/minuit2/Minuit2.html},
  urldate      = {2026-08-10}
}

@article{Rodrigues:2025,
    author = "Rodrigues, Ohana Benevides and others",
    title = "{Toward a Robust Confirmation or Refutation of the Sterile-Neutrino Explanation of Short-Baseline Anomalies}",
    eprint = "2503.13594",
    archivePrefix = "arXiv",
    primaryClass = "hep-ph",
    reportNumber = "FERMILAB-PUB-25-0098-T, MI-HET-852",
    doi = "10.1103/4fn7-svnh",
    journal = "Phys. Rev. Lett.",
    volume = "135",
    number = "8",
    pages = "081801",
    year = "2025"
}

@article{Dentler:2018sju,
    author = "Dentler, Mona and others",
    title = "{Updated Global Analysis of Neutrino Oscillations in the Presence of eV-Scale Sterile Neutrinos}",
    eprint = "1803.10661",
    archivePrefix = "arXiv",
    primaryClass = "hep-ph",
    reportNumber = "MITP-18-023, IFT-UAM-CSIC-18-033, MITP/18-023, FERMILAB-PUB-18-086-T, IFT-UAM/CSIC-18-033",
    doi = "10.1007/JHEP08(2018)010",
    journal = "JHEP",
    volume = "08",
    pages = "010",
    year = "2018"
}

@article{GEANT4:2002zbu,
    author = "Agostinelli, S. and others",
    collaboration = "GEANT4",
    title = "{GEANT4 - A Simulation Toolkit}",
    reportNumber = "SLAC-PUB-9350, FERMILAB-PUB-03-339, CERN-IT-2002-003",
    doi = "10.1016/S0168-9002(03)01368-8",
    journal = "Nucl. Instrum. Meth. A",
    volume = "506",
    pages = "250--303",
    year = "2003"
}

@article{Friedland:2018vry,
    author = "Friedland, Alexander and others",
    title = "{Understanding the energy resolution of liquid argon neutrino detectors}",
    eprint = "1811.06159",
    archivePrefix = "arXiv",
    primaryClass = "hep-ph",
    reportNumber = "SLAC-PUB-17352",
    doi = "10.1103/PhysRevD.99.036009",
    journal = "Phys. Rev. D",
    volume = "99",
    number = "3",
    pages = "036009",
    year = "2019"
}

@article{Castiglioni:2020tsu,
    author = "Castiglioni, W. and others",
    title = "{Benefits of MeV-scale reconstruction capabilities in large liquid argon time projection chambers}",
    eprint = "2006.14675",
    archivePrefix = "arXiv",
    primaryClass = "physics.ins-det",
    doi = "10.1103/PhysRevD.102.092010",
    journal = "Phys. Rev. D",
    volume = "102",
    number = "9",
    pages = "092010",
    year = "2020"
}

@article{HernandezMorquecho:2026hdk,
    author = "Hernandez Morquecho, Miguel and others",
    title = "{Neutron reconstruction via blips in liquid argon time projection chambers}",
    eprint = "2604.11774",
    archivePrefix = "arXiv",
    primaryClass = "hep-ex",
    reportNumber = "FERMILAB-PUB-26-0224-PPD",
    doi = "10.1103/dkbs-j6g5",
    journal = "Phys. Rev. D",
    volume = "114",
    number = "1",
    pages = "012013",
    year = "2026"
}

@article{NOvA:2018gge,
    author = "Acero, M. A. and others",
    collaboration = "NOvA",
    title = "{New constraints on oscillation parameters from $\nu_e$ appearance and $\nu_\mu$ disappearance in the NOvA experiment}",
    eprint = "1806.00096",
    archivePrefix = "arXiv",
    primaryClass = "hep-ex",
    reportNumber = "FERMILAB-PUB-18-223-ND",
    doi = "10.1103/PhysRevD.98.032012",
    journal = "Phys. Rev. D",
    volume = "98",
    pages = "032012",
    year = "2018"
}

@article{Feldman:1997qc,
    author = "Feldman, Gary J. and others",
    title = "{A Unified approach to the classical statistical analysis of small signals}",
    eprint = "physics/9711021",
    archivePrefix = "arXiv",
    reportNumber = "HUTP-97-A096",
    doi = "10.1103/PhysRevD.57.3873",
    journal = "Phys. Rev. D",
    volume = "57",
    pages = "3873--3889",
    year = "1998"
}

@article{NOvA:2024imi,
    author = "Acero, M. A. and others",
    collaboration = "NOvA",
    title = "{Dual-Baseline Search for Active-to-Sterile Neutrino Oscillations in NOvA}",
    eprint = "2409.04553",
    archivePrefix = "arXiv",
    primaryClass = "hep-ex",
    reportNumber = "FERMILAB-PUB-24-0555-PPD",
    doi = "10.1103/PhysRevLett.134.081804",
    journal = "Phys. Rev. Lett.",
    volume = "134",
    number = "8",
    pages = "081804",
    year = "2025"
}

@article{T2K:2019efw,
    author = "Abe, K. and others",
    collaboration = "T2K",
    title = "{Search for light sterile neutrinos with the T2K far detector Super-Kamiokande at a baseline of 295 km}",
    eprint = "1902.06529",
    archivePrefix = "arXiv",
    primaryClass = "hep-ex",
    doi = "10.1103/PhysRevD.99.071103",
    journal = "Phys. Rev. D",
    volume = "99",
    number = "7",
    pages = "071103",
    year = "2019"
}

@article{Kopp:2013vaa,
    author = "Kopp, Joachim and others",
    title = "{Sterile Neutrino Oscillations: The Global Picture}",
    eprint = "1303.3011",
    archivePrefix = "arXiv",
    primaryClass = "hep-ph",
    reportNumber = "IFT-UAM-CSIC-13-026",
    doi = "10.1007/JHEP05(2013)050",
    journal = "JHEP",
    volume = "05",
    pages = "050",
    year = "2013"
}

@article{Acero:2022wqg,
    author = "Acero, M. A. and others",
    title = "{White paper on light sterile neutrino searches and related phenomenology}",
    eprint = "2203.07323",
    archivePrefix = "arXiv",
    primaryClass = "hep-ex",
    reportNumber = "FERMILAB-PUB-22-318-ND-SCD-T",
    doi = "10.1088/1361-6471/ad307f",
    journal = "J. Phys. G",
    volume = "51",
    number = "12",
    pages = "120501",
    year = "2024"
}

@article{Kopeikin:2021ugh,
    author = "Kopeikin, V. and others.",
    title = "{Reevaluating reactor antineutrino spectra with new measurements of the ratio between U235 and Pu239 {\ensuremath{\beta}} spectra}",
    eprint = "2103.01684",
    archivePrefix = "arXiv",
    primaryClass = "nucl-ex",
    doi = "10.1103/PhysRevD.104.L071301",
    journal = "Phys. Rev. D",
    volume = "104",
    number = "7",
    pages = "L071301",
    year = "2021"
}

@article{RENO:2018pwo,
    author = "Bak, G. and others",
    collaboration = "RENO",
    title = "{Fuel-composition dependent reactor antineutrino yield at RENO}",
    eprint = "1806.00574",
    archivePrefix = "arXiv",
    primaryClass = "hep-ex",
    doi = "10.1103/PhysRevLett.122.232501",
    journal = "Phys. Rev. Lett.",
    volume = "122",
    number = "23",
    pages = "232501",
    year = "2019"
}

@article{Giunti:2017yid,
    author = "Giunti, C. and others",
    title = "{Reactor Fuel Fraction Information on the Antineutrino Anomaly}",
    eprint = "1708.01133",
    archivePrefix = "arXiv",
    primaryClass = "hep-ph",
    doi = "10.1007/JHEP10(2017)143",
    journal = "JHEP",
    volume = "10",
    pages = "143",
    year = "2017"
}

@article{Estienne:2019ujo,
    author = "Estienne, M. and others",
    title = "{Updated Summation Model: An Improved Agreement with the Daya Bay Antineutrino Fluxes}",
    eprint = "1904.09358",
    archivePrefix = "arXiv",
    primaryClass = "nucl-ex",
    doi = "10.1103/PhysRevLett.123.022502",
    journal = "Phys. Rev. Lett.",
    volume = "123",
    number = "2",
    pages = "022502",
    year = "2019"
}

@misc{SBNProgram,
    author = "Acciarri, R. and others",
    collaboration = "MicroBooNE, LAr1-ND, ICARUS-WA104",
    title = "{A Proposal for a Three Detector Short-Baseline Neutrino Oscillation Program in the Fermilab Booster Neutrino Beam}",
    eprint = "1503.01520",
    archivePrefix = "arXiv",
    primaryClass = "physics.ins-det",
    reportNumber = "FERMILAB-PUB-15-0726-AD-PPD",
    month = "3",
    year = "2015"
}

@article{DANSS:2018fnn,
    author = "Alekseev, I and others",
    collaboration = "DANSS",
    title = "{Search for sterile neutrinos at the DANSS experiment}",
    eprint = "1804.04046",
    archivePrefix = "arXiv",
    primaryClass = "hep-ex",
    doi = "10.1016/j.physletb.2018.10.038",
    journal = "Phys. Lett. B",
    volume = "787",
    pages = "56--63",
    year = "2018"
}

@article{NEOS:2016wee,
    author = "Ko, Y. J. and others",
    collaboration = "NEOS",
    title = "{Sterile Neutrino Search at the NEOS Experiment}",
    eprint = "1610.05134",
    archivePrefix = "arXiv",
    primaryClass = "hep-ex",
    doi = "10.1103/PhysRevLett.118.121802",
    journal = "Phys. Rev. Lett.",
    volume = "118",
    number = "12",
    pages = "121802",
    year = "2017"
}

@misc{DUNE:2016rla,
    author = "Acciarri, R. and others",
    collaboration = "DUNE",
    title = "{Long-Baseline Neutrino Facility (LBNF) and Deep Underground Neutrino Experiment (DUNE)}: {Conceptual Design Report, Volume 4 The DUNE Detectors at LBNF}",
    eprint = "1601.02984",
    archivePrefix = "arXiv",
    primaryClass = "physics.ins-det",
    reportNumber = "FERMILAB-DESIGN-2016-04",
    month = "1",
    year = "2016"
}

@article{MINERvA:2015nqi,
    author = "Park, J. and others",
    collaboration = "MINERvA",
    title = "{Measurement of Neutrino Flux from Neutrino-Electron Elastic Scattering}",
    eprint = "1512.07699",
    archivePrefix = "arXiv",
    primaryClass = "physics.ins-det",
    reportNumber = "FERMILAB-PUB-15-575-ND",
    doi = "10.1103/PhysRevD.93.112007",
    journal = "Phys. Rev. D",
    volume = "93",
    number = "11",
    pages = "112007",
    year = "2016"
}

@article{MINERvA:2022vmb,
    author = "Zazueta, L. and others",
    collaboration = "MINERvA",
    title = "{Improved constraint on the MINER{\ensuremath{\nu}}A medium energy neutrino flux using {\ensuremath{\nu}}{\textasciimacron}e-{\textrightarrow}{\ensuremath{\nu}}{\textasciimacron}e- data}",
    eprint = "2209.05540",
    archivePrefix = "arXiv",
    primaryClass = "hep-ex",
    reportNumber = "FERMILAB-PUB-22-701-ND",
    doi = "10.1103/PhysRevD.107.012001",
    journal = "Phys. Rev. D",
    volume = "107",
    number = "1",
    pages = "012001",
    year = "2023"
}

@misc{Hernandez-Garcia:2026asl,
    author = "Hern{\'a}ndez-Garc{\'\i}a, Josu and L{\'o}pez-Pav{\'o}n, Jacobo and Urrea, Salvador",
    title = "{Sharpening New Physics Searches in Neutrino Oscillations with DUNE-PRISM}",
    eprint = "2604.20951",
    archivePrefix = "arXiv",
    primaryClass = "hep-ph",
    month = "4",
    year = "2026"
}

@article{SBND:2025pjc,
    author = "Abratenko, P. and others",
    collaboration = "SBND",
    title = "{Sampling off-axis neutrino fluxes with the short-baseline near detector}",
    eprint = "2508.20239",
    archivePrefix = "arXiv",
    primaryClass = "hep-ex",
    reportNumber = "FERMILAB-PUB-25-0401-PPD",
    doi = "10.1103/hc3v-tqdb",
    journal = "Phys. Rev. D",
    volume = "113",
    number = "7",
    pages = "072007",
    year = "2026"
}

@article{Cadeddu:2026Gallium,
    author = "Cadeddu, M. and Cargioli, N. and Dordei, F. and Ferro, L. and Giunti, C. and Pitzalis, M.",
    title = "{Possible Solution to the Gallium Anomaly Moving beyond the Leptonic Wave-Function Factorization}",
    journal = "Phys. Rev. Lett.",
    volume = "137",
    number = "13",
    pages = "131805",
    year = "2026",
    doi = "10.1103/5pvf-mcr1",
    eprint = "2512.20560",
    archivePrefix = "arXiv",
    primaryClass = "hep-ph"
}

@article{DUNE:2020ypp,
    author = "{DUNE Collaboration}",
    title = "{Prospects for beyond the Standard Model physics searches at the Deep Underground Neutrino Experiment}",
    journal = "Eur. Phys. J. C",
    volume = "81",
    number = "4",
    pages = "322",
    year = "2021",
    doi = "10.1140/epjc/s10052-021-09007-w",
    eprint = "2008.12769",
    archivePrefix = "arXiv",
    primaryClass = "hep-ex"
}

@article{Belusevic:1988ab,
    author = "Belusevic, R. and Rein, D.",
    title = "{Neutrino Reactions in the Low Y Region}",
    doi = "10.1103/PhysRevD.38.2753",
    journal = "Phys. Rev. D",
    volume = "38",
    pages = "2753--2757",
    year = "1988"
}

@misc{Belusevic:1987rn,
    author = "Belusevic, R. and Rein, D.",
    title = "{Is There a Way to Measure the Deep Inelastic Cross-Section Using Wide Band Neutrino Beams?}",
    reportNumber = "FERMILAB-PUB-87-019-E",
    doi = "10.2172/6698002",
    month = "1",
    year = "1987"
}

@article{Bodek:2012uu,
    author = "Bodek, A. and Sarica, U. and Naples, D. and Ren, L.",
    title = "{Methods to Determine Neutrino Flux at Low Energies:Investigation of the Low $\nu$ Method}",
    eprint = "1201.3025",
    archivePrefix = "arXiv",
    primaryClass = "hep-ex",
    doi = "10.1140/epjc/s10052-012-1973-6",
    journal = "Eur. Phys. J. C",
    volume = "72",
    pages = "1973",
    year = "2012"
}

@inproceedings{Mishra:1990ax,
    author = "Mishra, S. R.",
    title = "{Probing Nucleon Structure with Neutrino - N Experiments}",
    booktitle = "{Workshop on Hadron Structure Functions and Parton Distributions}",
    reportNumber = "NEVIS-1426, FERMILAB-CONF-90-381",
    month = "4",
    year = "1990"
}

@phdthesis{Seligman:1997fe,
    author = "Seligman, William Glenn",
    title = "{A Next-to-Leading Order QCD Analysis of Neutrino - Iron Structure Functions at the Tevatron}",
    reportNumber = "FERMILAB-THESIS-1997-21, UMI-97-23849, NEVIS-292, CU-398, R-1257",
    doi = "10.2172/1421736",
    school = "Nevis Labs, Columbia U.",
    year = "1997"
}

@article{Auchincloss:1990tu,
    author = "Auchincloss, Priscilla S. and others",
    title = "{Measurement of the Inclusive Charged Current Cross-section for Neutrino and Anti-neutrino Scattering on Isoscalar Nucleons}",
    reportNumber = "NEVIS-1394, FERMILAB-PUB-90-0390",
    doi = "10.1007/BF01572022",
    journal = "Z. Phys. C",
    volume = "48",
    pages = "411--432",
    year = "1990"
}
\fi

\clearpage
\appendix
\section{L-E Dependency}
\label{apdx:LEdep}

\begin{figure}[h]
\centering
  \includegraphics[width=0.45\textwidth]{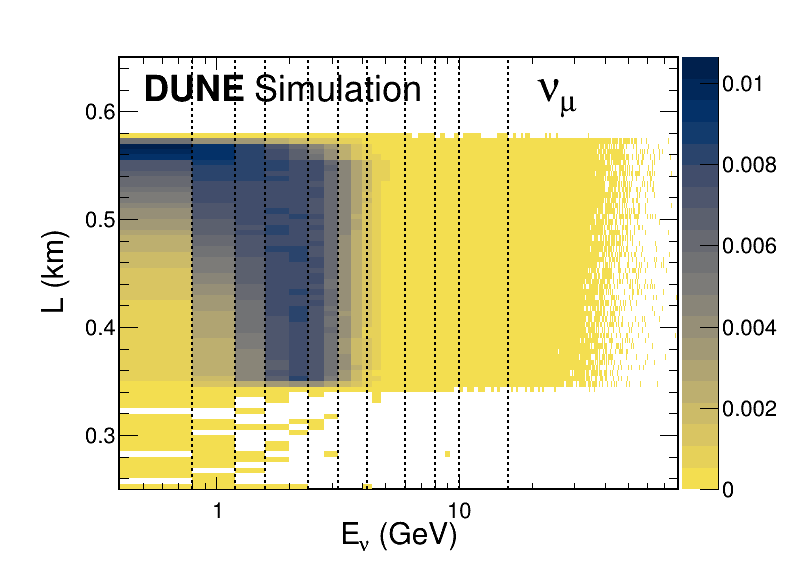}
  \includegraphics[width=0.45\textwidth]{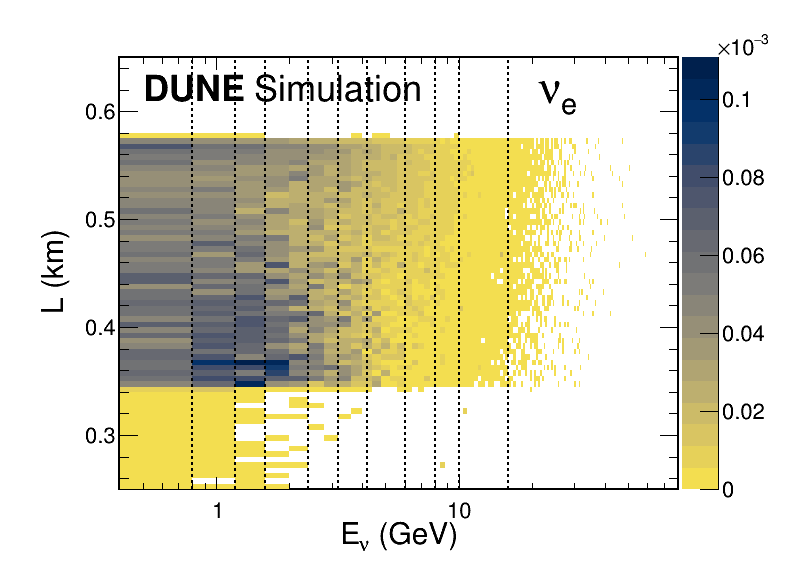}
\caption{Distributions in neutrino flight distance $L$ and true neutrino energy $E_\nu$ for neutrinos interacting in the DUNE ND, shown for the $\nu_\mu$ (left) and $\nu_e$ (right) samples. The nonuniform bin boundaries along the $E_\nu$ axis indicate the 12 energy bands used in the $L$-averaging procedure.}
\label{fig:LEdep}
\end{figure}

Equation~\ref{eqn:OscProb} shows that the oscillation probability depends on the neutrino flight distance $L$. In this analysis, the baseline used in the oscillation calculation is determined from the neutrino production position along the decay region relative to the nominal ND position at 574~m from the target. The finite extent of the neutrino production region is negligible for long-baseline oscillations in DUNE, for which the approximately 1300~km FD baseline is much larger than the length of the decay region. For the short-baseline oscillations considered here, however, this effect is non-negligible because the ND is located only approximately 574~m downstream of the target. Figure~\ref{fig:LEdep} shows the correlation between neutrino energy and this flight distance for the relevant neutrino species.

Ideally, the oscillation probability could be evaluated event by event using the production position of each simulated neutrino. At the event statistics considered in this analysis, repeatedly performing such a calculation during the oscillation fit would be computationally expensive. Instead, the $L$ dependence is incorporated through an energy-dependent averaging procedure.

The simulated two-dimensional $L$--$E_\nu$ distribution is first divided into fine bins in neutrino energy. The corresponding $L$ distribution is examined in each energy slice, and adjacent slices with similar $L$ distributions are combined into broader energy bands. This procedure results in 12 energy bands, within each of which a common baseline parameterization is used. The simulated $L$ distributions are represented using 60 equal-width bins over the interval $300$--$600~\mathrm{m}$, while the piecewise parameterization used in the oscillation calculation covers $340$--$600~\mathrm{m}$.

For each energy band $b$, the simulated $L$ distribution is first normalized before its shape is parameterized.The piecewise parameterization, denoted $\widetilde{f}_{b}(L)$, is defined as
\begin{equation}
\widetilde{f}_{b}(L)
=
\begin{cases}
p_{4,b},
& 0.34 \leq L < 0.35~\mathrm{km},
\\[4pt]
\begin{aligned}
&p_{0,b}+p_{1,b}L \\
&\quad +p_{2,b}L^2+p_{3,b}L^3
\end{aligned}
&
0.35 \leq L < 0.55~\mathrm{km},
\\[6pt]
p_{5,b},
& 0.55 \leq L \leq 0.60~\mathrm{km}.
\end{cases}
\label{eq:fittedFunction}
\end{equation}
The central region is therefore described by a cubic polynomial, while constant values are used in the two edge regions. The resulting parameterization contains six shape parameters for each energy band and neutrino species.

Although the underlying histogram is normalized before fitting, the piecewise approximation in Eq.~\ref{eq:fittedFunction} does not necessarily preserve unit normalization exactly. An explicit normalization factor is therefore calculated for each energy band,
\begin{equation}
    N_{b}
    =
    \left[
        \int_{0.34}^{0.60}
        \widetilde{f}_{b}(L)\,dL
    \right]^{-1},
    \label{eq:LEdep}
\end{equation}
so that the normalized flight-distance distribution used in the oscillation calculation is
\begin{equation}
    f_{b}(L)
    =
    N_{b}\widetilde{f}_{b}(L).
\end{equation}

For neutrinos with energies in the interval $E_\nu\in[E_{b-1},E_b]$, the oscillation probability is averaged over the corresponding flight-distance distribution,
\begin{equation}
    \overline{P}_{\alpha\beta}^{\,b}(E_\nu)
    =
    N_{b}
    \int_{0.34}^{0.60}
    \widetilde{f}_{b}(L)\,
    P_{\alpha\beta}(L,E_\nu)\,
    dL.
    \label{eq:OscProb_LE}
\end{equation}
Events within a given energy band therefore share the same parameterized baseline distribution, while the averaged oscillation probability retains its dependence on neutrino energy within that band.

The six parameters describing the flight-distance distribution and the corresponding normalization factor are determined in advance for each energy band and are stored separately for $\nu_\mu$ and $\nu_e$. During the oscillation fit, the appropriate energy band is identified from the neutrino energy, and these stored parameters are used to evaluate the corresponding $L$-averaged oscillation probability. Because the parameterized baseline distribution is piecewise constant or cubic in $L$, the integration over $L$ is performed analytically. The resulting probability is then averaged over the fine true-neutrino-energy bin used to construct the event prediction. This procedure provides a computationally efficient treatment of the neutrino production-position dependence while retaining the effect of the extended source on the oscillation probability.

\section{Binning and Spectra}
\label{apdx:binning}

The matrix inversion required for the $\chi^2$ calculation in Eq.~\ref{eqn:chi2} can become numerically unstable when the analysis bins span widely different event yields. This occurs within individual samples, between the high-statistics flux peak and lower-statistics tails, as well as across samples, particularly between the $\nu_\mu$-CC and $\nu e^-\to\nu e^-$ samples.

To improve the conditioning of the covariance matrix, finer reconstructed-lepton-energy bins are used near the flux peak and progressively coarser bins in the tails. The CC samples follow the same general binning strategy because they cover similar neutrino-energy ranges and use the same low-hadronic-energy selection, while the lower-statistics $\nu e^-\to\nu e^-$ sample is divided into fewer bins. The final analysis contains 41 bins for the $\nu_\mu$-CC sample, 54 bins for the $\nu_e$-CC sample, and 7 bins for the $\nu e^-\to\nu e^-$ sample. The bin definitions are summarized in Table~\ref{tab:binStats}.

\begin{figure*}
\centering
  \includegraphics[width=0.48\textwidth]{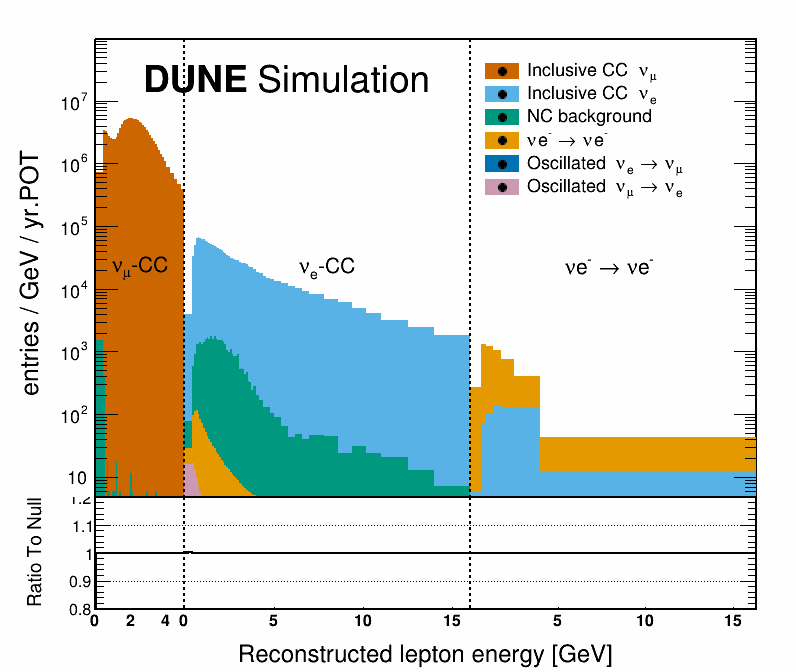}\hfill
  \includegraphics[width=0.48\textwidth]{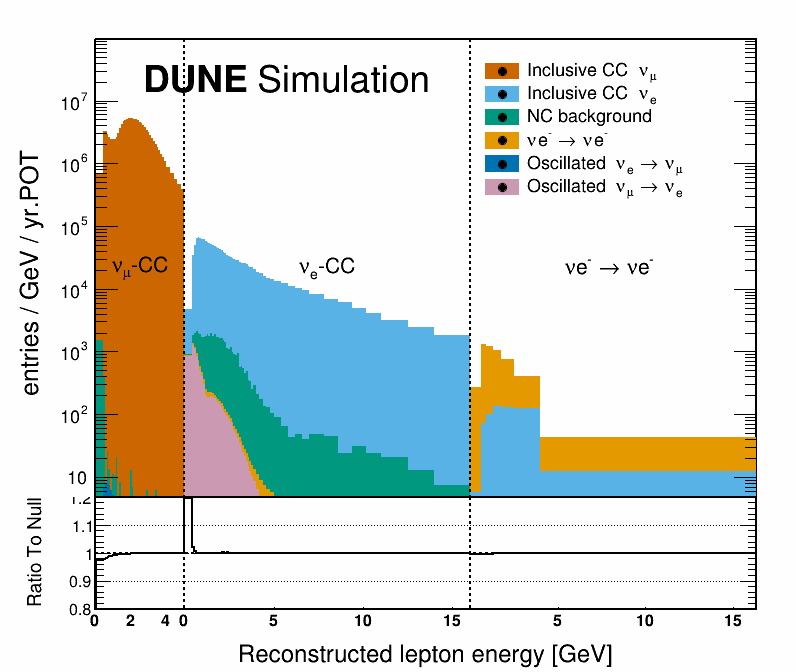}\\
  \includegraphics[width=0.48\textwidth]{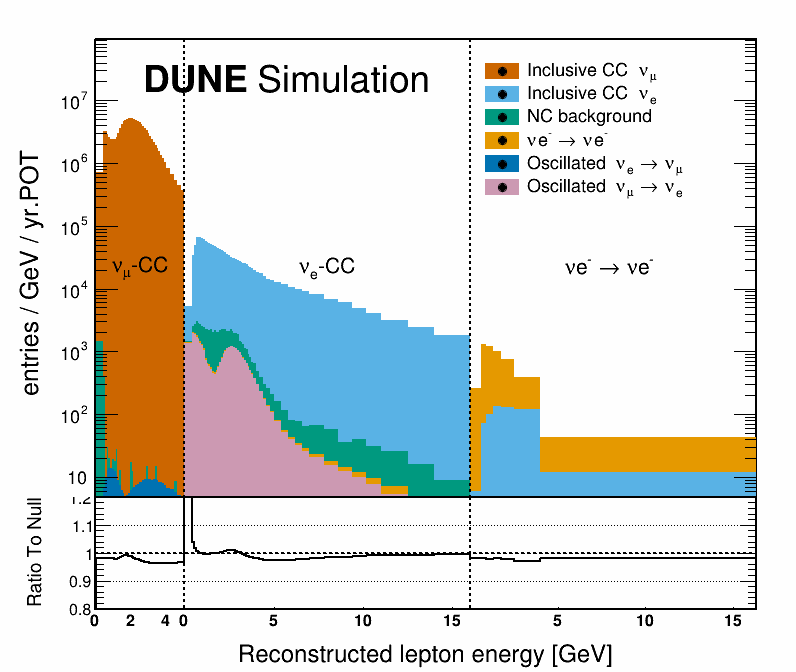}\hfill
  \includegraphics[width=0.48\textwidth]{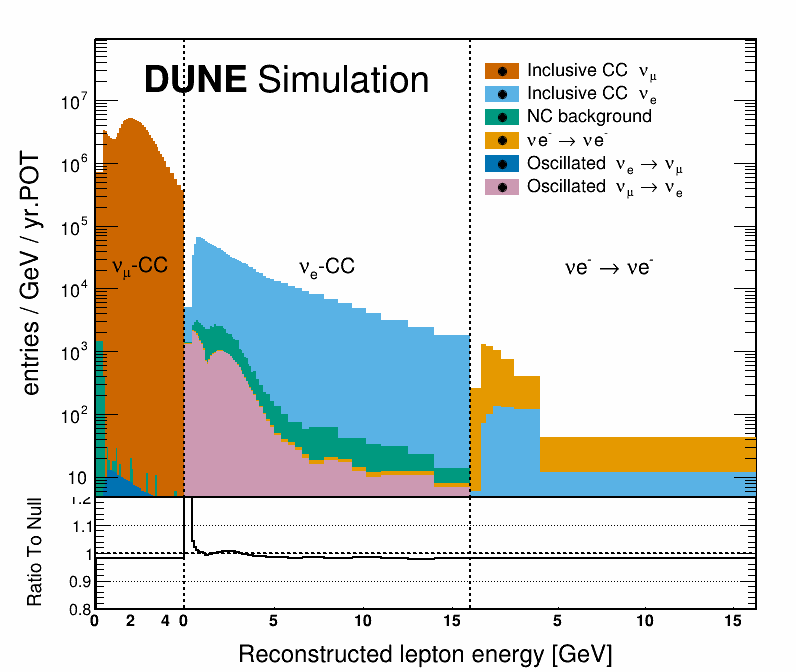}
\caption{Expected event spectra for one year-equivalent of on-axis DUNE ND exposure, corresponding to $1.1\times10^{21}$ POT, under representative sterile-neutrino oscillation hypotheses with fixed mixing parameters $(|U_{e4}|^2, |U_{\mu4}|^2, |U_{\tau4}|^2)=(0.01,\,0.01,\,0.0)$ and $\Delta m^2=0.1~\mathrm{eV}^2$ (top left), $1~\mathrm{eV}^2$ (top right), $10~\mathrm{eV}^2$ (bottom left), and $100~\mathrm{eV}^2$ (bottom right). The spectra are shown using the nominal selection requirements described in the text and illustrate the evolution from weak oscillation effects at low $\Delta m^2$, to pronounced spectral and rate distortions at intermediate $\Delta m^2$, and predominantly rate-like changes at large $\Delta m^2$.}
\label{fig:EventSpectraOtherHypotheses}
\end{figure*}

To illustrate how the oscillation signature evolves with the mass-squared splitting, Fig.~\ref{fig:EventSpectraOtherHypotheses} shows the full analysis event-rate vector for representative injected hypotheses with fixed mixing parameters and varying $\Delta m^2$. At low $\Delta m^2$, the oscillation effect is small over the ND baseline. At intermediate $\Delta m^2$, both spectral distortions and overall rate changes are visible across the analysis samples. At large $\Delta m^2$, the energy-dependent structure is largely averaged, leaving limited residual shape distortion while an overall rate difference remains.

\begin{table*}
\onecolumngrid
    \caption{Expected true signal event yields in each analysis bin for one year-equivalent of DUNE ND exposure under the no-oscillation hypothesis.}
    \tiny  
    \renewcommand{\arraystretch}{1.05} 
    \begin{tabular}{c c c} 
        \begin{tabular}{|c|c|c|}
            \hline
            \multicolumn{3}{|c|}{\textbf{$\nu_{\mu}$-CC}} \\ 
            \hline
            \textbf{Bin Number} & \textbf{Bin Edges (GeV)} & \textbf{Bin Content} \\ 
            \hline
            1	 & 	[0.00, 0.45]	 & 	324191.4	\\
            2	 & 	[0.45, 0.55]	 & 	336090.7	\\
            3	 & 	[0.55, 0.65]	 & 	328244.6	\\
            4	 & 	[0.65, 0.75]	 & 	299062.6	\\
            5	 & 	[0.75, 0.85]	 & 	269712.3	\\
            6	 & 	[0.85, 0.95]	 & 	250574.8	\\
            7	 & 	[0.95, 1.05]	 & 	244438.5	\\
            8	 & 	[1.05, 1.15]	 & 	248196.5	\\
            9	 & 	[1.15, 1.25]	 & 	265818.1	\\
            10	 & 	[1.25, 1.35]	 & 	309008.1	\\
            11	 & 	[1.35, 1.45]	 & 	370733.4	\\
            12	 & 	[1.45, 1.55]	 & 	426823.9	\\
            13	 & 	[1.55, 1.65]	 & 	466223.3	\\
            14	 & 	[1.65, 1.75]	 & 	491785.4	\\
            15	 & 	[1.75, 1.85]	 & 	509478.0	\\
            16	 & 	[1.85, 1.95]	 & 	521310.9	\\
            17	 & 	[1.95, 2.05]	 & 	524615.8	\\
            18	 & 	[2.05, 2.15]	 & 	522093.6	\\
            19	 & 	[2.15, 2.25]	 & 	513820.7	\\
            20	 & 	[2.25, 2.35]	 & 	500293.6	\\
            21	 & 	[2.35, 2.45]	 & 	482615.0	\\
            22	 & 	[2.45, 2.55]	 & 	461401.4	\\
            23	 & 	[2.55, 2.65]	 & 	434733.5	\\
            24	 & 	[2.65, 2.75]	 & 	405573.6	\\
            25	 & 	[2.75, 2.85]	 & 	376134.8	\\
            26	 & 	[2.85, 2.95]	 & 	345277.4	\\
            27	 & 	[2.95, 3.05]	 & 	312831.3	\\
            28	 & 	[3.05, 3.15]	 & 	282443.8	\\
            29	 & 	[3.15, 3.25]	 & 	253746.4	\\
            30	 & 	[3.25, 3.35]	 & 	226129.2	\\
            31	 & 	[3.35, 3.45]	 & 	202296.2	\\
            32	 & 	[3.45, 3.55]	 & 	180229.4	\\
            33	 & 	[3.55, 3.65]	 & 	160490.4	\\
            34	 & 	[3.65, 3.75]	 & 	142057.6	\\
            35	 & 	[3.75, 3.85]	 & 	125187.8	\\
            36	 & 	[3.85, 4.00]	 & 	161487.4	\\
            37	 & 	[4.00, 4.20]	 & 	175489.6	\\
            38	 & 	[4.20, 4.40]	 & 	140940.1	\\
            39	 & 	[4.40, 4.60]	 & 	113260.6	\\
            40	 & 	[4.60, 4.80]	 & 	92266.0	\\
            41	 & 	[4.80, 5.00]	 & 	76652.9	\\
            
            \hline
        \end{tabular}  

        & 

        \begin{tabular}{|c|c|c|}
            \hline
            \multicolumn{3}{|c|}{\textbf{$\nu_{e}$-CC}} \\ 
            \hline
            \textbf{Bin Number} & \textbf{Bin Edges (GeV)} & \textbf{Bin Content} \\ 
            \hline
            42	 & 	[0.00, 0.45]	 & 	1732.5	\\
            43	 & 	[0.45, 0.55]	 & 	3261.8	\\
            44	 & 	[0.55, 0.65]	 & 	4833.6	\\
            45	 & 	[0.65, 0.75]	 & 	6377.4	\\
            46	 & 	[0.75, 0.85]	 & 	6454.9	\\
            47	 & 	[0.85, 0.95]	 & 	6389.1	\\
            48	 & 	[0.95, 1.05]	 & 	6187.0	\\
            49	 & 	[1.05, 1.15]	 & 	5980.2	\\
            50	 & 	[1.15, 1.25]	 & 	5689.6	\\
            51	 & 	[1.25, 1.35]	 & 	5427.0	\\
            52	 & 	[1.35, 1.45]	 & 	5151.0	\\
            53	 & 	[1.45, 1.55]	 & 	4929.1	\\
            54	 & 	[1.55, 1.65]	 & 	4816.1	\\
            55	 & 	[1.65, 1.75]	 & 	4539.5	\\
            56	 & 	[1.75, 1.85]	 & 	4419.5	\\
            57	 & 	[1.85, 1.95]	 & 	4307.1	\\
            58	 & 	[1.95, 2.05]	 & 	4164.4	\\
            59	 & 	[2.05, 2.15]	 & 	3929.8	\\
            60	 & 	[2.15, 2.25]	 & 	3752.1	\\
            61	 & 	[2.25, 2.35]	 & 	3613.5	\\
            62	 & 	[2.35, 2.45]	 & 	3388.7	\\
            63	 & 	[2.45, 2.55]	 & 	3274.6	\\
            64	 & 	[2.55, 2.65]	 & 	3034.7	\\
            65	 & 	[2.65, 2.75]	 & 	2985.2	\\
            66	 & 	[2.75, 2.85]	 & 	2924.6	\\
            67	 & 	[2.85, 2.95]	 & 	2806.4	\\
            68	 & 	[2.95, 3.05]	 & 	2744.7	\\
            69	 & 	[3.05, 3.15]	 & 	2659.6	\\
            70	 & 	[3.15, 3.25]	 & 	2550.1	\\
            71	 & 	[3.25, 3.35]	 & 	2540.2	\\
            72	 & 	[3.35, 3.45]	 & 	2540.8	\\
            73	 & 	[3.45, 3.55]	 & 	2430.8	\\
            74	 & 	[3.55, 3.65]	 & 	2268.9	\\
            75	 & 	[3.65, 3.75]	 & 	2222.3	\\
            76	 & 	[3.75, 3.85]	 & 	2166.4	\\
            77	 & 	[3.85, 4.00]	 & 	3013.1	\\
            78	 & 	[4.00, 4.20]	 & 	3640.9	\\
            79	 & 	[4.20, 4.40]	 & 	3525.0	\\
            80	 & 	[4.40, 4.60]	 & 	3205.9	\\
            81	 & 	[4.60, 4.80]	 & 	2915.9	\\
            82	 & 	[4.80, 5.00]	 & 	2810.5	\\
            83	 & 	[5.00, 5.40]	 & 	5232.5	\\
            84	 & 	[5.40, 5.80]	 & 	4890.1	\\
            85	 & 	[5.80, 6.20]	 & 	4392.7	\\
            86	 & 	[6.20, 6.60]	 & 	4074.2	\\
            87	 & 	[6.60, 7.00]	 & 	3728.3	\\
            88	 & 	[7.00, 7.80]	 & 	6555.0	\\
            89	 & 	[7.80, 8.60]	 & 	5428.2	\\
            90	 & 	[8.60, 9.40]	 & 	4849.9	\\
            91	 & 	[9.40, 10.20]	 & 	3978.1	\\
            92	 & 	[10.20, 11.00]	 & 	3304.3	\\
            93	 & 	[11.00, 12.50]	 & 	4794.0	\\
            94	 & 	[12.50, 14.00]	 & 	3677.6	\\
            95	 & 	[14.00, 16.00]	 & 	3633.3	\\

            \hline
        \end{tabular}    

        & 

        \begin{tabular}{|c|c|c|}
            \hline
            \multicolumn{3}{|c|}{\textbf{$\nu e^-\to\nu e^-$}} \\ 
            \hline
            \textbf{Bin Number} & \textbf{Bin Edges (GeV)} & \textbf{Bin Content} \\ 
            \hline
            96	 & 	[0.30, 0.60]	 & 	157.7	\\
            97	 & 	[0.60, 0.92]	 & 	396.8	\\
            98	 & 	[0.92, 1.30]	 & 	425.8	\\
            99	 & 	[1.30, 1.75]	 & 	410.1	\\
            100	 & 	[1.75, 2.45]	 & 	447.8	\\
            101	 & 	[2.45, 3.90]	 & 	409.4	\\
            102	 & 	[3.90, 16.00]	 & 	380.4	\\
            \hline
        \end{tabular}    
    \end{tabular} 
    \label{tab:binStats}
\end{table*}

\section{Flux Uncertainties}
\label{apdx:fluxCov}

\begin{figure*}
\centering
  \includegraphics[width=0.48\textwidth]{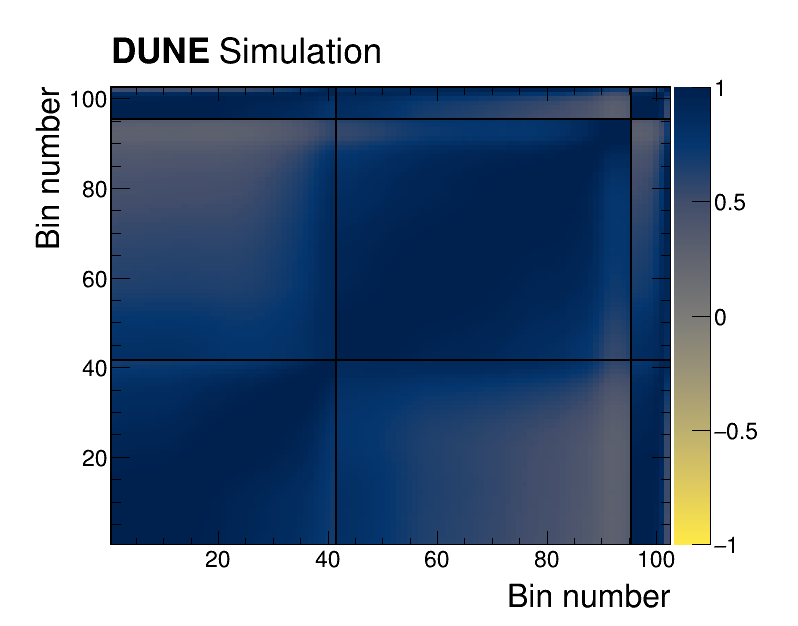}\hfill
  \includegraphics[width=0.48\textwidth]{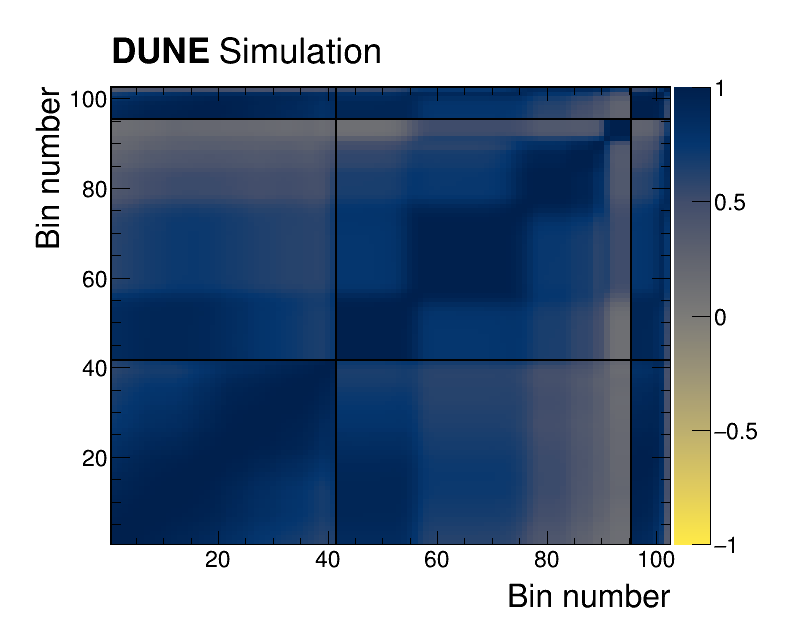}
\caption{Flux correlation matrices without a reconstructed-energy-transfer requirement (left) and for the nominal selection $\nu^{\mathrm{reco}}<0.3~\mathrm{GeV}$ (right). Analysis bins 1--41 correspond to the CC$\nu_\mu$ sample, bins 42--95 to the CC$\nu_e$ sample, and bins 96--102 to the $\nu e^-\to\nu e^-$ sample.}
\label{fig:fluxCor}
\end{figure*}

Figure~\ref{fig:fluxCor} shows the flux correlation matrices with and without the nominal low-$\nu^{\mathrm{reco}}$ requirement. Within the nominal interaction and detector model, the low-$\nu^{\mathrm{reco}}$ selection strengthens the relationship between true neutrino energy and reconstructed lepton energy. As a result, neutrino-energy-dependent flux variations affect a more localized region of reconstructed-energy space, reducing correlations between widely separated reconstructed-energy bins and producing a more diagonal correlation matrix.

\begin{figure*}
\centering
  \includegraphics[width=0.48\textwidth]{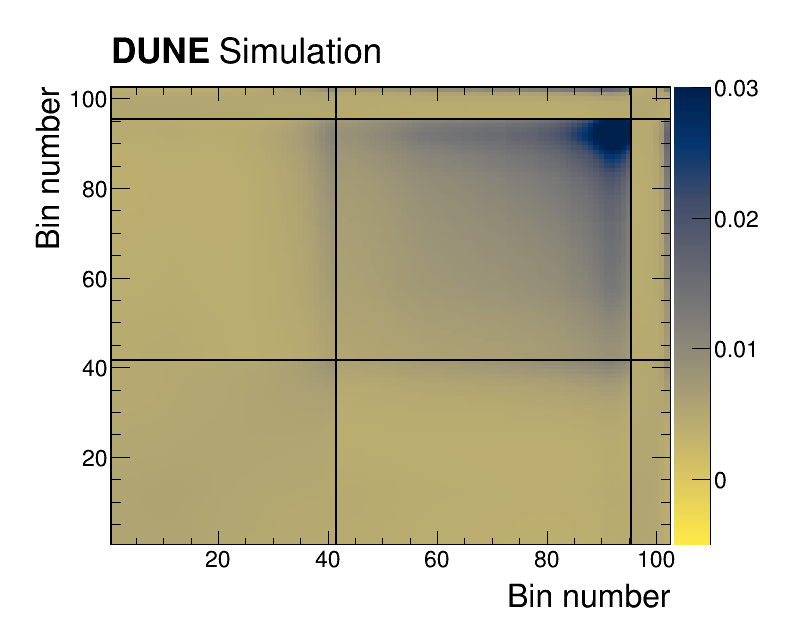}\hfill
  \includegraphics[width=0.48\textwidth]{figures/flux_frcov3_trim_paper.png}
\caption{Fractional flux covariance matrices without a reconstructed-energy-transfer requirement (left) and for the nominal selection $\nu^{\mathrm{reco}}<0.3~\mathrm{GeV}$ (right). Analysis bins 1--41 correspond to the CC$\nu_\mu$ sample, bins 42--95 to the CC$\nu_e$ sample, and bins 96--102 to the $\nu e^-\to\nu e^-$ sample.}
\label{fig:fluxCov_nucuts}
\end{figure*}

The same behavior is visible in the fractional flux covariance shown in Fig.~\ref{fig:fluxCov_nucuts}, where the low-$\nu^{\mathrm{reco}}$ requirement reduces the off-diagonal structure.

\section{Detector Uncertainties}
\label{apdx:detCov}

\begin{figure*}
    \includegraphics[width=0.48\textwidth]{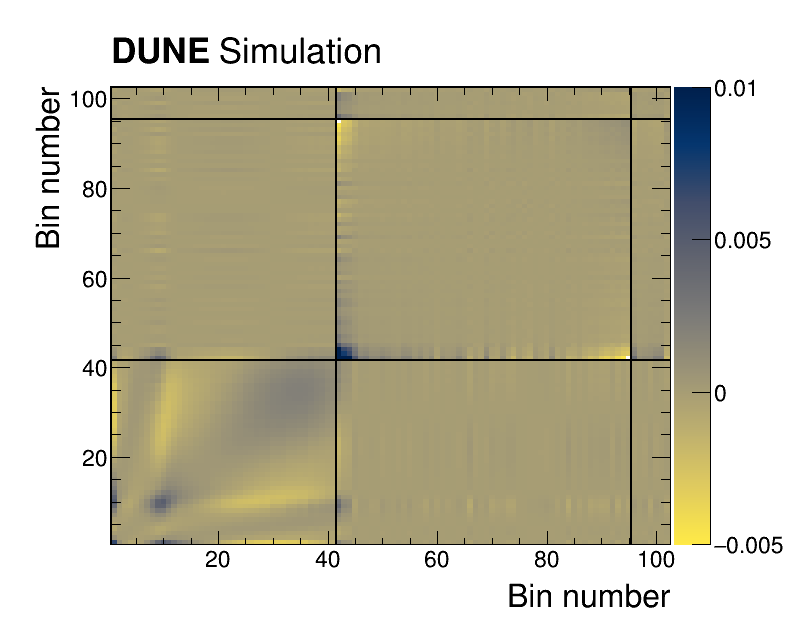}\hfill
    \includegraphics[width=0.48\textwidth]{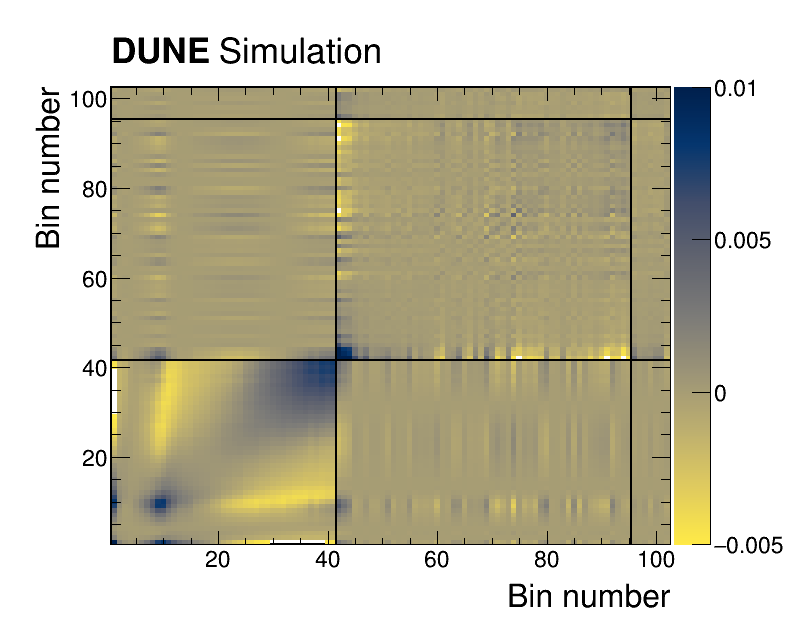}
\caption{Summed fractional covariance matrices associated with lepton-response uncertainties without a reconstructed-energy-transfer requirement (left) and for the nominal selection $\nu^{\mathrm{reco}}<0.3~\mathrm{GeV}$ (right). Analysis bins 1--41 correspond to the CC$\nu_\mu$ sample, bins 42--95 to the CC$\nu_e$ sample, and bins 96--102 to the $\nu e^-\to\nu e^-$ sample.}
\label{fig:hlepCov}
\end{figure*}

\begin{figure*}
    \includegraphics[width=0.48\textwidth]{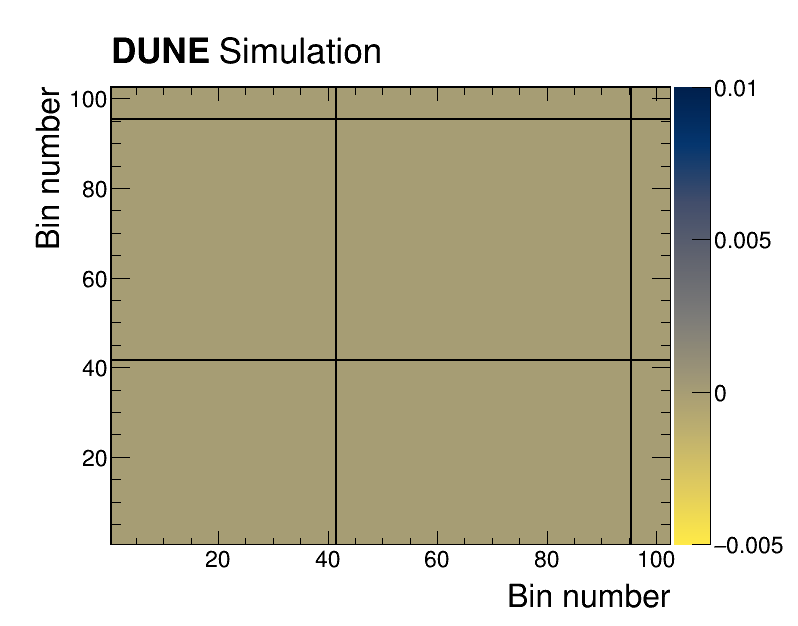}\hfill
    \includegraphics[width=0.48\textwidth]{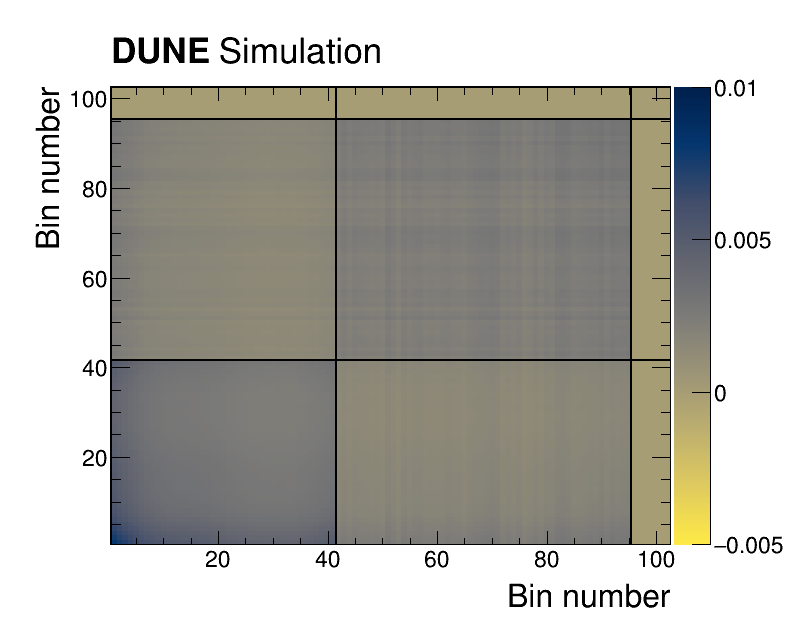}
\caption{Summed fractional covariance matrices associated with hadronic-response uncertainties without a reconstructed-energy-transfer requirement (left) and for the nominal selection $\nu^{\mathrm{reco}}<0.3~\mathrm{GeV}$ (right). Analysis bins 1--41 correspond to the CC$\nu_\mu$ sample, bins 42--95 to the CC$\nu_e$ sample, and bins 96--102 to the $\nu e^-\to\nu e^-$ sample.}
\label{fig:hhadCov}
\end{figure*}

Detector-response uncertainties are parameterized using three variations affecting the reconstructed lepton response and five variations affecting the reconstructed hadronic response. The lepton-response variations correspond to the muon energy scale, electron energy scale, and lepton energy resolution. The hadronic-response variations correspond to the overall hadronic-energy scale entering the $\nu^{\mathrm{reco}}$ selection, the proton response, charged-pion response, neutron response, and an additional low-energy neutron response. Figures~\ref{fig:hlepCov} and~\ref{fig:hhadCov} show the summed fractional covariance matrices associated with the lepton- and hadronic-response uncertainties, respectively, with and without the nominal $\nu^{\mathrm{reco}}<0.3~\mathrm{GeV}$ requirement.

\section{Interaction-Model Uncertainty in the Low-$\nu$ Samples}
\label{apdx:xSModel}

To assess how the low-$\nu$ selections reduce sensitivity to neutrino interaction modeling, we decompose the GENIE systematic variations used in the analysis into normalization and shape components. The covariance treatment is unchanged: two-sided nuisance parameters use the average of the positive and negative variations, while one-sided nuisance parameters use the corresponding single variation.

For the shape-only comparison, each shifted prediction is renormalized to the nominal event yield over the fitted energy range before its deviation from the nominal spectrum is evaluated. The comparison therefore uses the same bins that enter the oscillation fit.

\begin{figure*}[t]
  \centering
  \includegraphics[width=0.48\textwidth]{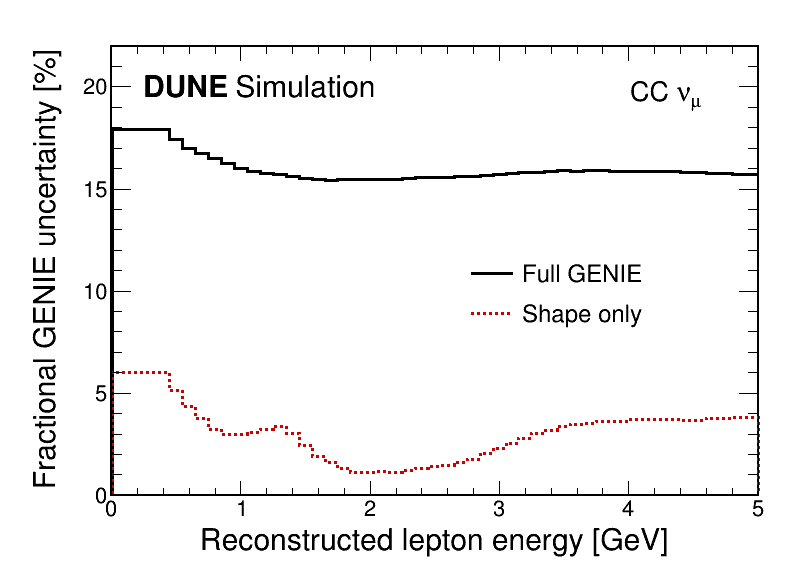}
  \hfill
  \includegraphics[width=0.48\textwidth]{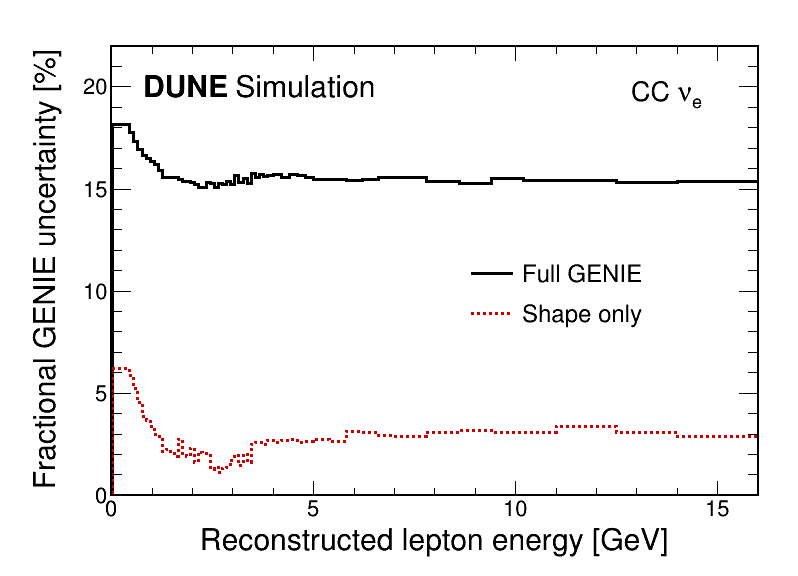}
  \caption{
    Fractional GENIE systematic uncertainties for the selected CC $\nu_{\mu}$ (left) and CC $\nu_{e}$ (right)  samples with $\nu^{\mathrm{reco}} < 0.3$~GeV. The solid curves show the full GENIE uncertainty used in the covariance construction. The dashed curves show the residual shape-only component after each shifted prediction is normalized to the nominal event yield over the fitted reconstructed-energy range.
  }
  \label{fig:genieShape}
\end{figure*}

Figure~\ref{fig:genieShape} shows the resulting full and shape-only fractional GENIE uncertainties for the selected CC $\nu_{\mu}$ and CC $\nu_{e}$ samples. The full variations correspond to integrated normalization uncertainties of approximately 15.6\% and 15.5\%, respectively. The bin-by-bin uncertainty reaches about 18\% in the lowest reconstructed-energy bin for both samples. After the normalization component is removed, the residual shape uncertainty is substantially smaller, remaining at the few-percent level over most of the fitted range and reaching approximately 6\% in the lowest-energy bin.

Within the GENIE nuisance model used in this analysis, the uncertainty in the selected low-$\nu$ samples is therefore dominated by normalization changes rather than large distortions of the reconstructed-energy spectra. The low-$\nu$ selection reduces, but does not eliminate, sensitivity to interaction-model uncertainties.

\begin{figure}[t]
\centering
  \includegraphics[width=0.48\textwidth]{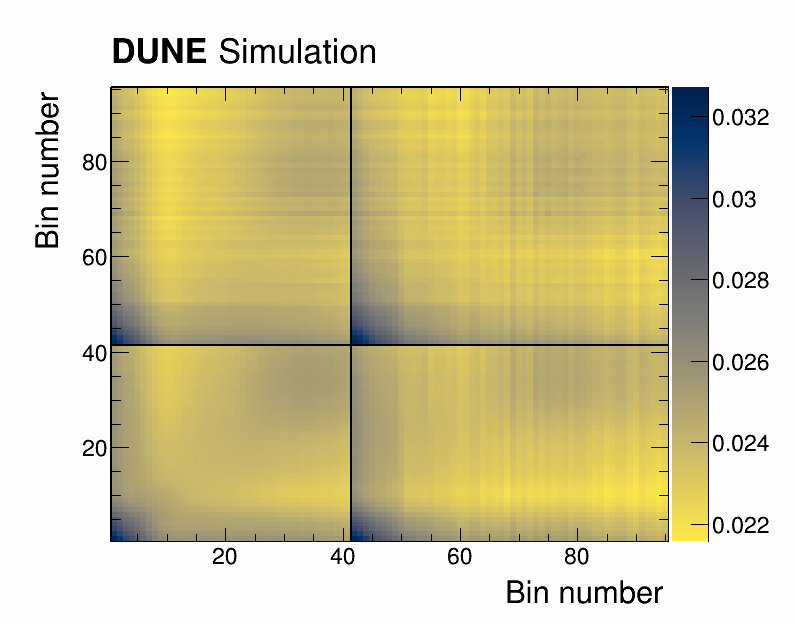}
\caption{Fractional interaction-model covariance matrix restricted to the two CC samples for the nominal selection. Bins 1--41 correspond to the CC$\nu_\mu$ sample and bins 42--95 to the CC$\nu_e$ sample.}
\label{fig:xScov_CC}
\end{figure}

To make the covariance structure within and between the two CC samples more visible, Fig.~\ref{fig:xScov_CC} shows the CC-only block of the fractional interaction-model covariance matrix. This contains the same covariance entries used in the full analysis, restricted to bins 1--95 corresponding to the CC$\nu_\mu$ and CC$\nu_e$ samples.

\section{Numerical Minimization}
\label{apdx:FCfit}

The four-dimensional fit in $|U_{e4}|^2$, $|U_{\mu4}|^2$, $|U_{\tau4}|^2$, and $\Delta m^2$ contains multiple local minima, particularly in $\Delta m^2$. To improve identification of the global minimum and reduce dependence on the initial seed, we use a three-stage fitting procedure based on the MIGRAD algorithm implemented in {\sc Minuit2}~\cite{Minuit2}. In the exploratory and intermediate stages, independent minimizations are performed from multiple seeds, and the converged solution with the lowest $\chi^2$ is retained. A final refined fit is then initialized from the best intermediate-stage solution.

In the first stage, exploratory fits are initialized from four points chosen to sample distinct regions of the allowed parameter space:
\begin{equation}
\begin{split}
\vec{x}^C_0 &= (0.000,\,0.000,\,0.000,\,0.0), \\
\vec{x}^C_1 &= (0.394,\,0.489,\,0.718,\,100.0), \\
\vec{x}^C_2 &= (0.000,\,0.000,\,0.000,\,100.0), \\
\vec{x}^C_3 &= (0.394,\,0.489,\,0.000,\,0.0),
\end{split}
\end{equation}
where the entries correspond to
$(|U_{e4}|^2,|U_{\mu4}|^2,|U_{\tau4}|^2,\Delta m^2)$ and $\Delta m^2$ is expressed in $\mathrm{eV}^2$. This stage uses a maximum of 500 iterations and 5000 function evaluations, with initial step sizes of 0.03 for the three mixing parameters and $10.0~\mathrm{eV}^2$ for $\Delta m^2$, and a MIGRAD tolerance of 1.0.

{\sc Minuit2} assesses convergence using the estimated distance to minimum (EDM), with MIGRAD terminating when
\begin{equation}
    \mathrm{EDM}
    <
    10^{-3}
    \times \mathrm{tolerance}
    \times \mathrm{Up},
\end{equation}
where $\mathrm{Up}=1$ for the $\chi^2$ objective used here.

The best-fit mixing parameters from the exploratory stage are then used to initialize eight intermediate fits with $\Delta m^2$ seeds at
\begin{equation}
\left\{
0.1,\,
0.4,\,
1.0,\,
3.0,\,
8.0,\,
20.0,\,
50.0,\,
100.0
\right\}
~\mathrm{eV}^2.
\end{equation}
This stage uses a maximum of 200 iterations and 2000 function evaluations, a tolerance of 0.1, and a mixing-parameter step size of 0.01. The $\Delta m^2$ step sizes are $1.0~\mathrm{eV}^2$ for seeds from $0.1$ through $8.0~\mathrm{eV}^2$, $2.0~\mathrm{eV}^2$ at $20.0~\mathrm{eV}^2$, $5.0~\mathrm{eV}^2$ at $50.0~\mathrm{eV}^2$, and $10.0~\mathrm{eV}^2$ at $100.0~\mathrm{eV}^2$.

Finally, a refined fit is initialized from the best solution obtained in the intermediate stage. The step sizes are reduced to 0.001 for the three mixing parameters and $0.1~\mathrm{eV}^2$ for $\Delta m^2$. The fit uses a tolerance of 0.01, a maximum of 200 iterations, and 1000 function evaluations. Table~\ref{tab:FC_FittingAlgorithm} summarizes the three-stage configuration.

\begin{table*}
\onecolumngrid
\centering
\caption{Configuration settings used in each stage of the fitting procedure.}
\begin{tabular}{|l|c|c|c|}
\hline
\textbf{Setting}
& \textbf{Exploratory Fit}
& \textbf{$\Delta m^2$ Scan}
& \textbf{Refined Fit} \\
\hline
Maximum Iterations
& 500 & 200 & 200 \\
Maximum Function Calls
& 5000 & 2000 & 1000 \\
Tolerance
& 1.0 & 0.1 & 0.01 \\
Mixing-Parameter Step Size
& 0.03 & 0.01 & 0.001 \\
$\Delta m^2$ Step Size
& $10.0~\mathrm{eV}^2$ & Seed dependent & $0.1~\mathrm{eV}^2$ \\
Initialization
& Four parameter-space seeds
& Best-fit mixing parameters + eight $\Delta m^2$ seeds
& Best fit from scan stage \\
\hline
\end{tabular}
\label{tab:FC_FittingAlgorithm}
\end{table*}

Two penalty terms are included in all three stages to restrict excursions outside the parameter region relevant to the analysis. For the mass-squared splitting, define
\begin{equation}
y \equiv
\log_{10}
\left(
\frac{\Delta m^2_{41}}{\mathrm{eV}^2}
+10^{-8}
\right),
\qquad
y_{\min}\equiv \log_{10}(0.05).
\end{equation}
The penalty is
\begin{equation}
\Pi_{\Delta m^2} =
\begin{cases}
\left(\dfrac{y-y_{\min}}{0.3}\right)^2,
& \Delta m^2_{41}<0.05~\mathrm{eV}^2, \\[6pt]
0,
& 0.05\leq\dfrac{\Delta m^2_{41}}{\mathrm{eV}^2}\leq500, \\[6pt]
\left(\dfrac{\Delta m^2_{41}/\mathrm{eV}^2-500}{1}\right)^2,
& \Delta m^2_{41}>500~\mathrm{eV}^2.
\end{cases}
\end{equation}
This penalty vanishes for $0.05\leq\Delta m^2_{41}/\mathrm{eV}^2\leq500$.

A second penalty enforces the physical constraint implied by normalization of the fourth column of the mixing matrix. Defining
\begin{equation}
S_U
\equiv
|U_{e4}|^2
+
|U_{\mu4}|^2
+
|U_{\tau4}|^2,
\end{equation}
the corresponding penalty is
\begin{equation}
\Pi_{\mathrm{unit}}
=
\lambda
\left[
\max(0,S_U-1)
\right]^2,
\qquad
\lambda=10^6.
\end{equation}
The objective function minimized in each stage is therefore
\begin{equation}
\chi^2_{\mathrm{obj}}
=
\chi^2
+
\Pi_{\Delta m^2}
+
\Pi_{\mathrm{unit}}.
\end{equation}

\begin{figure}[h]
\centering
  \includegraphics[width=0.5\textwidth]{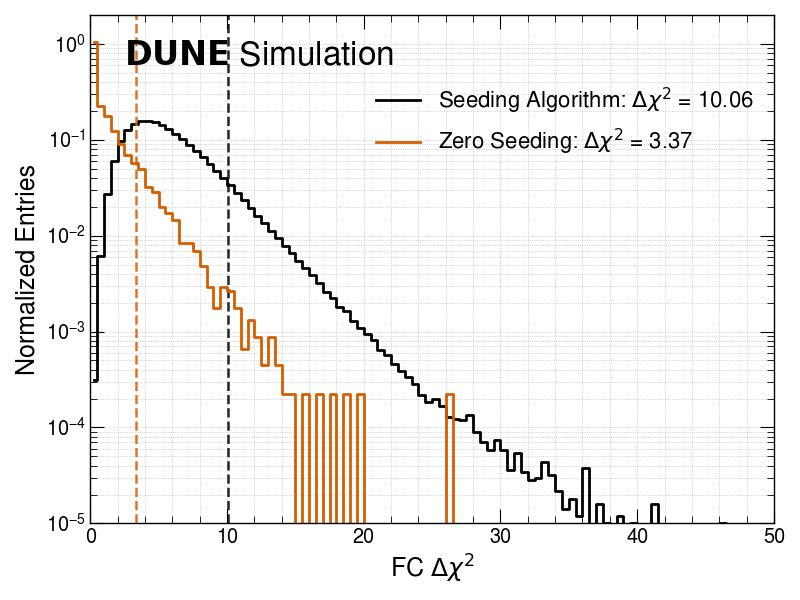}
\caption{Feldman--Cousins $\Delta\chi^2$ distributions obtained using two different fit-seeding schemes, with the corresponding 90\% C.L. critical values.}
\label{fig:FC_seedings}
\end{figure}

Figure~\ref{fig:FC_seedings} compares the $\Delta\chi^2$ distributions obtained using initialization only at the no-oscillation point and using the multi-stage procedure. Because the no-oscillation point lies within the fitted parameter space, the global minimum must satisfy $\chi^2_{\mathrm{BF}}\leq\chi^2_{\mathrm{NULL}}$. Negative values of $\Delta\chi^2=\chi^2_{\mathrm{NULL}}-\chi^2_{\mathrm{BF}}$ therefore indicate failure of the numerical minimization to locate the global minimum. Such values occur in 984 of 10{,}000 pseudo-experiments when fitting from the no-oscillation seed alone, but in none of the 1{,}000{,}000 pseudo-experiments evaluated with the multi-stage procedure.

\begin{figure*}
\centering
  \includegraphics[width=0.48\textwidth]{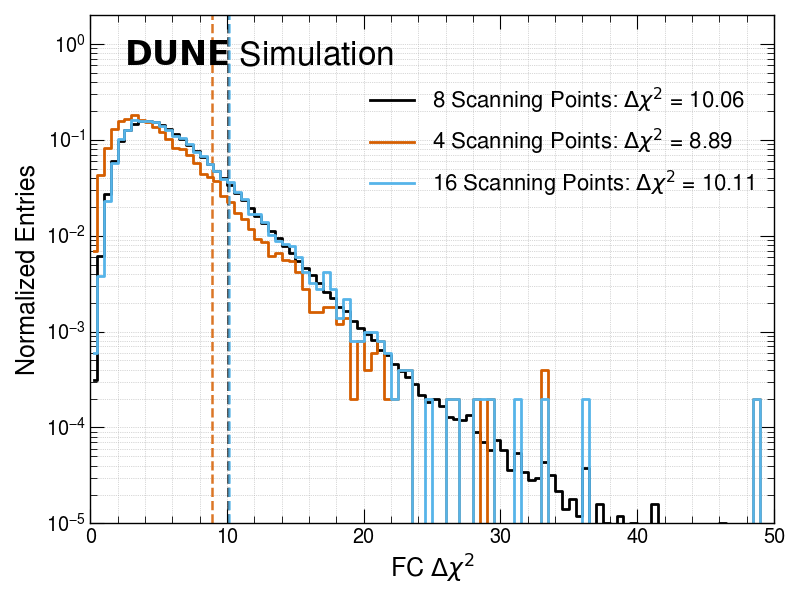}\hfill
  \includegraphics[width=0.48\textwidth]{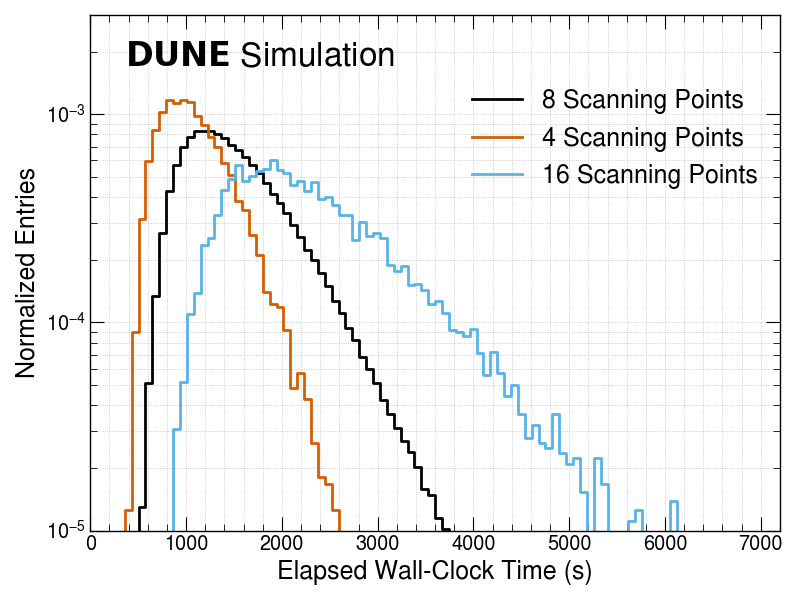}
\caption{Feldman--Cousins $\Delta\chi^2$ distributions (left) and fitting elapsed times (right) for different numbers of scan points along the $\Delta m^2$ axis. The corresponding 90\% C.L. critical $\Delta\chi^2$ values are indicated in the left panel.}
\label{fig:FC_nPointsDm2}
\end{figure*}

We also examine the number of $\Delta m^2$ seeds used in the intermediate stage. Figure~\ref{fig:FC_nPointsDm2} compares configurations with 4, 8, and 16 scan points. The four-point configuration produces a lower critical $\Delta\chi^2$, indicating insufficient sampling of the local-minimum structure, whereas the 8- and 16-point configurations give nearly identical distributions. Because increasing from 8 to 16 points substantially increases the computation time with negligible change in the result, eight points are used in the nominal fit.

\begin{figure*}
\centering
  \includegraphics[width=0.48\textwidth]{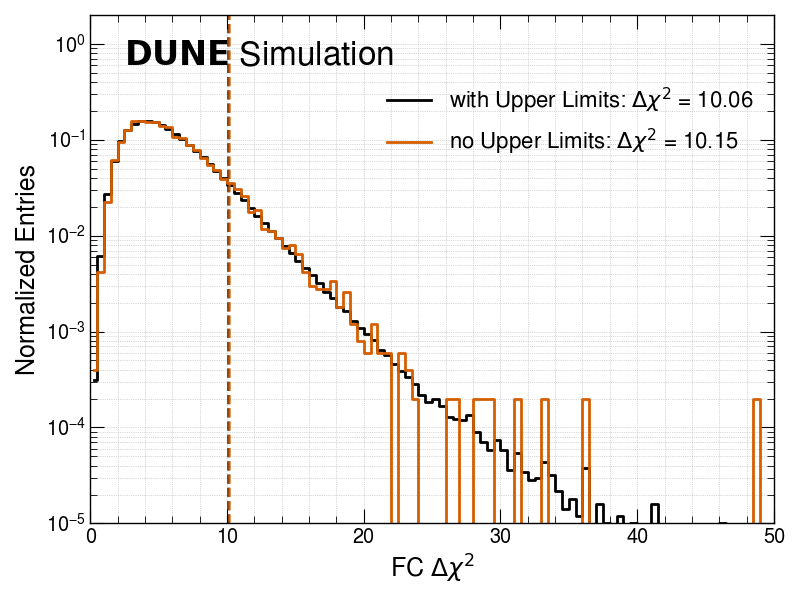}\hfill
  \includegraphics[width=0.48\textwidth]{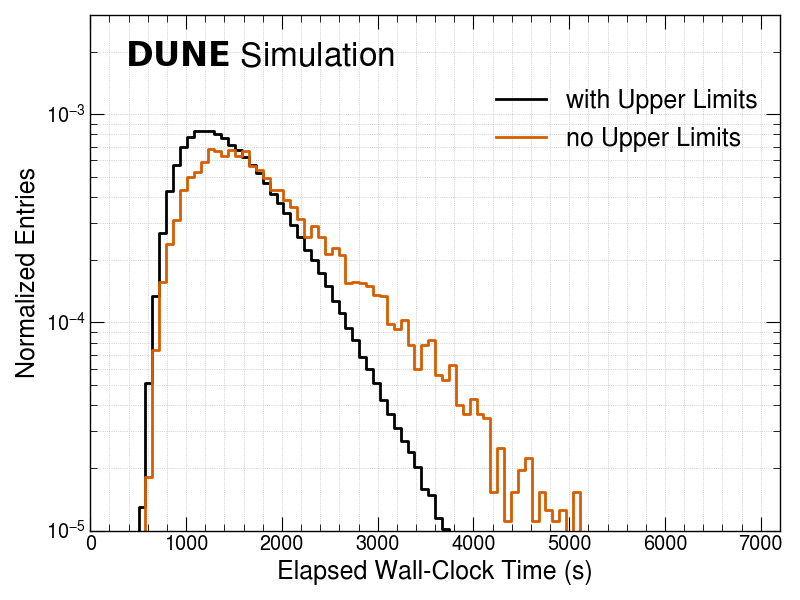}
\caption{Feldman--Cousins $\Delta\chi^2$ distributions (left) and fitting elapsed times (right) for different upper bounds on the mixing parameters. The corresponding 90\% C.L. critical $\Delta\chi^2$ values are indicated in the left panel.}
\label{fig:FC_upperLimits}
\end{figure*}

Finally, we test the upper bounds on the three mixing parameters in Eq.~\ref{eq:OscPara_Ranges_dchi2Fit}. Replacing the nominal bounds with an upper bound of 1 for each parameter leaves the $\Delta\chi^2$ distribution essentially unchanged while substantially increasing the computation time, as shown in Fig.~\ref{fig:FC_upperLimits}. The nominal bounds are therefore retained.

These studies demonstrate that the three-stage procedure provides stable best-fit results with substantially lower computational cost than more exhaustive configurations and is therefore used for the FC ensemble presented in Sec.~\ref{sec:results}.

\section{Sensitivity Contours}
\label{apdx:contours}

This appendix presents additional fixed-parameter slices of the sterile-neutrino sensitivity described in Sec.~\ref{sec:results}. To illustrate the dependence of the sensitivity on the oscillation parameters, two-dimensional sensitivity contours are shown for several representative fixed values of the parameters not displayed.

\begin{figure*}
\centering
      \includegraphics[width=0.33\textwidth]{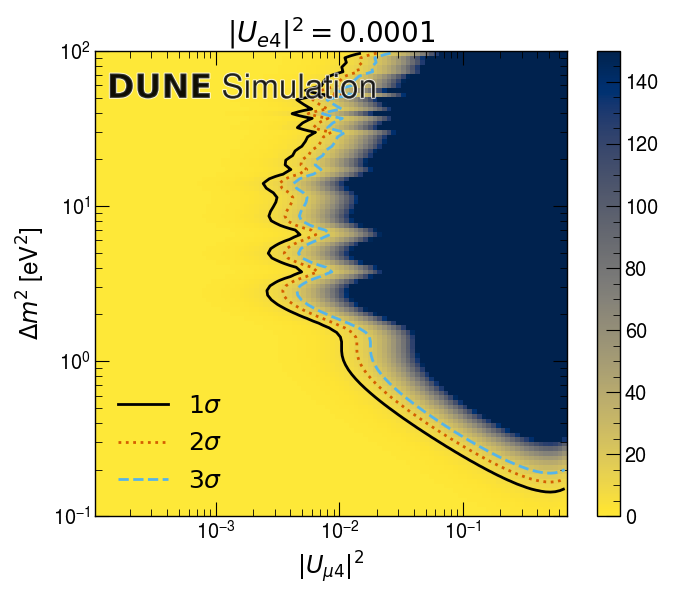}%
      \includegraphics[width=0.33\textwidth]{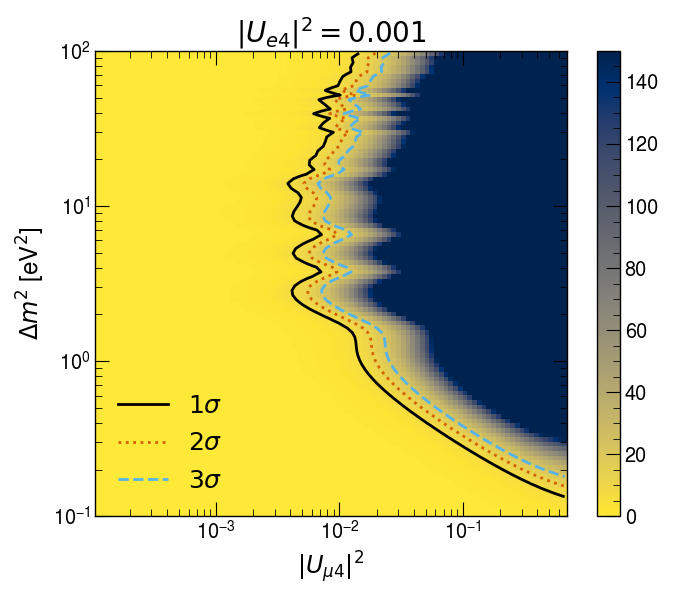}%
      \includegraphics[width=0.33\textwidth]{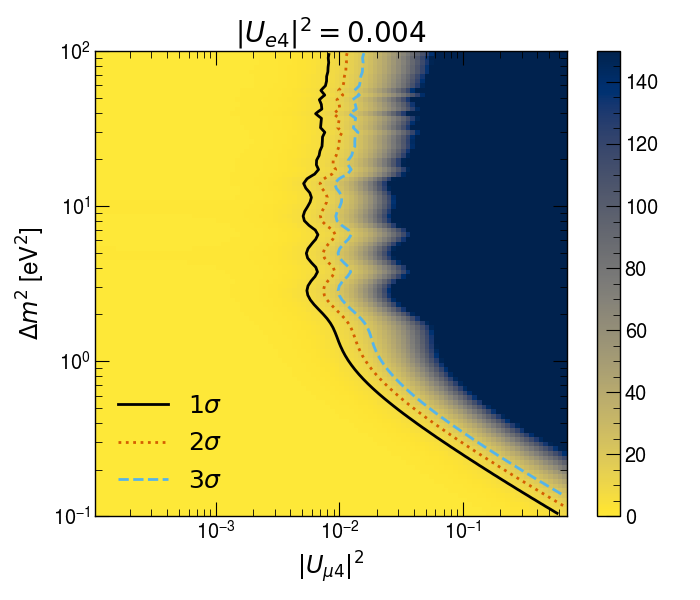}\\
      \includegraphics[width=0.33\textwidth]{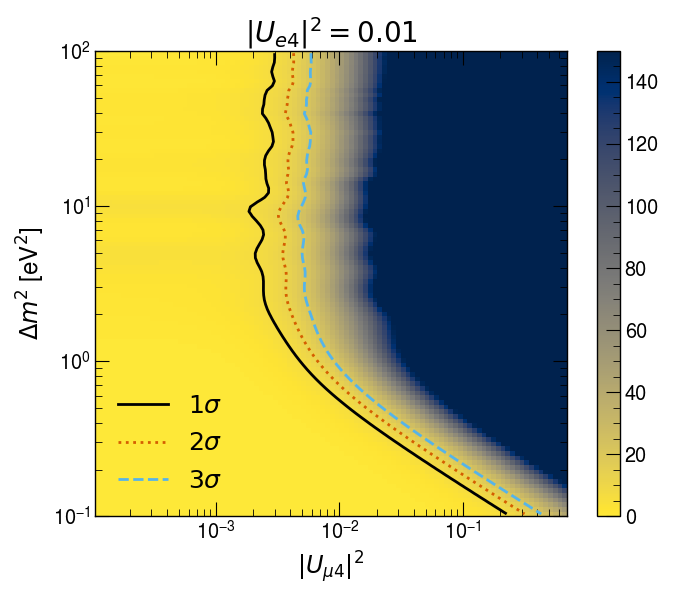}%
      \includegraphics[width=0.33\textwidth]{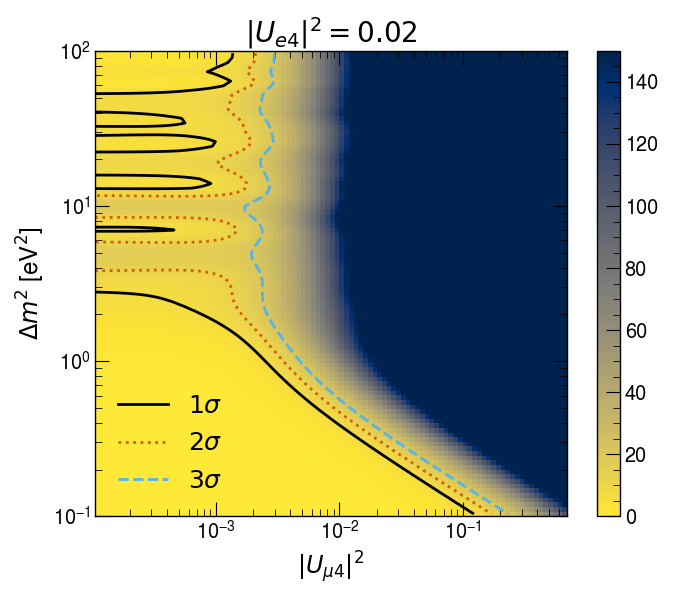}%
      \includegraphics[width=0.33\textwidth]{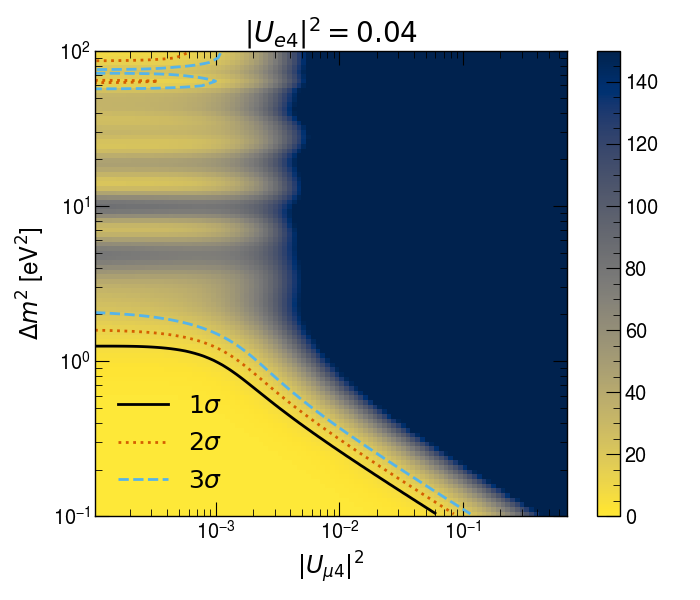}\\
      \includegraphics[width=0.33\textwidth]{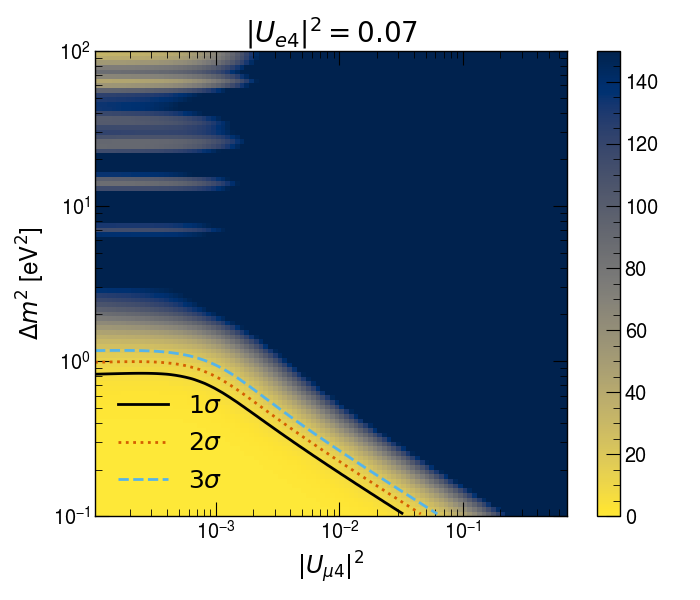}%
      \includegraphics[width=0.33\textwidth]{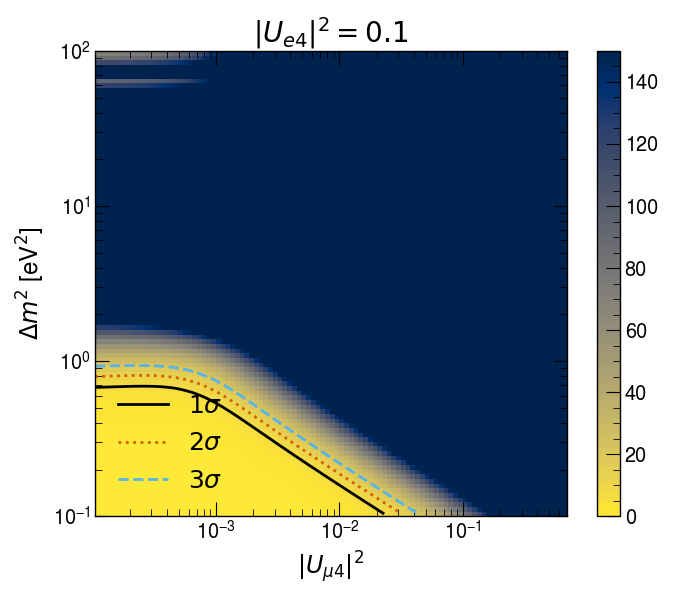}%
      \includegraphics[width=0.33\textwidth]{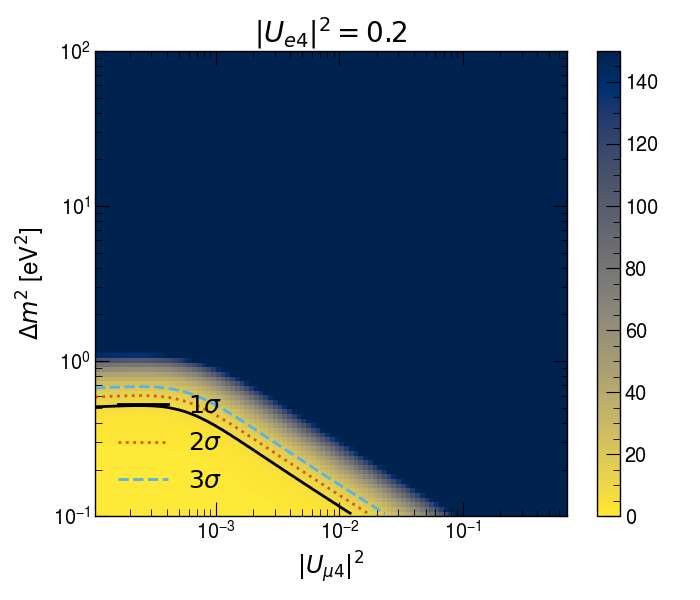}\\
\caption{DUNE sensitivity in the $(\Delta m^2,|U_{\mu4}|^2)$ plane for nine fixed values of $|U_{e4}|^2$, with $|U_{\tau4}|^2=0$. The contours correspond to the $1\sigma$, $2\sigma$, and $3\sigma$ confidence levels.}
\label{fig:contours_fixedUe42}
\end{figure*}

\begin{figure*}
\centering
  \includegraphics[width=0.33\textwidth]{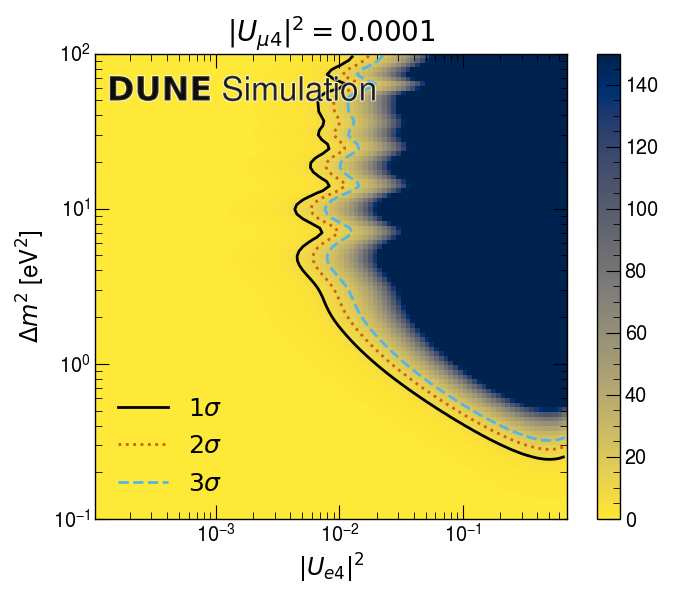}%
  \includegraphics[width=0.33\textwidth]{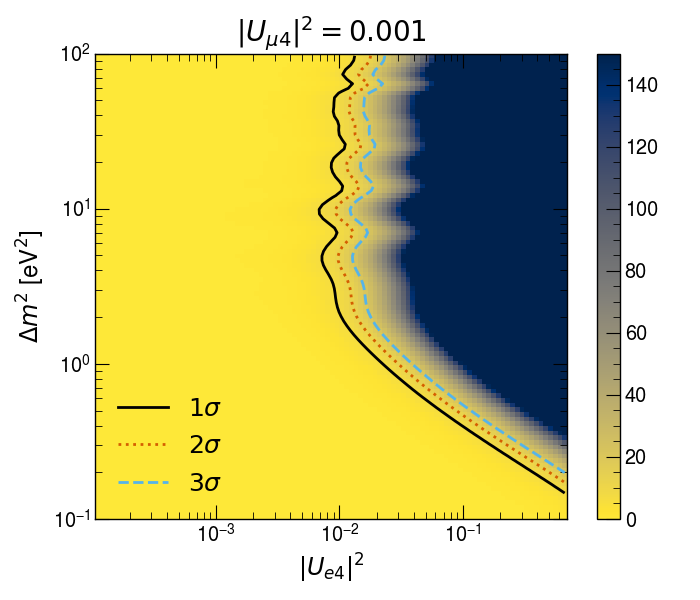}%
  \includegraphics[width=0.33\textwidth]{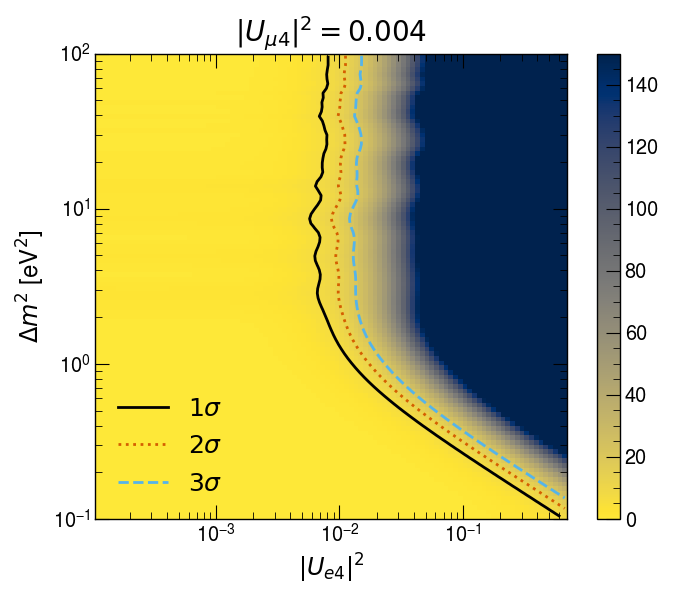}\\
  \includegraphics[width=0.33\textwidth]{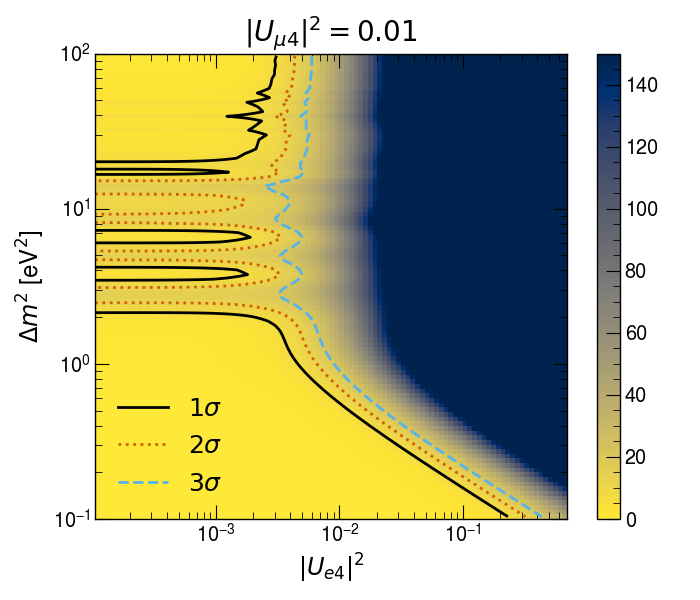}%
  \includegraphics[width=0.33\textwidth]{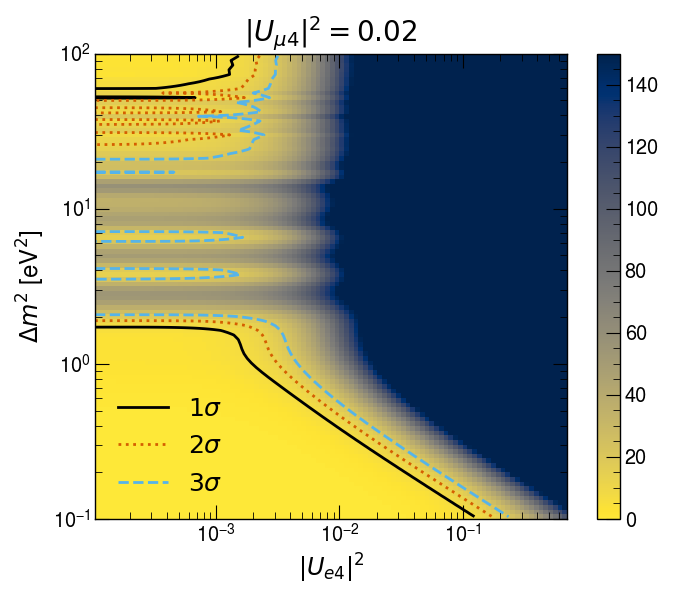}%
  \includegraphics[width=0.33\textwidth]{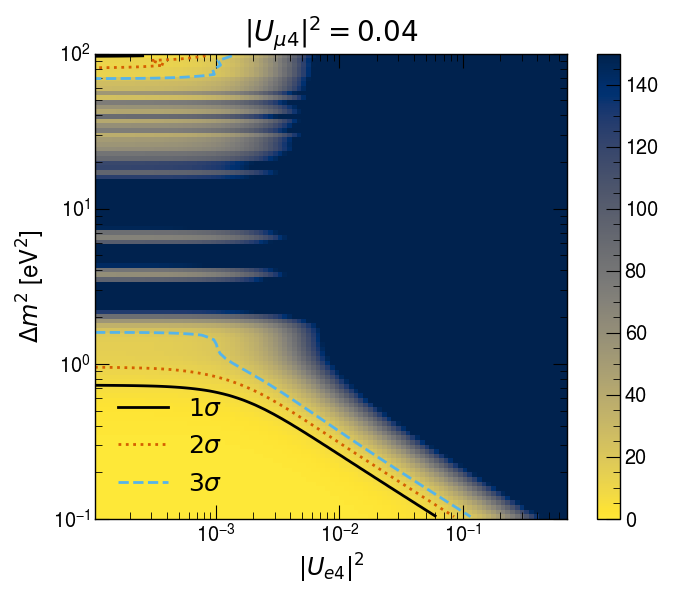}\\
  \includegraphics[width=0.33\textwidth]{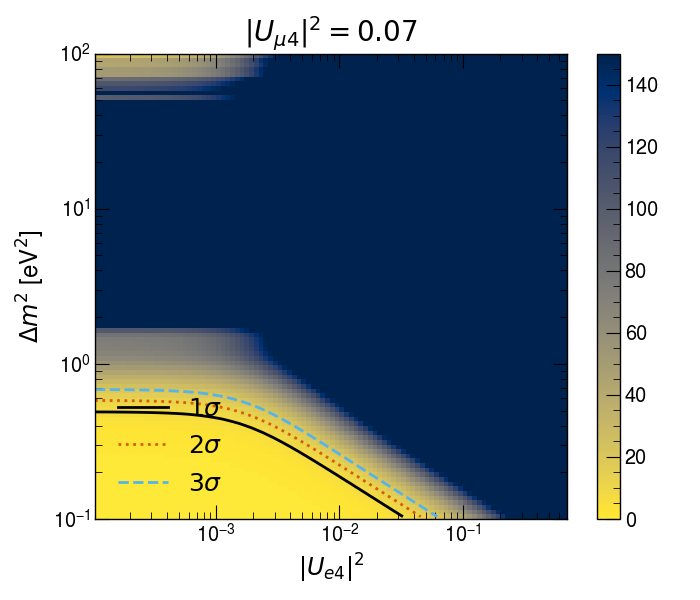}%
  \includegraphics[width=0.33\textwidth]{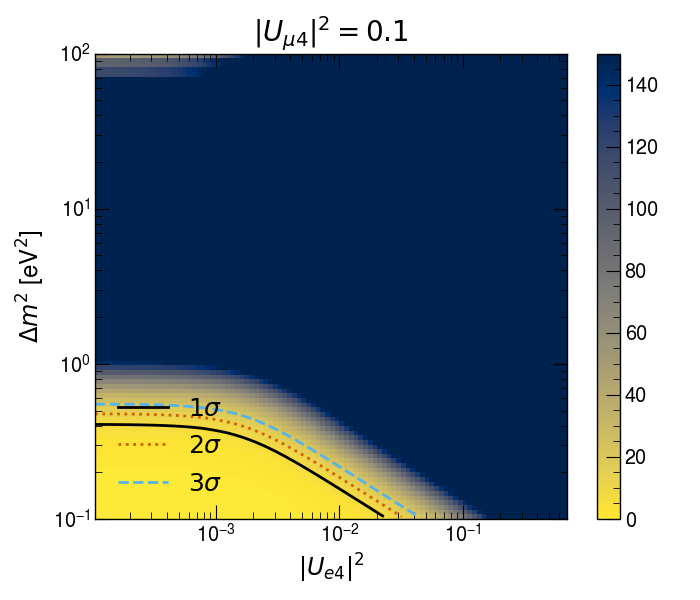}%
  \includegraphics[width=0.33\textwidth]{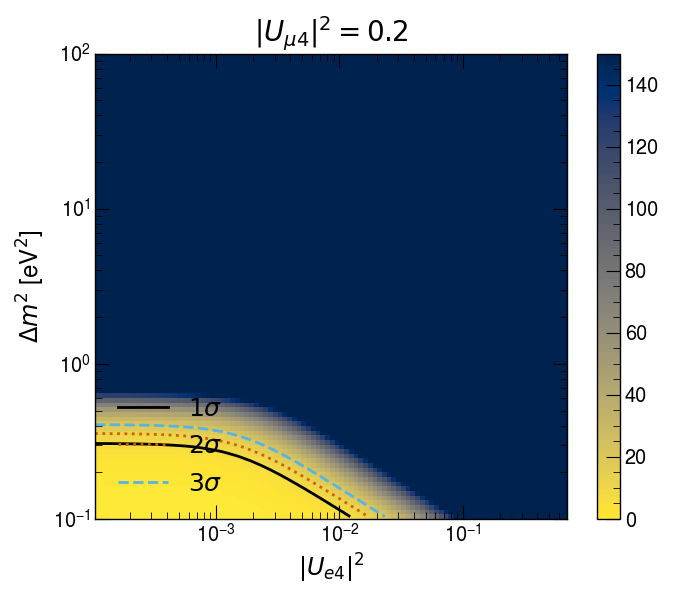}\\
\caption{DUNE sensitivity in the $(\Delta m^2,|U_{e4}|^2)$ plane for nine fixed values of $|U_{\mu4}|^2$, with $|U_{\tau4}|^2=0$. The contours correspond to the $1\sigma$, $2\sigma$, and $3\sigma$ confidence levels.}
\label{fig:contours_fixedUm42}
\end{figure*}

\begin{figure*}
\centering
  \includegraphics[width=0.33\textwidth]{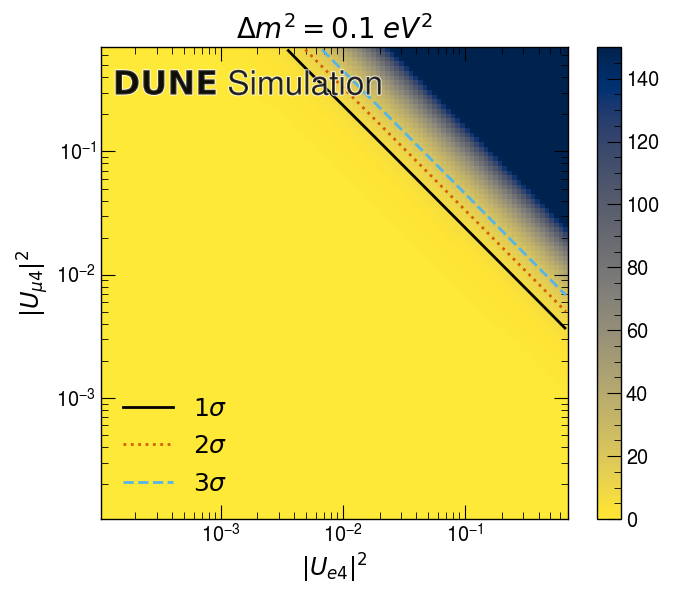}%
  \includegraphics[width=0.33\textwidth]{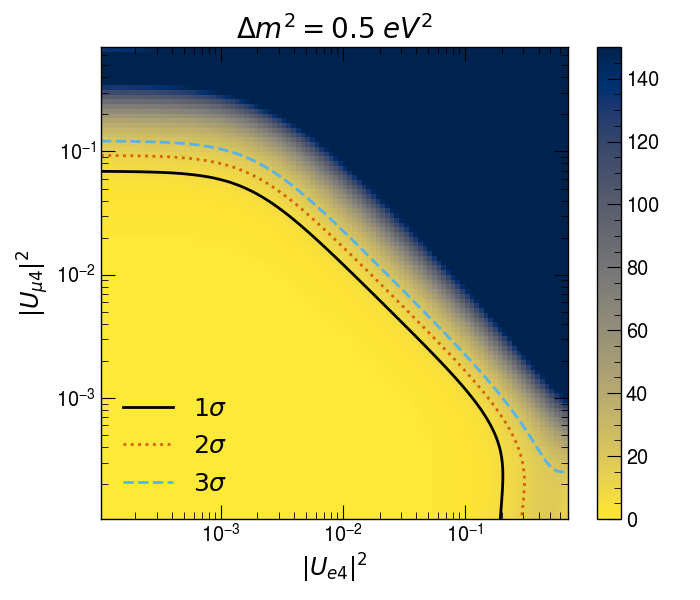}%
  \includegraphics[width=0.33\textwidth]{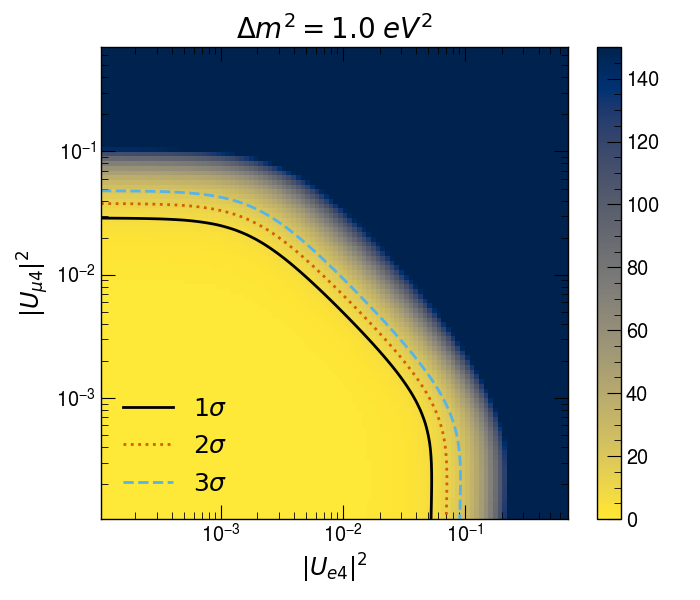}\\
  \includegraphics[width=0.33\textwidth]{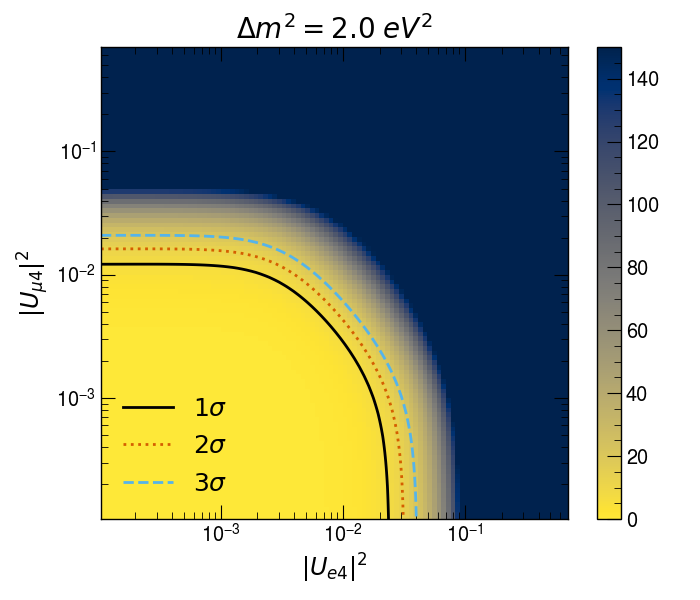}%
  \includegraphics[width=0.33\textwidth]{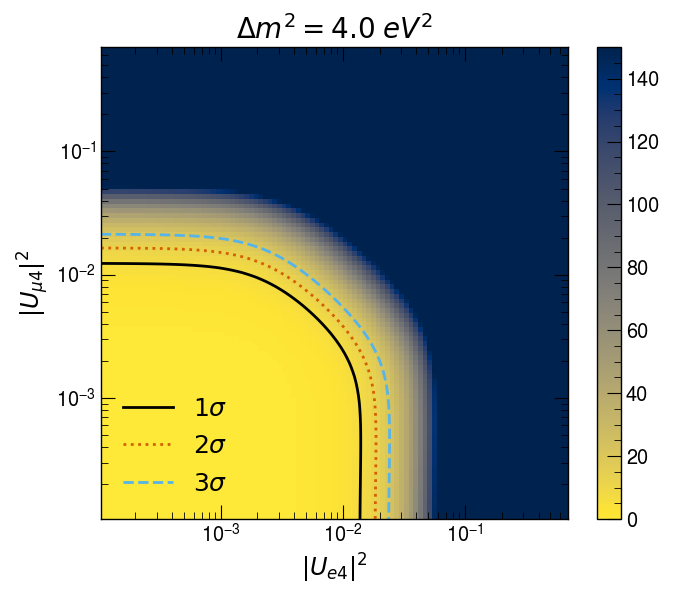}%
  \includegraphics[width=0.33\textwidth]{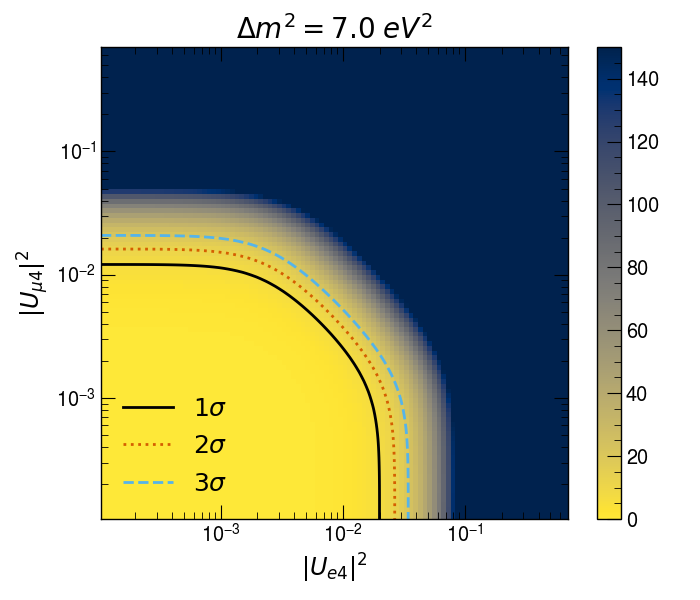}\\
  \includegraphics[width=0.33\textwidth]{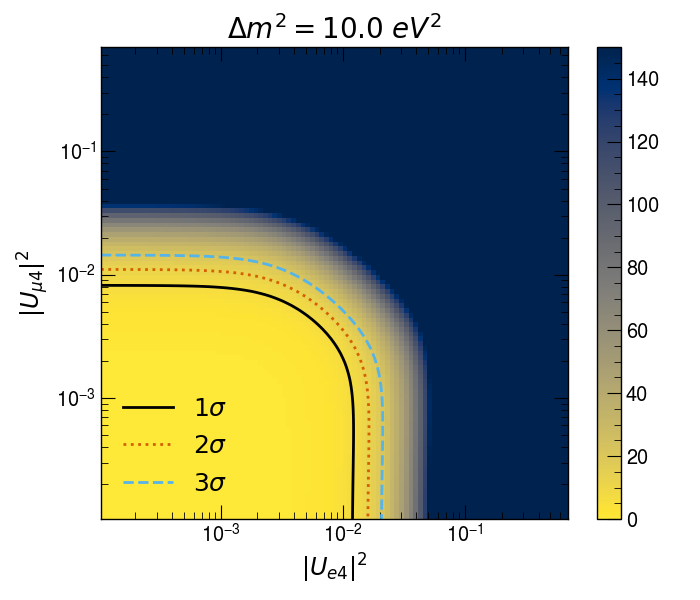}%
  \includegraphics[width=0.33\textwidth]{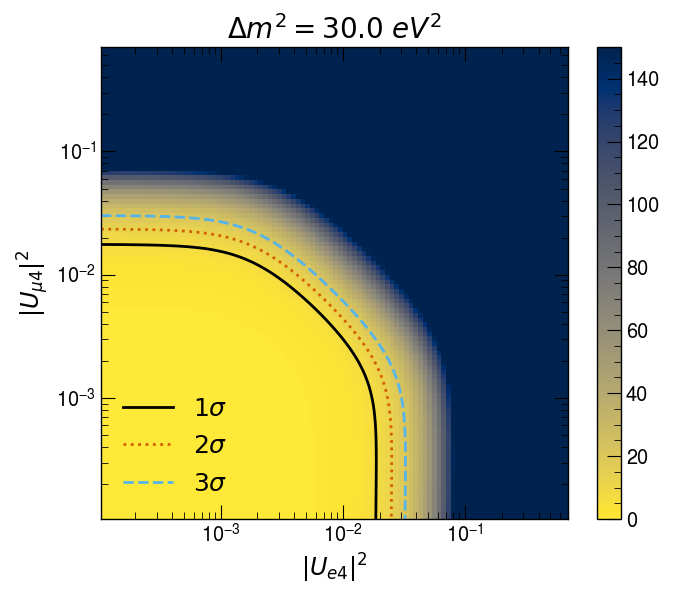}%
  \includegraphics[width=0.33\textwidth]{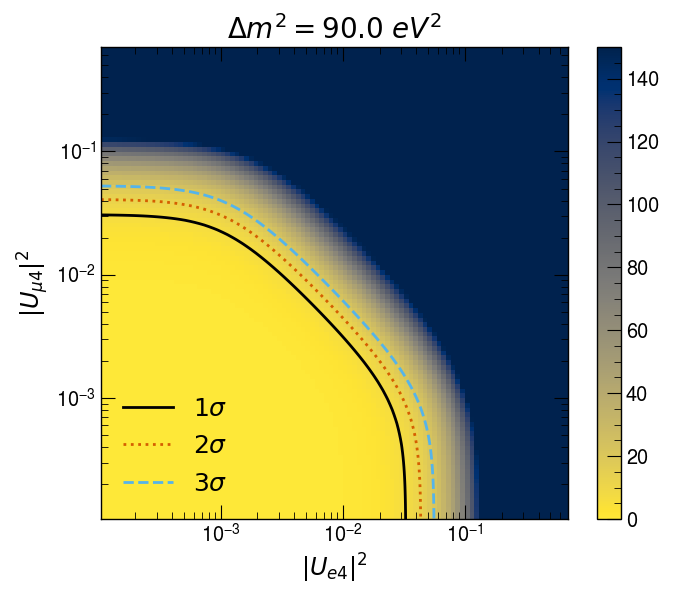}\\
\caption{DUNE sensitivity in the $(|U_{\mu4}|^2,|U_{e4}|^2)$ plane for nine fixed values of $\Delta m^2$, with $|U_{\tau4}|^2=0$. The contours correspond to the $1\sigma$, $2\sigma$, and $3\sigma$ confidence levels.}
\label{fig:contours_fixedDm2}
\end{figure*}

Figures~\ref{fig:contours_fixedUe42} and~\ref{fig:contours_fixedUm42} show the sensitivity in the $(\Delta m^2,|U_{\mu4}|^2)$ and $(\Delta m^2,|U_{e4}|^2)$ planes for fixed values of $|U_{e4}|^2$ and $|U_{\mu4}|^2$, respectively. Figure~\ref{fig:contours_fixedDm2} shows the sensitivity in the $(|U_{\mu4}|^2,|U_{e4}|^2)$ plane for fixed values of $\Delta m^2$. In all cases, $|U_{\tau4}|^2=0$. These slices illustrate the correlations among the oscillation parameters and the evolution of the expected sensitivity across the explored mass-squared-splitting range.

\end{document}